%% file: ms.tex
\documentclass[iop]{emulateapj}
\usepackage{float}
\usepackage{perpage}
\usepackage{morefloats}

\begin{document}

\def\msun   {{$M_{\odot}$}}
\def\etal   {{et al. }}
\def\teff   {{$T_{\rm eff}$}}

\title{A Tale of Two Anomalies: Depletion, Dispersion, and the Connection Between the Stellar Lithium Spread and Inflated Radii on the Pre-Main Sequence}

\author{Garrett Somers and Marc H. Pinsonneault}

\begin{abstract}
We investigate lithium depletion in standard stellar models (SSMs) and main sequence (MS) open clusters, and explore the origin of the Li dispersion in young, cool stars of equal mass, age and composition. We first demonstrate that SSMs accurately predict the Li abundances of solar analogs at the zero-age main sequence (ZAMS) within theoretical uncertainties. We then measure the rate of MS Li depletion by removing the [Fe/H]-dependent ZAMS Li pattern from three well-studied clusters, and comparing the detrended data. MS depletion is found to be mass dependent, in the sense of more depletion at low mass.  A dispersion in Li abundance at fixed \teff\ is nearly universal, and sets in by $\sim$200 Myr. We discuss mass and age dispersion trends, and the pattern is mixed. We argue that metallicity impacts the ZAMS Li pattern, in agreement with theoretical expectations but contrary to the findings of some previous studies, and suggest Li as a test of cluster metallicity. Finally, we argue that a radius dispersion in stars of fixed mass and age, during the epoch of pre-MS Li destruction, is responsible for the spread in Li abundances and the correlation between rotation and Li in young cool stars, most well known in the Pleiades. We calculate stellar models, inflated to match observed radius anomalies in magnetically active systems, and the resulting range of Li abundances reproduces the observed patterns of young clusters. We discuss ramifications for pre-MS evolutionary tracks and age measurements of young clusters, and suggest an observational test.
\end{abstract}

\section{Introduction}

The lithium content of stars is an important quantity for a variety of astrophysical measurements. First, Li is a powerful tracer of mixing in stars. It is destroyed efficiently at $\sim$ 2.5 million K ($T_{LB}$), and as a result can only survive in the outer layers of stars. When a star is undergoing deep mixing, Li-depleted stellar material is transported from depths that surpass $T_{LB}$ to the surface, diluting the observed Li abundance (A(Li) = 12 + [Li/H]). The evolution of Li in stellar atmospheres is therefore a direct consequence of mixing, which in turn affects the surface composition and main-sequence (MS) lifetimes of stars across the stellar mass function (e.g. Pinsonneault 1997). Second, the evolution of Li abundances on the pre-MS and MS contains information about stellar ages (e.g. Jeffries 2000), and may inform our knowledge about their rotational history (Pinsonneault 1990). Finally, the Li content of the universe is a strong prediction of big bang nucleosynthesis (Boesgaard \& Steigman 1985), and can be probed by measuring the initial Li abundance of very old stars in the Galaxy (e.g. Spite \& Spite 1982; Cyburt \etal 2008).

Interiors models make strong predictions about the timescales of mixing, and thus the evolution of surface Li, as a function of mass, age, and composition. Because the sole mixing mechanism in standard stellar models (SSMs) is convection, the surface Li abundance of a star is predicted to decrease only when the temperature at the base of the surface convection zone ($T_{BCZ}$) is greater than $T_{LB}$. This occurs on the pre-MS for stars of mass 0.5\msun\ - 1.3\msun\ at solar metallicity, but not on the MS (Iben 1965). The general theoretical expectations of pre-MS Li depletion in this mass range are well established, and a qualitative explanation follows. Pre-MS stars have deep convective envelopes, and heat up as they contract, causing $T_{BCZ}$ to eventually surpasses $T_{LB}$ and inducing Li depletion. In fully convective stars (FCSs; \msun\ $<$ 0.35\msun), Li is completely destroyed in only a few Myr once $T_{LB}$ is reached. This occurs earlier for higher mass objects, creating a boundary between FCSs which have depleted Li, and lower mass FCSs which remain Li-rich. The location of this boundary is age dependent, permitting its use as a method for dating clusters; this is called the lithium depletion boundary (LDB) technique (Basri \etal 1996; Bildsten \etal 1997). Stars less massive than $\sim$0.06\msun\ never reach $T_{LB}$ in their interior, and so retain their initial Li abundance forever.

\begin{figure*}
\includegraphics[width=7.0in]{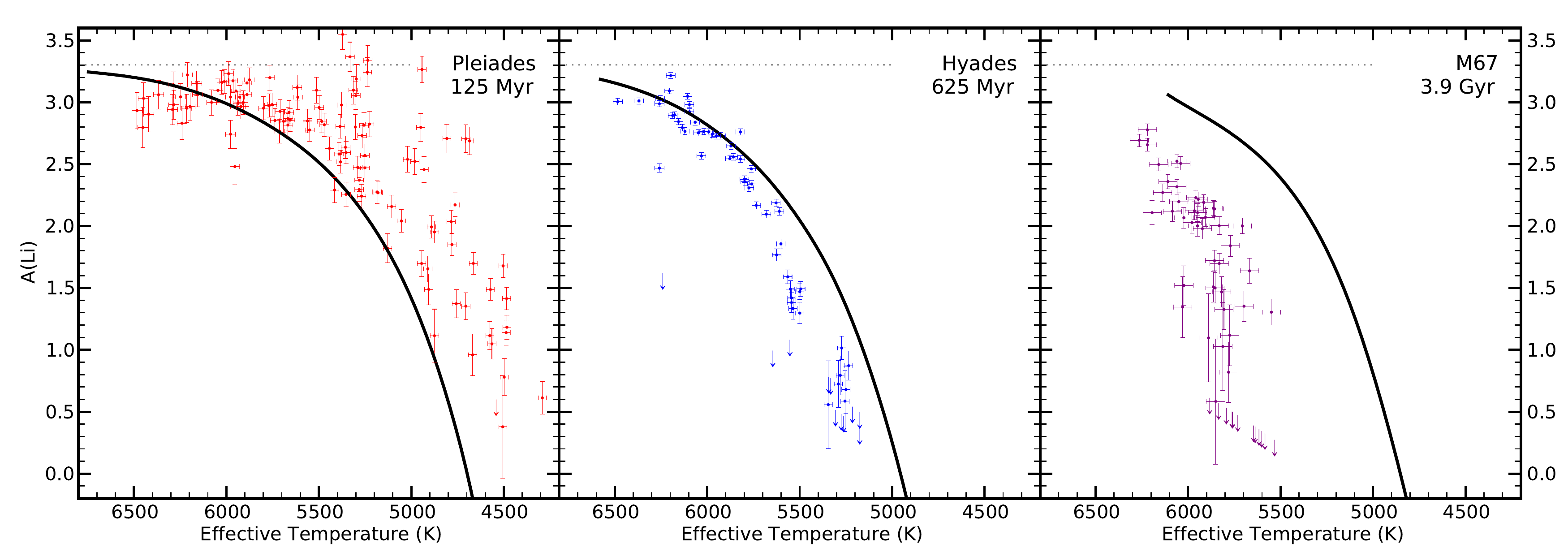}
\caption{Lithium data for the Pleiades, Hyades, and M67 are shown alongside standard stellar model lithium patterns calculated for their respective cluster parameters. The solid lines represent the SSM prediction for each cluster, and the dashed line in the top right panel represents the M67 LDP at the age of the Pleiades, demonstrating the MS evolution of LDPs in the standard model. This is almost entirely due to the increasing \teff\ on the MS, but also includes minor gravitational settling. Arrow denote upper limits. }
\end{figure*}

For stars more massive than 0.35\msun, the convective envelope begins to retreat on the late pre-MS; this causes $T_{BCZ}$ to once again cool below $T_{LB}$, and terminates Li depletion. Lower mass stars take longer to reach $T_{LB}$, but remain in the burning phase for longer. This results in greater depletion factors in these objects. Li also burns more rapidly in metal-rich stars, as the higher resulting opacity deepens the convective envelope, increasing $T_{BCZ}$. The result is a strongly mass and metallicity dependent lithium depletion pattern (LDP) on the zero-age MS (ZAMS) with no dispersion at fixed mass, in accordance with the Vogt-Russell theorem.

For a solar metallicity cluster, Li has been completely destroyed in stars $\lesssim$0.6\msun\ by the zero-age MS (ZAMS), but only about $\sim$ 20\% has been destroyed for 1.3\msun. Once on the MS, the Li depletion zone resides inside the radiative core, and SSMs predict no more mixing-related Li depletion until the MS turn-off. The LDP will continue to change on Gyr timescales, but this is not a result of mixing, and is instead due to the evolving temperature of MS stars, and gravitational settling in late F and early G dwarfs. 

Many Li data sets exist in the literature, but the important features of the general observational picture can be accurately represented by three well-studied clusters. The top panel of Fig. 1 shows empirical Li data for the Pleiades, Hyades, and M67, alongside SSM LDPs calculated for their respective cluster parameters (solid black lines; see $\S$2.1-2.3 and Table 1 for details). As illustrated by this figure, there are several inconsistencies between standard predictions and observed cluster patterns:

\textit{(i)} The median abundance is over-predicted by a few tenths of a dex above 6100K, and under-predicted by a few tenths of dex to greater than an order-of-magnitude below 6100K. While the general trend of greater depletion in cooler stars is accurately predicted, this SSM marginally fails to predict the median of solar analogs, and catastrophically fails to predict the median of cool stars.

\textit{(ii)} There is a significant scatter in surface Li in the cool (\teff\ $<$ 5500K) Pleiads, where the width of this distribution can be as large as a factor of 100. This implies that additional physical parameters, which can vary between equal-\teff\ stars, affects pre-MS depletion by \textit{orders-of-magnitude} in this temperature regime. Furthermore, the fastest rotating cool Pleiads are on average the most Li-rich stars at their respective temperatures (Soderblom \etal 1993a; S93 hereafter). This strongly implies a connection between rotation and Li depletion.

\textit{(iii)} Fig. 1 shows a strong temporal evolution of the median Li abundance at all masses, occurring on a much shorter timescale than MS evolution. From left to right, SSMs first under-predict, then over-predict, then greatly over-predict the median pattern at 100 Myr, 600 Myr, and 4 Gyr. This implies that mechanisms other than convection are able to mix stellar material on the MS. Furthermore, the rate of depletion decreases at advanced ages (e.g. Sestito \& Randich 2005; SR05 hereafter).

\textit{(iv)} By the age of M67, a large Li scatter has developed in solar analogs. Such a scatter is not present in the Pleiades, and so likely develops during the MS. This demonstrates that the Li abundance of a given star depends on factors other than just mass, age, and metallicity. Additional physical parameters that vary between equal-mass stars must induce this relative depletion.

These inconsistencies demonstrate that, in contrast to standard theory, MS LDPs are the product of two distinct processes: a pre-MS process that imparts a strongly mass-dependent median trend with a variable width, and a longer timescale processes on the MS that causes the median and dispersion to evolve with time. In this paper we will deal, in some part, with both of these processes. For the former, we will validate the accuracy of SSMs in warm stars, argue the importance of metallicity in shaping ZAMS Li patterns, and propose an explanation for the dispersion in cool stars. For the latter, we will produce an empirical measure of MS Li depletion that has been corrected for metallicity effects, and study the timescales of the emergence of Li dispersion on the MS. These can be used both to anchor mixing calculations on empirical data and to place constraints on proposed mixing mechanisms. In all cases, we will refer to the difference between SSM predictions, and the empirical abundance of stars, as the \textit{lithium anomaly}.

Li depletion on the MS has be known about for several decades (e.g. Herbig 1965; Zappala \etal 1972; Balachandran 1995; Pinsonneault 1997; SR05), but the mechanism, or mechanisms, responsible have yet to be definitively established. Suspects include mixing driven by rotation and angular momentum (AM) loss (Pinsonneault \etal 1989; Zahn 1992; Chaboyer \etal 1995), mixing driven by internal gravity waves (Press 1981; Montalban \& Schatzman 2000), dilution of the envelope through mass loss (Swenson \& Faulkner 1992), and microscopic diffusion (Richer \& Michaud 1993). Rotational mixing is a particularly promising explanation for two reasons. First, stellar rotation slows over time (Skumanich 1972), naturally explaining the decay of Li depletion rates described in \textit{(iii)}. Second, stellar rotation rates show a large dispersion at ZAMS (e.g. Stauffer \etal 1984), providing the necessary variant between stars of equal mass described in \textit{(iv)}. We will present updated models of rotationally-induced mixing in a forthcoming paper (Somers \& Pinsonneault 2014b, in prep; Paper II hereafter). However, before we can perform precision tests, accurate measurements of this depletion must be obtained.

\input{Tab01.txt}

While previous authors have measured MS Li depletion by comparing the abundances of different-aged MS clusters (e.g. SR05), these studies have not accounted for one crucial element: higher metallicity stars are expected to deplete greater amounts of Li during the pre-MS. This effect can severely bias comparisons in absolute space, as the ZAMS abundance at a given \teff\ may differ between clusters by up to an order of magnitude ($\S$3.1). To address this, we present a novel method in $\S$4 for quantifying the lithium anomaly that develops on the MS. We will argue that the MS anomaly signal can be isolated from an empirical MS Li pattern by subtracting a SSM LDP from the data. This removes the relative, [Fe/H]-dependent pre-MS depletion signal, leaving behind the depletion induced by non-standard MS mixing. Although some authors have claimed that this metallicity effect is not supported by observational evidence, we present a case in $\S$6.1 that composition is indeed central in shaping ZAMS Li patterns.

The efficacy of this method hinges on the quality of SSM LDP predictions, which we know from \textit{(i)} can be poor. Therefore, we must first reconcile our theoretical predictions with the data. To do this, we explore the possibility that errors in model input physics account for this discrepancy. The extreme sensitivity of the rate of Li burning to the surrounding temperature ($\propto T^{20}$; Bildsten \etal 1997) implies that minute changes in $T_{BCZ}$ on the pre-MS may have large effects on the magnitude of Li depletion predicted in SSMs. $T_{BCZ}$ in a stellar model may be affected by the assumed physics, so in order to validate this method, we first address the following question: can SSMs accurately predict the magnitude of pre-MS Li depletion within the errors of our adopted input physics? If this is so, we can use empirical data to calibrate our SSMs, and produce accurate predictions of pre-MS Li depletion ($\S$4).

We also investigate one of the key outstanding problems in our understanding of pre-MS depletion: the cool star Li dispersion in the Pleiades and other young systems (S93). It is unlikely that long timescale mechanisms such as rotationally induced mixing are responsible for this dispersion, given its early onset. However, the rotation-Li correlation in young Pleiads, described in \textit{(ii)}, suggests either a causal or corollary relationship between rotation and the efficiency of early convective depletion. Another effect known to correlate with rotation in stars is the so-called $\textit{radius anomaly}$. This describes a discrepancy of $\sim$5-15\% the observed radii of some stars, and their SSM predictions. The radius anomaly has been observed in detached eclipsing binaries (DEBs; Popper 1997; Torres \& Ribas 2002; Ribas 2003; L\'opez-Morales \& Ribas 2005, L\'opez-Morales 2007; Torres \etal 2010; Kraus \etal 2011; Irwin \etal 2011; Feiden \& Chaboyer 2012, Stassun \etal 2012), and may be present in interferometric radius measurements of single field stars (Berger \etal 2006; Boyajian \etal 2008; Boyajian \etal 2012; but, see Demory \etal 2009). This effect has also been reported in solar analogs (e.g. Clausen \etal 2009), and in pre-MS stars (Stassun \etal 2006; Stassun \etal 2007). The latter authors discovered a brown dwarf binary system where the more massive object has a lower \teff. Temperature anomalies correlate with radius anomalies, definitively showing that coeval objects can be differentially affected by non-standard stellar parameters. This radius effect may be caused by accretion from a circumstellar disk (Palla \& Stahler 1992), unidentified sources of opacity (Berger \etal 2006), or inhibition of convection by magnetic activity (Mullan \& MacDonald 2001; Chabrier \etal 2007; Morales \etal 2008; MacDonald \& Mullan 2012; Feiden \& Chaboyer 2013). Although the two fields of radius anomalies and open cluster Li data have not previously intersected, we reveal a surprising connection between them in $\S$6.2. 

\input{Tab02.txt}

The rest of the text is organized as follows. In $\S$2.1, we describe the open cluster Li data sets used for our analysis. $\S$2.2 describes the equivalent width and photometric data we use to infer stellar parameters of our benchmark clusters, and the abundance analysis we employ. $\S$2.3 describes the stellar evolution code we used to generate theoretical LDPs, and enumerates the physics in our fiducial calculations. We then begin our exploration of uncertainties inherent to the detrending process. In $\S$3.1, we quantify the impact of [Fe/H] errors in our benchmark clusters, and conclude that extremely well-constrained composition is necessary to accurately predict the pre-MS signal. In $\S$3.2, we perform a systematic study of the effects of various physical inputs on SSM LDPs. We then constrain the input physics in our models with the empirical Pleiades Li pattern, and describe our method of cluster detrending in $\S$4. In $\S$5.1, we measure the lithium anomaly in our benchmark clusters, and compare our results to previous calculations. This measurement will ultimately anchor the mixing included in our rotating stellar models (Paper II). We then revisit the data collected in SR05 with our detrending methodology in $\S$5.2, and explore the evolution of both median Li abundances and Li dispersion along the MS. In $\S$6.1, we argue that composition is an important factor in shaping ZAMS LDPs, validate our anomaly measurements, and suggest Li as a precision test of the metallicity of clusters. Finally, we test an explanation in $\S$6.2 for the cool star spread in the Pleiades, related to the radii of young, low-mass stars. We conclude by summarizing our findings and suggesting directions for future studies in $\S$7.

\section{Methods}

\subsection{Cluster Selection and Parameters}

Cluster Li data will serve several purposes in this paper. First, we will use three benchmark clusters to precisely measure the rate of MS Li depletion ($\S$2.1.1). The Pleiades will be used to calibrated the physics in our theoretical models, by requiring that its SSM Li pattern agree with its empirical Li pattern at the ZAMS, and the relative abundances of the Hyades and M67 will be used to infer the MS lithium anomaly, by comparing their detrended patterns to those expected from the Pleiades calibration, corrected for their metallicities. Second, we will reanalyze the data of SR05, and examine the timescales of Li depletion and dispersion ($\S$5.2). Third, we will argue that metallicity plays an important role in pre-MS Li destruction by comparing similar-aged clusters of dissimilar composition ($\S$6.1). Finally, we will test our pre-MS Li depletion models by comparing their predictions to several additional young clusters ($\S$6.2).

\subsubsection{Benchmark Clusters}

The literature hosts a wealth of Li data for FGK dwarfs in open clusters (see SR05 and refs. therein; Torres \etal 2006; Sacco \etal 2007; Prisinzano \& Randich 2007; Pasquini \etal 2008; Randich \etal 2009; Jeffries \etal 2009; Anthony-Twarog \etal 2009; Cargile \etal 2010; Balachandran \etal 2011; Cummings \etal 2012; Pace \etal 2012; Fran\c{c}ois \etal 2013). Li depletion calculations are exquisitely sensitive to composition, so we restrict our potential choices to clusters with small [Fe/H] errors ($<$ 0.05 dex) to minimize uncertainties. This excludes all but the most well-studied clusters. Furthermore, large Li data sets are required to minimize errors resulting from shot noise, because dispersion is a ubiquitous feature. 

With these considerations in mind, we select the Pleiades, Hyades, and M67 as our benchmark clusters. These are well-suited for this investigation because they are exceptionally well studied, thus minimizing errors associated with photometry, extinction, binarity, membership, and most importantly, composition. The Pleiades is 125 $\pm$ 5 Myr old (Stauffer \etal 1998; see Table 1), making it our near ZAMS cluster, The Hyades is 625 $\pm$ 25 Myr old (Perryman \etal 1998), and M67 is 3.9 $\pm$ 0.6 Gyr old (Castro \etal 2011). This level of temporal coverage allows us to characterize the relative strengths of early and late-time mixing. 

\subsubsection{Additional Clusters}

In $\S$5.2, we will revisit the clusters examined by SR05 through the use of our detrending analysis (see their Table 1 for details). For both the benchmark clusters and those considered in $\S$6.1, we will adopt the data sets described in this text. For the rest, we adopt the photometry and Li equivalent widths (EWs) reported by SR05, and use the analysis techniques described in $\S$2.2. We adopt the cluster reddening, ages, and Fe abundances reported by SR05, except for the following cases, where we have substituted higher resolution metal abundances: [Fe/H] = -0.03 $\pm$ 0.04 for IC 4665 (Shen \etal 2005), [Fe/H] = 0.00 $\pm$ 0.01 for IC 2602 (D'Orazi \etal 2009), [Fe/H] = -0.01 $\pm$ 0.02 for IC 2391 (D'Orazi \etal 2009), [Fe/H] = 0.04 $\pm$ 0.02 for Blanco 1 (Ford \etal 2005), [Fe/H] = 0.01 $\pm$ 0.07 for NGC 2516 (Terndrup \etal 2002) and [Fe/H] = 0.03 $\pm$ 0.02 for NGC 6475 (Villanova \etal 2009).

\begin{figure*}
\includegraphics[width=7.0in]{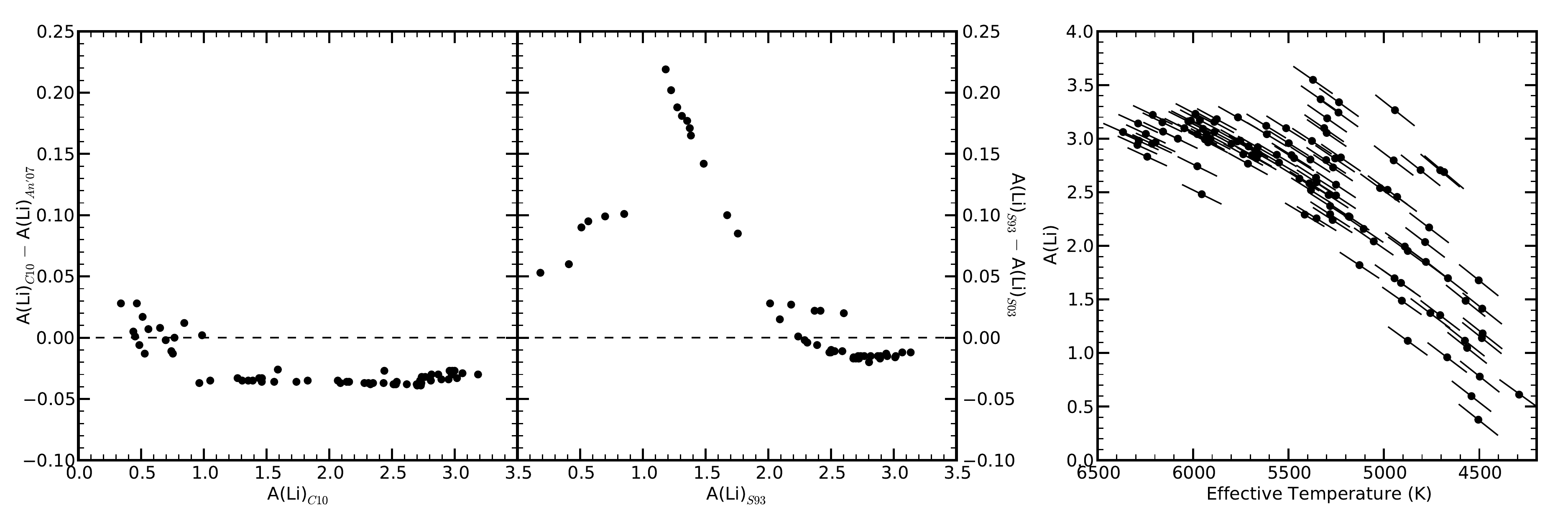}
\caption{Graphic illustrations of the potential systematic errors in our Li abundance analysis (see $\S$2.2 for details). \textit{Left}: Hyades abundances derived with the Casagrande \etal (2010) \teff\ scale versus the offset between that sample and Hyades abundances derived with the An \etal (2007) \teff\ scale. Both use S93 curves of growth. \textit{Center:} Hyades abundances derived with the S93 curves of growth versus the offset between that sample and Hyades abundances derived with the S03 curves of growth. Both use Casagrande \etal (2010) \teff\ scale. \textit{Right:} The effect of \teff\ errors on derived abundances. The line shows the direction and magnitude of data movement resulting from $\pm$100K errors in the data. The correlation between \teff\ and A(Li) causes the points to move diagonally, largely preserving the LDP shape in the cool star regime.}
\end{figure*}

In $\S$6.2, we describe a framework for predicting the evolution of the upper and lower envelopes of the Li dispersion in young systems. As a test of our models, we compare their predictions with the Li patterns of a number of young clusters and associations. These clusters are NGC 2264, $\beta$ Pictoris, IC 2602, NGC 2451 A+B, $\alpha$ Persei, and Blanco 1. Effective temperatures and A(Li)s were taken directly from the literature for these clusters. Ages and [Fe/H]s were obtained from various sources, for calculating their respective model predictions. These sources are listed in Table 2. NGC 2451 A and NGC 2451 B are two different clusters along the same line of sight, but since they appear to have similar ages and compositions, we combine their data into a single set. Each age comes from the LDB technique, except that of NGC 2451 A+B, which comes from fitting isochrones to the MS turn-off, and that of NGC 2264. The age of NGC 2264 is a contentious topic; previous studies place it between 0.1~Myr and 10~Myr (see Dahm 2008 for a thorough discussion), but most authors agree there is a substantial age spread within the cluster population. We adopt the age of $6 \pm 3$~Myr, to roughly bracket the range of literature ages. Each quoted Fe abundance was measured with high-resolution spectroscopy, though we caution that they were not derived uniformly. Furthermore, each author employed their own methodology for deriving effective temperatures and Li abundances. We consider this level of precision acceptable, since these clusters will be used to seek qualitative agreement rather than quantitative rigor.

\subsection{Abundance Analysis}

For each benchmark cluster, we drew $\lambda$6707.8 \ion{Li}{1} EWs, and photometric BV measurements, from various literature sources (see Table 1). The Pleiades EWs and photometry come from S93. Hyades EWs come from Thorburn \etal (1993) and Hyades photometry comes from Johnson \& Knuckles (1955). M67 EWs come from Pasquini \etal (2008) and M67 photometry comes from Montgomery \etal (1993). To maximize the internal consistency of our data sets, we did not merge multiple catalogs of Li EWs or photometry. We applied the reddening corrections referenced in Table 1 to these data, calculated effective temperatures using the BV polynomial fit of Casagrande \etal (2010; C10 hereafter), and derived A(Li)s with the curves of growth (CoG) of S93. These CoG are valid between 4000K, where total depletion on the pre-MS occurs, and 6500K, where significant additional non-standard mixing occurs on short time scales (the lithium dip $-$ Boesgaard \& Tripicco 1986; Balachandran 1995), so we discard stars that lie outside these bounds. This does not affect our conclusions, as our main concern is solar analogs.

The Li absorption line suffers from blending with a nearby \ion{Fe}{1} line located at 6707.4\AA. Although the resolution of the Hyades spectra of Thorburn \etal (1993) was high enough to directly remove the blend, the resolution of the Pleiades and M67 spectra was not. The authors therefore removed the blend contribution using the method suggested by S93: the \ion{Fe}{1} contribution is calculated as EW($\lambda 6707.4$ \ion{Fe}{1}) = $20(B-V)-3$ m\AA, and subtracted from the measured EW. They estimate that this relation is accurate to $3-5$m\AA, significantly smaller than the errors on the Pleiades EWs, but comparable to the M67 EWs. This may impact the inferred abundances of stars with low EWs, for which blends are naturally harder to remove. However, the [Fe/H] of M67 is similar to the cluster this relation was calibrated on (the Pleiades), so the errors are likely on the low end of the range. Furthermore, systematic offsets should not affect comparisons of Pleiades and M67 stars, given that their EWs were obtained with the same deblending process. Finally, this may introduce minor systematic errors between these data and the Hyades sample (see Thorburn \etal 1993 for a discussion).

\begin{figure*}
\includegraphics[width=7.0in]{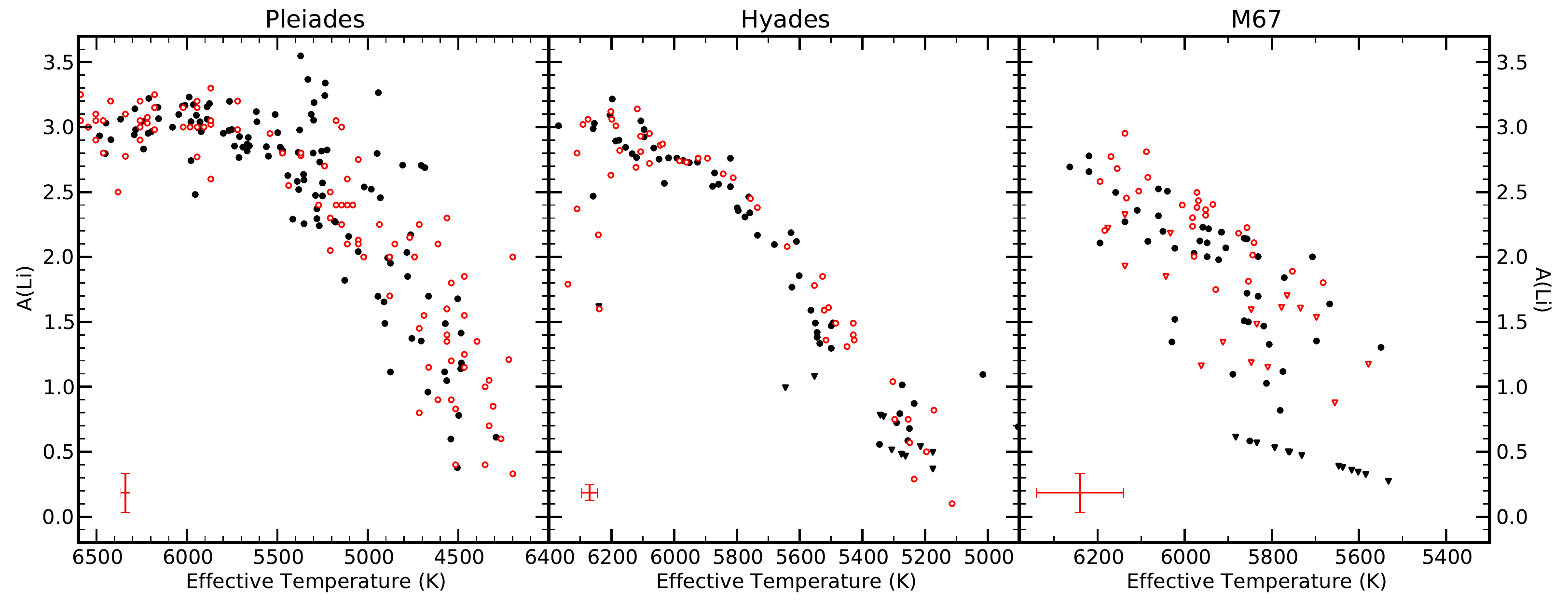}
\caption{A comparison of our benchmark data to alternative data set choices from the literature. Our adopted data, as described in $\S$2.1-2.2, are shown in filled black. The alternative choices are in empty red: Margheim (2007) for the Pleiades, Takeda \etal (2013) for the Hyades, and Jones \etal (1999) for M67. Typical error bars for the comparison samples are shown in the bottom left.}
\end{figure*}

To estimate the accuracy of these abundances, we compare Hyades A(Li)s derived from the C10 \teff\ scale to abundances derived using temperatures obtained with the An \etal (2007) \teff\ scale. This comparison is shown in the left panel of Fig. 2. Both derivations used the CoG of S93, to isolate the effect of \teff\ systematics on our final results. The abundances agreed to better than 0.05 dex for all stars. Errors in \teff\ do not have a strong impact on LDPs, because \teff\ and A(Li) are correlated such that \teff\ errors move stars diagonally along the pattern (Fig. 2; right panel). Though systematic offsets may affect the relative amounts of stars in each \teff\ and A(Li) bin, the median of the pattern is largely unaffected. Next, we combine the Hyades \teff\ and Li EWs presented in Steinhauer (2003; S03 hereafter) with the S93 CoG, and compared the resulting abundances to those presented by S03, who used his own CoG. This is illustrated in the center panel of Fig. 2. There is good agreement for stars with A(Li) $\gtrsim$ 2, but the derived abundances for Li-poor stars differ by up to 0.22 dex between the CoGs. This is a potentially significant systematic error source, so we perform the analysis with both CoGs and compare the results in $\S$5.1. 

Finally, we compare the derived data for each cluster to alternative abundances patterns from the literature. This is shown in Fig. 3. The black points represent the data we use in this paper, as described above, and the red points represent temperatures and abundances taken directly from literature sources. The comparison data were drawn from Margheim (2007) for the Pleiades, Takeda \etal (2013) for the Hyades, and SR05 for M67. The first two authors derived Li parameters from new spectroscopy of their samples, and SR05 used the M67 \ion{Li}{1} EWs of Jones \etal (1999), a distinct data set from the one employed in our analysis. This figure demonstrates a key point: a consistent analysis method is crucial for controlling systematic effects. Non-uniform parameter derivation leads to systematic offsets between samples; for example, our data is shifted, on average, blue-ward compared to the alternative samples for the Pleiades and the Hyades, and red-ward compared to the alternative M67. However, when placed on the same temperature scale, and the same CoG employed, much of this systematic jitter is removed. Ultimately, our Pleiades and Hyades sets are quite similar to the alternative choices, particularly in the 5500-6100K range, where we desire the cleanest sample ($\S$4). Our data is more dissimilar with M67, but this is largely due to the improved data quality of Pasquini \etal (2008) relative to that of Jones \etal (1999). Some detections in this sample are for lower abundances than the strictest upper limits from SR05, so we believe that our chosen sample is superior.

\subsection{Stellar Models}

To calculate the SSM predictions of the LDPs of open clusters, we use the Yale Rotating Evolution Code (see Pinsonneault \etal 1989 for a discussion of the mechanics of the code). We adopt a Grevesse \& Sauval (1998) proto-solar metal abundance ($Z$/$X$ = 0.025293; this is larger than the current solar surface abundance $Z$/$X$ = 0.02292 due to gravitational settling), and choose the solar hydrogen mass fraction and the mixing length ($\alpha$) such that a solar mass model reproduces the solar luminosity and radius at 4.57 Gyr. The calibrated values are $\alpha$ = 1.88269, $X$ = 0.71304, the helium mass fraction $Y$ = 0.26882, and metal mass fraction $Z$ = 0.018035 for [Fe/H] = 0.0. Our models use the 2006 OPAL equation of state (Rogers \etal 1996, Rogers \& Nayfonov 2002), atmospheric initial conditions from Kurucz (1979), high temperature opacities from the opacity project (Mendoza \etal 2007), low temperature opacities from Ferguson \etal (2005), and the $^7$Li$(p,\alpha)\alpha$ cross section of Lamia \etal (2012). In each of our models, we assume a initial Li abundance equal to the proto-solar abundance: A(Li) = 3.31 (Anders \& Grevesse 1989). We return to this assumption when evaluating the accuracy of our measurements, but for now mention that because Li depletion is logarithmic, dA(Li)/dt does not depend on the abundance.

To obtain the chemical mixture for arbitrary [Fe/H], we adopt a big bang helium mass fraction $Y_{BB}$ = 0.2484 (Cyburt \etal 2004), and assume a linear evolution of $Y$ with metals:

\begin{equation}
Y = Y_{BB} + \frac{\Delta Y}{\Delta Z} Z
\end{equation}

$\Delta Y$/$\Delta Z$ is solved for by calculating the slope between the big bang mixture ($Y$,$Z$) = (0.2484,0.0), and our calibrated solar mixture ($Y$,$Z$) = (0.26882,0.018035). We derive $\Delta Y$/$\Delta Z$ = 1.13, consistent with recent estimates (Casagrande \etal 2007). We use this method to create abundance mixtures for the adopted [Fe/H] of the clusters in this study, and evolve forward models of mass 0.5-1.3\msun, in steps of 0.05\msun, for each.

\section{Sources of Error}

We first consider sources of error that could affect the accuracy of our ZAMS Li predictions. These fall into two categories: errors affecting the relative predictions between cluster LDPs, and errors affecting the absolute predictions of all cluster LDPs. The relative error budget is dominated by uncertainties in cluster [Fe/H], which shift the predicted LDPs of clusters relative to one another, impacting the inferred lithium anomaly ($\S$3.1). This error has been minimized in our analysis through the selection of clusters with exquisitely measured [Fe/H]. Absolute errors are dominated by uncertainties in the physics adopted in our models, the physics of Li burning, and the proto-solar abundance, which affect the LDP predictions of all clusters simultaneously. Although these uncertainties are systematic, they can impact the relative predictions of clusters, and thus the inferred anomaly. We describe the effect of these errors in $\S$3.2, and account for their impact in $\S$4. 

\subsection{Metallicity Errors}

\begin{figure}
\includegraphics[width=3.3in]{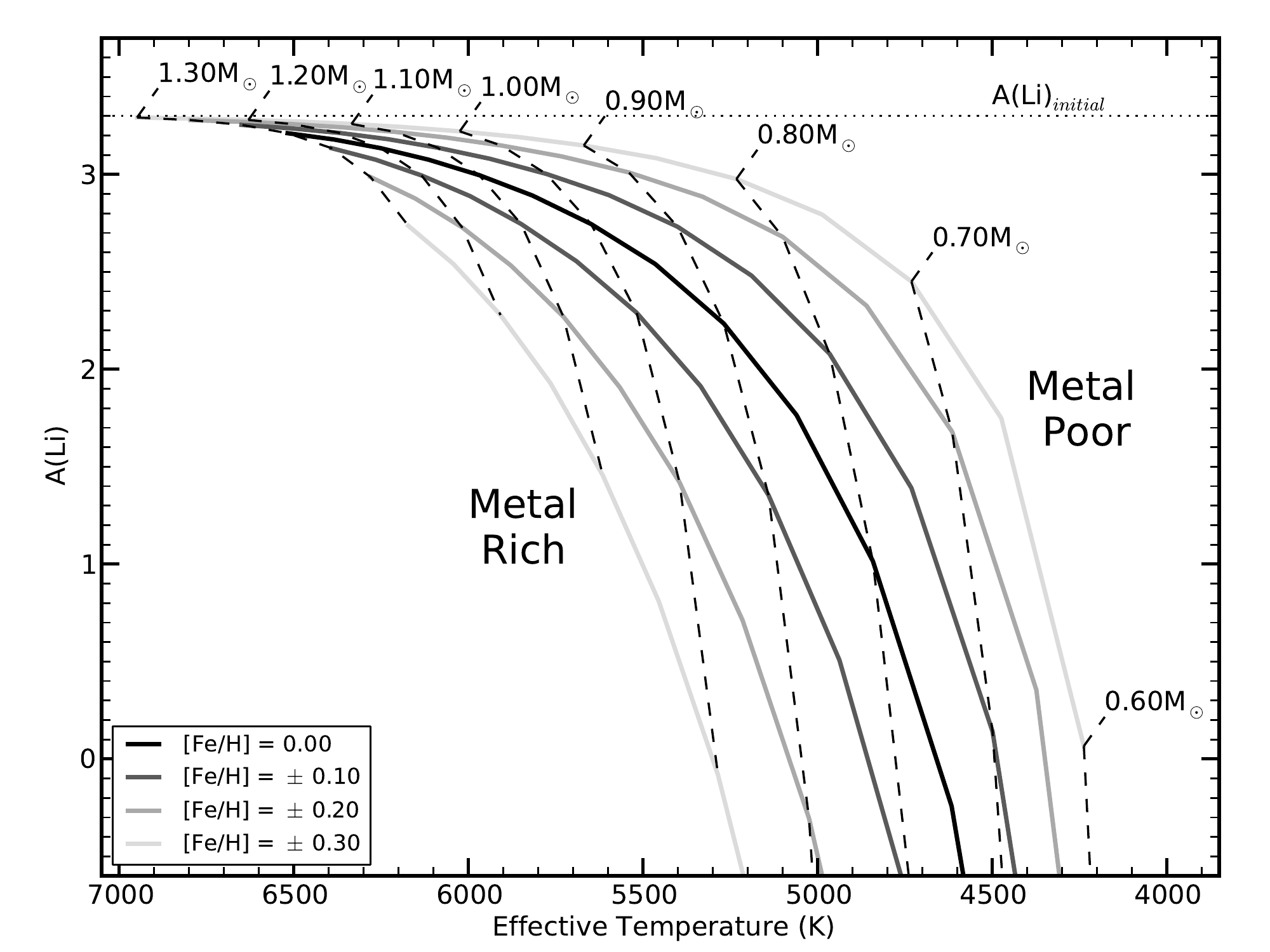}
\caption{SSM Li predictions at the ZAMS for a variety of metallicities, which illustrate the strong dependence of pre-MS Li depletion on composition. The lowest metallicity pattern is the top-most curve, showing that increased metallicity drastically increases the Li depletion rate at early stages of stellar evolution. X and Z values are calculated with respect to a Grevesse \& Sauval (1998) proto-solar abundance. Dashed lines display curves of constant mass, and the dotted line represents the initial lithium abundance.}
\end{figure}

\begin{figure}
\begin{centering}
\includegraphics[width=3.3in]{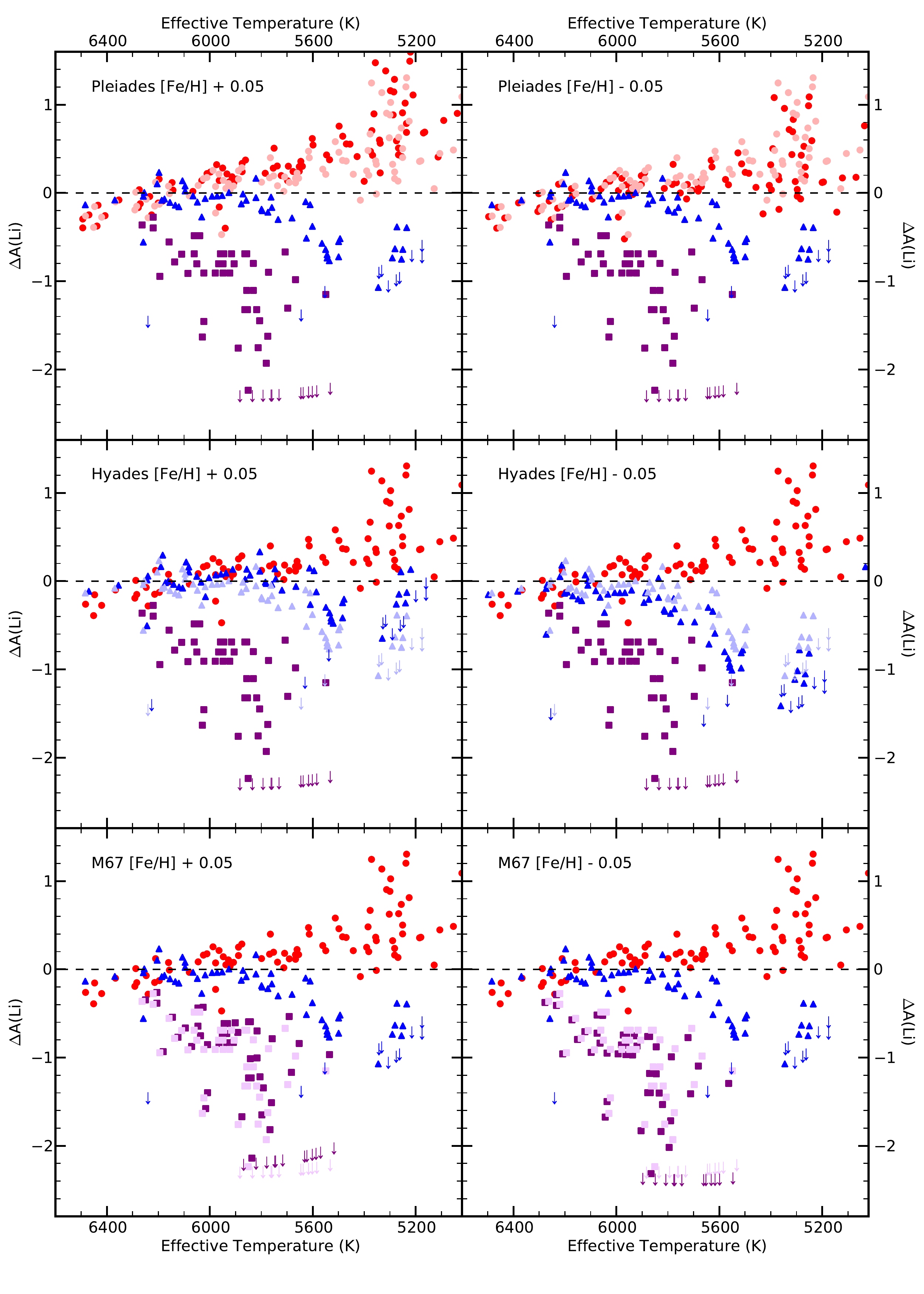}
\caption{The impact of metallicity errors on the relative locations of cluster Li patterns in Li anomaly space. In the top row, the transparent red circles represent the anomaly inferred with with original [Fe/H], given in Table 1, and the solid red points are Pleiades anomalies inferred with a different assumed [Fe/H], as shown in the top left of each panel. The pattern is shifted significantly if an erroneous composition is assumed. The blue and purple points represent Hyades and M67 anomaly data. The middle and bottom rows repeat the same exercise for the Hyades and M67 respectively. Arrows denote upper limits.}
\end{centering}
\end{figure}

\begin{figure*}
\begin{centering}
\includegraphics[width=7.0in]{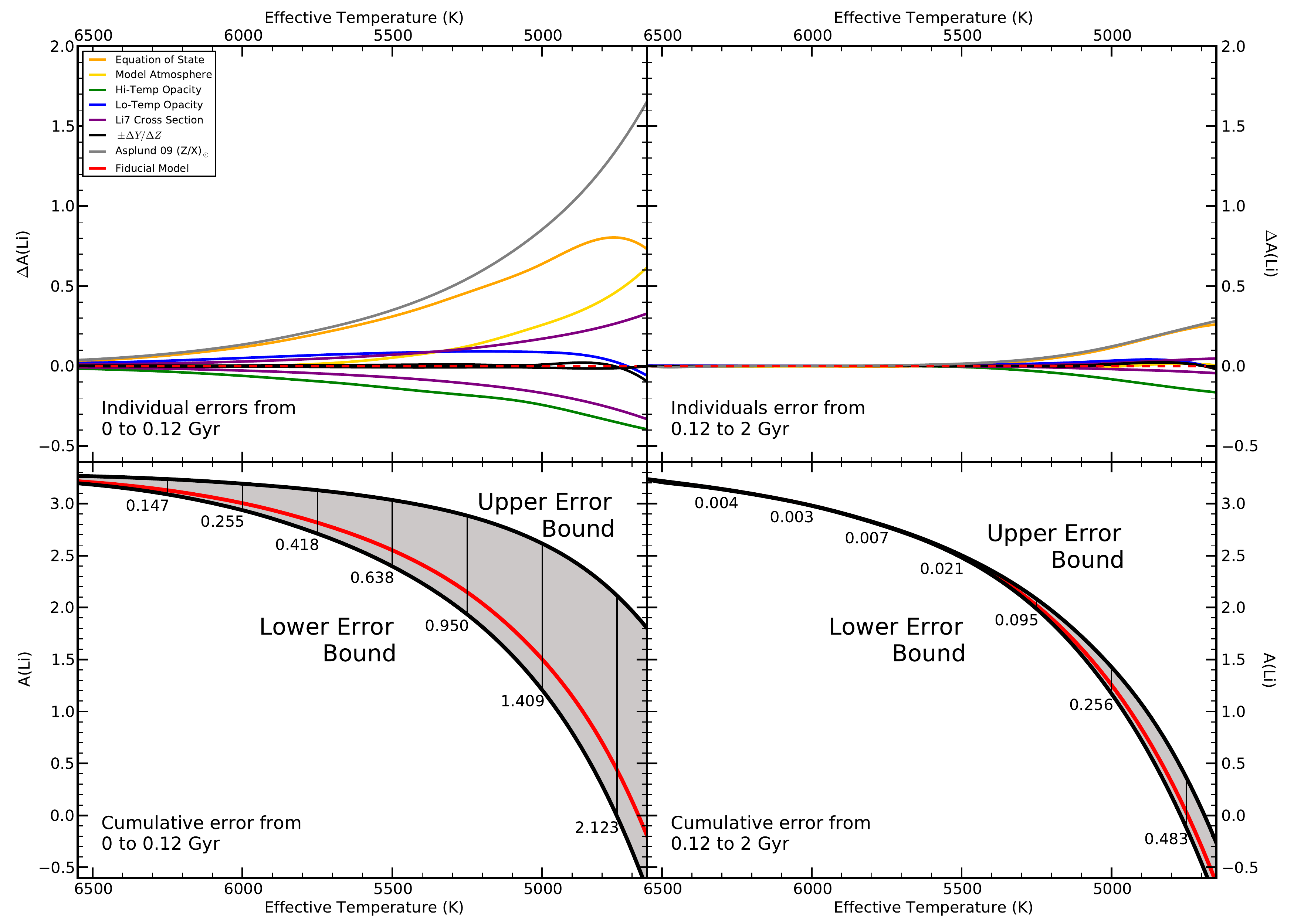}
\caption{The Li depletion errors induced by uncertainties in the adopted model during the pre-MS (left), and during the MS (right). \emph{Top}: The difference between the fiducial Li pattern and the Li pattern resulting from exchanging one physical input for another commonly adopted in the literature. Colors correspond to different physical inputs, as enumerated in the key. The errors in the left panel develop during the first 0.12 Gyr, and those on the right develop between 0.12 and 2 Gyr. Depletion on the MS is insensitive to physical errors,  \emph{Bottom}: The quadrature sum of the errors as a function of \teff. Any Li curve lying in the shaded region is a statistically acceptable Li depletion pattern.}
\end{centering}
\end{figure*}

ZAMS LDPs are extremely sensitive to composition. This is demonstrated by Fig. 4, which shows several Pleiades-age patterns calculated with a range of [Fe/H]s. As can be seen, the abundances vary greatly at fixed \teff\ and fixed mass (dashed lines) depending on the composition. More metal-rich clusters are progressively more depleted, and in some cases, a deviation of 0.1 dex in [Fe/H] produces a deviation $\gtrsim$ 1 dex in A(Li)! This strong dependence results from a deeper CZ in metal-rich stars, which increases $T_{BCZ}$ and drastically increases the burning rate. By contrast, the rate of depletion on the MS is insensitive to composition. Regardless of metallicity, $T_{BCZ}$ on the MS is much less than $T_{LB}$, suggesting that the rate of Li destruction cannot depend on the thermal properties of the envelope. Therefore, once the [Fe/H]-dependent pre-MS signal is removed, comparisons between MS LDPs become stable to metallicity errors.

The sensitivity of the benchmark cluster anomalies on the adopted [Fe/H] is seen in Fig. 5. Here, we show the cluster data subtracted from a variety of SSM LDPs. In the top left and right, the solid red circles show the distribution of stars had we assumed an [Fe/H] for the Pleiades that was 0.05 dex higher or lower, respectively. The ghosted red circles show the pattern produced by assuming the adopted [Fe/H] in Table 1, and the Hyades (blue triangles) and M67 (purple squares) are the same as in the bottom row of Fig. 1. The middle row shows the same effect but for the Hyades, and the bottom row for M67. Fig. 5 demonstrates that metallicity errors move detrended cluster LDPs relative to one another. A very small logarithmic shift in [Fe/H] can introduce a large fractional shift between two detrended clusters, significantly altering the inferred magnitude of MS depletion. The formal [Fe/H] error for each of our benchmark clusters is less than 0.05 dex, so composition-related errors are well controlled in our measurements, but these effects will be important when comparing clusters with less precisely determined relative metallicities. Furthermore, metallicity errors are far more important for low-mass stars. This suggests higher-mass stars are more stable to [Fe/H] uncertainties.

\subsection{Theoretical Systematics}

Lithium depletion is extremely sensitive to the adopted stellar physics (D'Antona \& Mazzitelli 1994; Piau \& Turck-Chi\`{e}ze 2002; Tognelli \etal 2012). There are several physical inputs that could potentially impact Li predictions, the most important of which we list in Table 3. To estimate the theoretical errors associated with these components on the pre-MS, we adopted the SSM prediction for the Pleiades as a fiducial model. We then varied each source of uncertainty in turn, computed the resulting LDP, and compared it to our fiducial pattern. The results can be seen in the left column of Fig. 6. The top panel shows the differences between the fiducial pattern and each alternative pattern as function of \teff, and the bottom panel shows the quadrature sum of the uncertainties, with the width of the error band enumerated at periodic intervals. We treat the sum of systematic differences between inputs, such as distinct equation of state tables or opacity calculations, as effective 2$\sigma$ errors.

Our theoretical errors are asymmetric. This is largely because changes in the solar heavy element mixture systematically reduce Li burning; adopting the Asplund \etal (2009) abundance ratios has an effect analogous to reducing [Fe/H] by $\sim$ 0.1 dex. The equation of state is the second largest effect, because a different relationship between $T$ and $p$ can change the depth of the surface convection zone (CZ), altering the temperature at its base and the rate of Li destruction. Other significant effects include the choice of model atmosphere, though this is a large effect only for stars with MS \teff\ $<$ 5000K, and the $^7$Li$(p,\alpha)\alpha$ cross-section, which produces symmetrical LDPs about the fiducial choice. There is a larger dynamic range in the model uncertainties for cool stars than for hot stars, similar to errors induced by metallicity uncertainties, because of their lengthier pre-MS burning phase.

We repeated the above exercise to investigate the impact of theoretical uncertainties on the minor SSM depletion occurring on the MS. To do this, we evolved our fiducial and alternate stellar models from the age of the Pleiades to 2 Gyrs, and measured the \textit{additional} discrepancy that develops during this time period. This can be see in the right column of Fig. 6. The impact on hot stars is almost nonexistent. The base of the CZ above 5500K is so cool during this period that small changes to the thermal structure of the envelope do not result in substantial changes to the rate of Li destruction. Minor changes are seen in cool stars, whose CZ bases are still somewhat warm, but at a much lower level than the uncertainties developing on the pre-MS. This demonstrates that once a depletion pattern has been corrected for theoretical errors arising on the pre-MS, the inferred MS depletion is stable. 

\input{Tab03.txt}

\begin{figure*}
\begin{centering}
\includegraphics[width=7.0in]{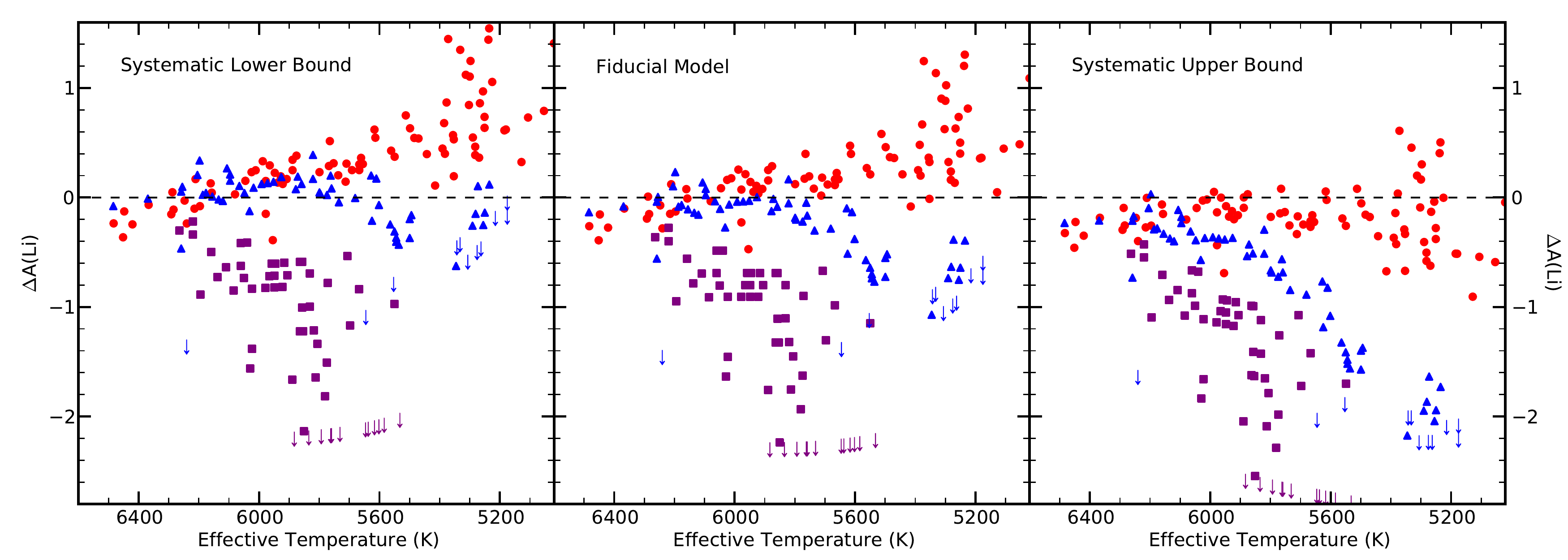}
\caption{The impact of uncertainties in our adopted model physics on the relative locations of clusters in anomaly space. Lithium data for the Pleiades (red circles), Hyades (blue triangles), and M67 (purple squares) are transformed into anomaly space using SSM Li predictions from the bottom envelope of the systematic error band (left), the fiducial model (center), and the top envelope of the systematic error band (right), as illustrated by Fig. 6. The absolute difference between cluster anomaly patterns is sensitive to the adopted model physics, so the models must be calibrated using empirical data. Arrows denote upper limits.}
\end{centering}
\end{figure*}

Fig. 7 shows data from our benchmark clusters detrended with respect to the bottom and top of the pre-MS systematic error band, and the fiducial model. While [Fe/H] errors move individual clusters with respect to one another, theoretical uncertainties move the clusters up and down in tandem. Uncertainties inherent to our models are predominately an absolute bias level in the magnitude of Li depletion, rather than a relative error between the clusters. Nevertheless, the inferred difference between clusters can change by a few tenths of a dex at fixed \teff\ depending on the choice of systematics, because more metal-rich clusters have a wider systematic error band. For instance, the anomaly measurement at 5800K between the Pleiades and Hyades is different in the left and right panels of Fig. 7. This demonstrates the need for guidance in the selection of physical inputs.

\section{Calibrating Standard Stellar Model Physics}

We have shown that Li depletion is sensitive to a large number of physical inputs. In the absence of compelling information about which inputs are most in error, we cannot pinpoint which parameter, or parameters, should be changed to reconcile our fiducial theoretical models with nature. However, since MS depletion is insensitive to physics in SSMs, and because there is great theoretical freedom in the ZAMS pattern, it is reasonable to adopt an empirical fit to the absolute depletion of a ZAMS cluster. This gives us an \textit{ad hoc} calibration of the ensemble physics in our models. Once this has been done for one cluster, SSMs predict the relative scaling required to obtain the calibrated predictions of other clusters. Although this scaling is highly sensitive to errors in metallicity, if the composition of a cluster is reliable, analysis on detrended warm cluster stars will not be biased by significant failures of standard models. 

\begin{figure}
\includegraphics[width=3.3in]{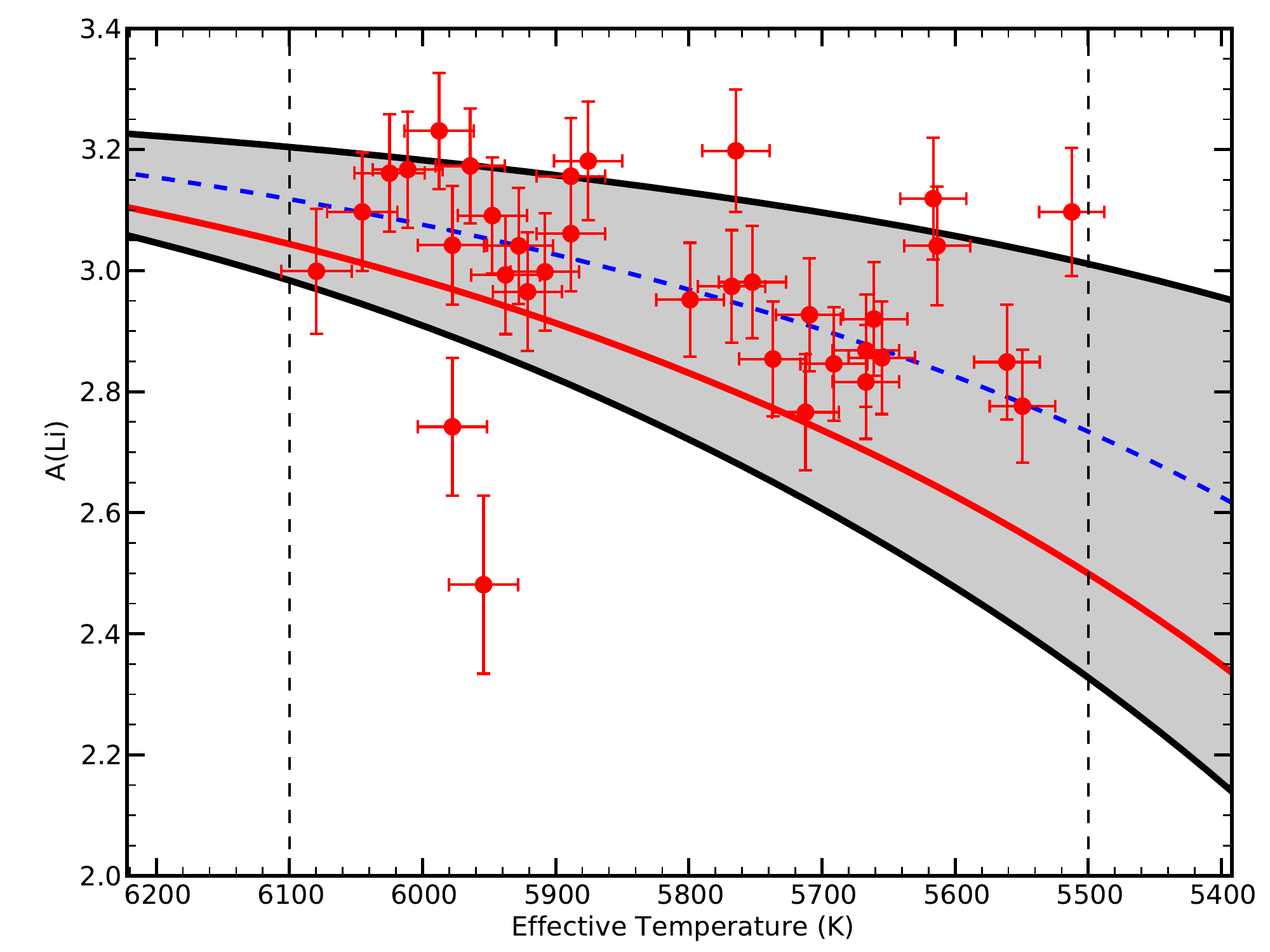}
\caption{Pleiades systematic error band (Fig. 6), over-plotted with Pleiades data in the range 5500K-6100K. Although any LDP lying in the shaded region is statistically acceptable given the uncertainties in our model physics, the dashed blue line represents the best fit to the data. This represents a calibration of the input physics in our models.}
\end{figure}

\begin{figure*}
\includegraphics[width=7.0in]{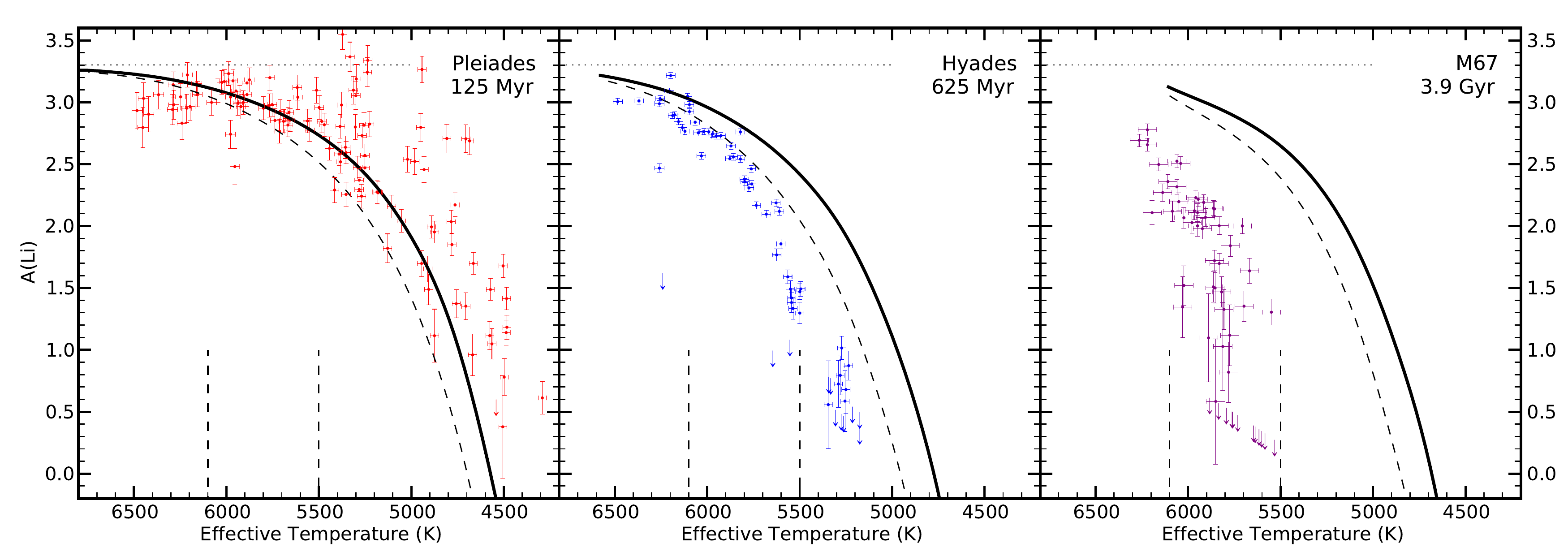}[!hb]
\caption{Solid black lines represent SSM Li patterns, whose input physics has been calibrated on the Pleiades ($\S$4). Dashed black Li patterns represent the fiducial LDPs. The vertical dashed lines represent the \teff\ regime employed to calibrate our model physics. Red, blue and purple data points are the same as in Fig. 1.}
\end{figure*}

Lithium patterns of young clusters, such as the Pleiades, are expected to closely mimic the true ZAMS LDP, as they have undergone minimal MS depletion. We can therefore use the empirical Pleiades pattern to guide our selection of the theoretical LDP that best reflects the true ZAMS distribution. Before we can do this, we must choose a suitable \teff\ range to use for this calibration. As described in $\S$3, errors induced by both observational and theoretical uncertainties are more significant for cool stars than for hot stars. We therefore restrict our model calibration to stars with \teff\ $>$ 5500K. At this \teff, the width of the theoretical error band is still large ($\Delta$A(Li) $\sim$ 0.7 dex), but the random errors due to uncertainties in [Fe/H] are small ($\Delta$A(Li) $\lesssim$ 0.2 dex; see Fig. 4). This is beneficial, since we have great leverage in calibrating the input physics due to the large theoretical uncertainties, and relatively small relative errors due to [Fe/H] uncertainties. Additionally, we do not include stars with \teff\ $>$ 6100K in this analysis, because early mixing could be impacting their abundances (Margheim 2007). Our final analysis regime is therefore 5500-6100K.

We then interpolate to each theoretically allowed ZAMS LDP inside the 2$\sigma$ error bounds in the top left panel of Fig. 6, and determine how well it matches the data by computing the median absolute deviation (MAD) of the empirical Pleiades Li distribution around it. We then select the LDP that produces the best goodness-of-fit within the calibration regime. The resulting function is shown in Fig. 8. Here, the black and red lines are the same as in Fig. 6, and the blue line represents the best fit model. The empirical Pleiades data follow this curve well, with the exception of a few outliers, which may reflect true scatter. We adopt this as our calibrated Pleiades SSM, and as the function we will use to detrend out pre-MS depletion from Pleiades data.

The calibrated patterns for other clusters will lie the same fractional distance between their fiducial LDPs and their upper theoretical error envelopes as the calibrated Pleiades SSM does. To calculate these models, we derive a scale factor from the Pleiades system by dividing the absolute depletion of the detrending function by the absolute depletion of the fiducial model. The scale factor is nearly invariant within the \teff\ analysis domain, allowing us to approximate it as a constant. Scaling the Hyades fiducial model by this value produced a close approximation to a detrending function generated by fully calculating the theoretical errors for Hyades parameters and interpolating to the same location as in the Pleiades. This demonstrates that this method can be used to generate systematics-corrected LDPs for any given set of cluster parameters. Our calibrated models for the benchmark clusters are shown as solid black lines in Fig. 9, alongside the fiducial SSMs, which are represented by dashed lines. As can be seen, the tension between the median depletion predicted at ZAMS and solar analogs in the Pleiades has been resolved. This figure also demonstrates that the lower envelope of the cool star dispersion in the Pleiades is well approximated by SSMs. This implies that whatever mechanism is inducing this spread does so by \textit{suppressing} Li depletion in rapid rotators, and not by inducing additional Li depletion in slow rotators ($\S$6.2).

We note that the metal content of our models has been determined by scaling the proto-solar abundance by the measured [Fe/H] of each cluster. In actuality, [Fe/H]s are reported relative to the solar \textit{photospheric} abundance. The true metallicity of stars during the pre-MS was larger than their current abundances by a factor equal to the amount of gravitational settling occurring during their lifetime. This is a mass and age dependent effect, in the sense that older and more massive stars undergo more settling. The magnitude of the effect for solar analogs in M67 is $\Delta$[Fe/H] = -0.06, leading to a 0.04 dex change in the predicted pre-MS Li depletion. This effect is smaller for lower mass stars, because settling is negligible when the CZ is deep, and smaller for higher mass stars as well, because the rate of Li depletion is insensitive to composition in this regime. The corresponding effect is significantly lower for the Hyades, since it is younger by a factor of 8, and non-existent for the Pleiades, since we have calibrated our models on empirical data for this cluster. Given the complex, and in general unknown, dependence of settling on mass and age, we do not include this effect in our calculations. Given the small magnitude of the effect, it does not significantly impact the accuracy of our measurements.

\section{Results}

\begin{figure*}
\includegraphics[width=7.0in]{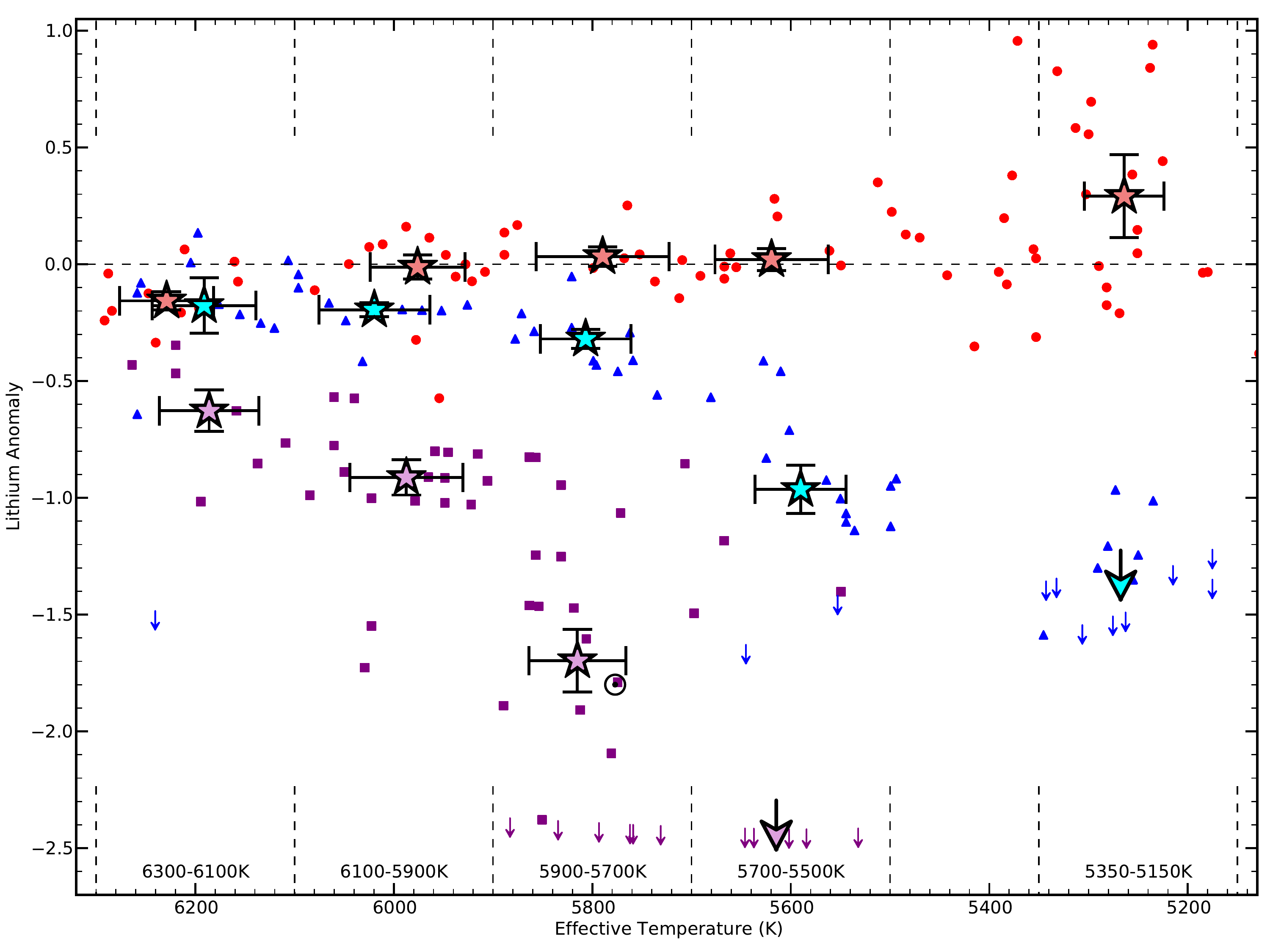}
\caption{The final Li anomaly measurements for our benchmark clusters. The Pleiades (red circles), Hyades (blue triangles) and M67 (purple squares) data have been detrended with respect to their calibrated SSM LDPs, described in $\S$4. Light red, light blue, and light purple stars are the average \teff\ and median lithium anomaly in each bin (denoted by black dashed lines) for the Pleiades, Hyades and M67, respectively. An inverted triangle denotes an upper limit to the median anomaly. The solar lithium abundance is also shown, and arrows denote upper limits.}
\end{figure*}

Surface Li destruction proceeds rapidly on the pre-MS, and much slower thereafter. To empirically measure the rate of Li depletion on the MS, we must accurately predict the amount of depletion that occurs for stars on the pre-MS, so it can be subtracted from their present day abundances. However, the amount of Li destruction occurring on the pre-MS in SSMs is highly sensitive to errors in both the assumed metallicity of the cluster and the physical inputs in our models.  The former can be controlled by considering clusters with well known composition, so we have selected three well-studied clusters for a precision measurement. The latter can be controlled by selecting the input physics which best reproduces the observed Li pattern of a ZAMS cluster, which we do in an ensemble fashion in $\S$4. Once our physics is calibrated, the Li destruction occurring on the pre-MS can be detrended out of an empirical Li pattern, thus isolating the depletion occurring on the MS. The rate of MS Li depletion can then be inferred by comparing the average Li anomalies of different aged clusters.

In this section, we use this methodology to obtain the MS anomaly for several clusters. First, we will detrend the Hyades and M67, and infer from their anomaly patterns the rate of MS Li depletion, as a function of mass and age. We find a strong mass trend in the rate of MS depletion in both clusters, and confirm that the average rate of depletion decreases at advanced ages. We also discuss a few caveats to this method, including the effects of the initial Li abundances of clusters, and the possibility of cosmic variance differentially impacting cluster LDPs, but conclude that our results are robust. We then detrend and analyze a large sample of open clusters, previously studied by SR05, to explore the timescales of MS depletion and dispersion. We find that the rate of depletion of solar analogs is unchanged by the transformation to anomaly space, but the mass dependence of depletion is stronger in absolute space. Finally, we measure the Li dispersion at fixed \teff\ in each cluster, and explore the timescales of its development. We find that dispersion sets in early in many clusters, and can increase, decrease, or remain steady over time depending on the temperature bin. Furthermore, dispersion is not a simple function of age, suggesting cosmic variance in cluster initial conditions. 

\subsection{Benchmark Clusters}
	
\subsubsection{The Lithium Anomaly}

\input{Tab04.txt}

Our final detrended benchmark clusters are shown in Fig. 10. Here, we have divided our analysis regime into three bins of 200K width. This provides us sufficiently low shot noise to calculate the median Li depletion without introducing too much error due to the \teff\ dependence of the pattern. We also include two additional \teff\ regimes that were excluded in $\S$4: 6300-6100K and 5350-5150K. Although these temperature ranges were not suitable for calibrating our models, the anomaly remains a useful measure of mixing. The light red, light blue, and light purple stars represent the average \teff\ and the median lithium anomaly of the Pleiades, Hyades, and M67 within each bin (illustrated by the vertical dashed lines). We have ignored the \teff\ range 5500-5350K, due to a gap in the Hyades data in this regime.

Between the Pleiades and the Hyades, we find the following lithium anomalies, where the first quoted error is due to Poisson noise, and the second is due to [Fe/H] uncertainties: 0.020 $\pm$ 0.124 $\pm$ 0.018 dex at 6200K, 0.183 $\pm$ 0.057 $\pm$ 0.018 dex at 6000K, 0.353 $\pm$ 0.055 $\pm$ 0.025 dex at 5800K, 0.984 $\pm$ 0.111 $\pm$ 0.021 dex at 5600K, and $\gtrsim$ 1.636 dex at 5250K. Between the Pleiades and M67, we find anomalies of 0.471 $\pm$ 0.094 $\pm$ 0.025 at 6200K, 0.901 $\pm$ 0.082 $\pm$ 0.041 dex at 6000K, 1.730 $\pm$ 0.128 $\pm$ 0.058 dex at 5800K, and $\gtrsim$ 2.436 dex at 5600K. There is no anomaly for the 5250K bin here, since no M67 Li data exists in the literature in this range. The Hyades anomaly at 5250K and the M67 anomaly at 5600K are upper limits, due to the preponderance of upper limits in this bin. These measurements are collected in Table 4. Our choice of benchmarks with well-constrained [Fe/H] has caused the uncertainties to be dominated by shot noise. 

In $\S$2.2, we compared the S93 CoG used in this paper with the S03 CoG, and found substantial differences in final abundance for some stars. To evaluate the impact of this uncertainty on our final answer, we compare our answers to anomaly values derived using the S03 CoG. These are also shown in Table 4. For all but the coolest \teff\ bin, the relative anomalies between the Pleiades and Hyades agree with our original values at $\sim 1 \sigma$ or better. Between the Pleiades and M67, the values agree at $\sim 1.5 \sigma$ or better. This increases our confidence that our results are robust. The uncertainties are somewhat larger for M67, since fractional errors increase when EWs are small. The lower limits derived with S03 are $\sim$0.2 dex lower for the 5250K bin, reflecting the sensitivity of derived abundances on EWs in Li-poor stars. This error is not too worrisome, as 0.2 dex represents $<$15\% of the total anomaly at this temperature. 

An important conclusion about MS Li depletion in FGK dwarfs can be drawn from this plot: low mass stars deplete Li more rapidly than high mass stars on the MS. While this effect has been seen previously in absolute space, our work confirms that the effect persists when the additional depletion suffered by low-mass stars on the pre-MS has been removed. This result holds regardless of the choice of theoretical bias and CoG. This provides a stringent constraint on mechanisms seeking to explain the MS lithium anomaly in open clusters. A second conclusion we can draw from this plot is that the average rate of MS depletion is higher for the Hyades than for M67. This can be seen in Fig. 10 by the ratio of the anomaly at the age of M67 to the anomaly at the age of the Hyades. If the depletion rate were constant for these two clusters, this ratio should be $\sim$ 7.6, equal to the ratio of time the clusters have spent on the MS. However, in the 5800K and 6000K bins, the ratio is 4.8 and 5.1 respectively. This depletion plateau has been seen before (i.e. SR05), but we have shown that the result persists even when differential metallicity effects are accounted for.

These arguments are strengthened by the cosmic evolution of lithium. The Li content of the interstellar medium has increased over time (Spite \& Spite 1982), and a correlation between cluster Fe abundance and initial Li abundance has also been reported (Cummings 2011), so it is possible in principle that our benchmark clusters began their lives with different Li abundances. The initial M67 Li content is likely close to solar, since it is nearly equal to the Sun in both age and metallicity. The initial Pleiades Li abundance has also been measured to be near solar (e.g. Cummings 2011). However, the Hyades may have been born with a higher Li abundance than we have assumed. If this is correct, the true magnitude of MS depletion for the Hyades will be $\textit{greater}$ than we have measured. This would reduce the difference between the Hyades and M67 medians, decrease the ratio described above, and therefore increase the tension between our measurement and the putative expectations of constant logarithmic depletion. Furthermore, changing the assumed initial abundance of a cluster moves each of its members up and down in tandem, and so will not change the mass-dependent pattern revealed through detrending. We therefore believe that shifts in the initial cluster abundance could impact the precision of our anomaly measurements, but will not impact either of the conclusions stated above. A rigorous evaluation of the Hyades initial cluster abundances can be undertaken by analyzing the Li content of stars blue-ward of the Li gap, as they will have suffered minimal pre-MS and MS depletion.

It is plausible that additional cluster parameters could impact the Li pattern, jeopardizing the generality of our measured anomalies. For example, rotation is expected to drive deep mixing flows through meridional circulation and shear instabilities, and induce non-standard Li depletion during the MS as a result (e.g. Pinsonneault \etal 1997, and refs. therein). The rate of depletion increases with faster rotation, so clusters with different rotation distributions will eventually develop different LDPs. Environment could therefore impact the Li depletion properties of clusters.

The early (0-10 Myr) rotation evolution of a star is dictated by its circumstellar disk, which locks magnetically to the star and efficiently drains AM from the envelope (e.g. Koenigl 1991; Rebull \etal 2006). If a T Tauri star has a close interaction with another cluster member, its circumstellar disk may be subject to early disruption through the interaction. This will truncate the timescale of the disk-locking phase, and the star will retain the AM it would have otherwise lost, appearing as a rapid rotator at the ZAMS. Such interactions are more likely in dense stellar environments, so this process could induce a correlation between the number density of a cluster at birth, and the fraction of rapidly rotating stars. Assuming a connection between rapid rotation and MS Li depletion, dense clusters would be expected to host a commensurately large fraction of Li-poor stars in the solar regime. This would impact both the width of the Li distribution, and the median anomaly. Although this explanation is qualitatively sensible, Bouvier \etal (1997) found that the rotation rates of primaries in binary systems are statistically indistinguishable from rotation rates of single stars in the Pleiades. This suggests that physics local to the star sets rotation rates, and not environmental factors.

\subsubsection{Literature Comparison}

\begin{figure}
\begin{centering}
\includegraphics[width=3.3in]{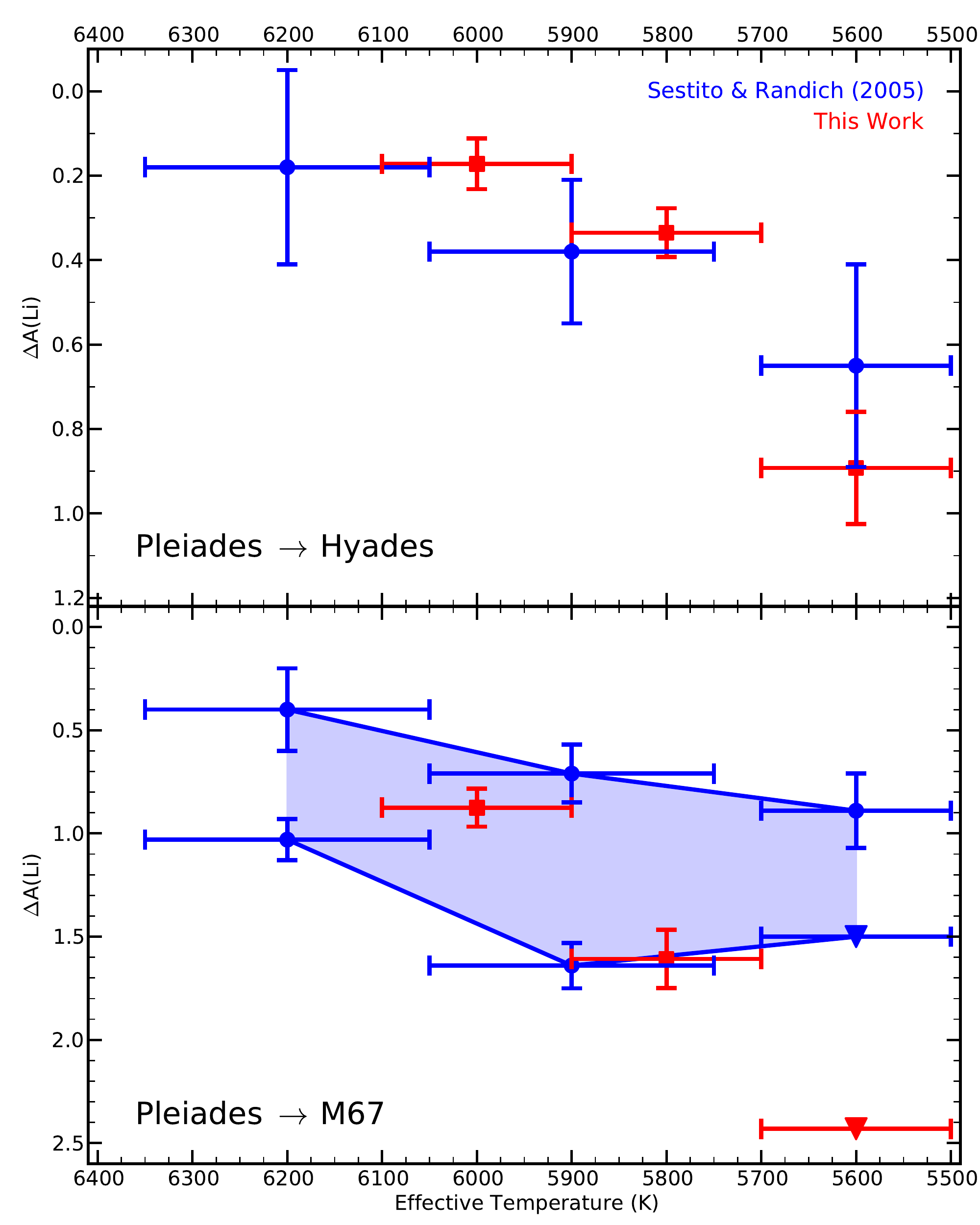}
\caption{A comparison of the lithium anomaly measurements of this work with the measurements of SR05. Red squares represent this work's calculation of the lithium anomaly for the Hyades (top) and M67 (bottom). Blue circles represent the difference between the SR05 bin containing the Pleiades and the bins containing the Hyades (top) and M67 (bottom). SR05 reported the average of the upper and lower envelopes for M67, so we include both sets of points, and show the region interior. Triangles represent upper limits.}
\end{centering}
\end{figure}

We now compare our results to SR05, who calculated the average Li abundance in three \teff\ bins at a variety of ages along the MS. It should be noted that there are two important differences between our analysis and that of SR05. First, we measured relative abundances in lithium anomaly space, whereas SR05 worked in absolute abundance space. The practical effect is that SR05 did not correct for differential depletion on the pre-MS due to composition differences. Second, we analyzed only one cluster at a time, whereas SR05 grouped several similarly-aged clusters together and calculated an ensemble average. Their method tends to wash out both random errors in \teff\ and A(Li), and the effect of different metallicities.

With these precautions, we compare results in Fig. 11. In the top panel, the red points are our lithium anomaly measurements for the Hyades, and the blue points represent the difference between the bins containing the Hyades and the bins containing the Pleiades in SR05. The measurements are generally consistent with one another. Between 5700K and 6200K, we measure a marginally smaller Hyades depletion than SR05. This is because the Hyades is a metal-rich cluster, and thus less MS depletion is inferred when pre-MS effects are considered. However, our measurement is lower than the SR05 measurement at 5600K. Given the uncertainties inherent to CoG analysis in low abundance regimes, this may reflect differences in the abundance derivation rather than a true difference in measurements. Alternatively, given the steepness of the LDP in this \teff\ range, this difference may be due to random errors in our sample. More stars populate the warm side than the cool side of this bin, biasing our answer towards greater depletion.

The red points in the bottom panel of Fig. 11 represent our lithium anomaly measurements for M67. SR05 did not bin M67 with other clusters, but instead measured the average of the M67 upper and lower envelopes separately. We therefore show the range in which the global average of their sample could reside. Our measurement at 6000K is in good agreement with SR05. This is not surprising, since M67 and Pleiades are very similar in [Fe/H], and thus pre-MS corrections are minimal. Our 5800K and 5600K measurement are large compared to SR05, but this is likely due to the different data samples used in the two studies. The sample of SR05 is dominated by Li measurements from Jones \etal (1999), which had a lower detection threshold than Pasquini \etal (2008), used by this work. We thus find a lower average at 5800K, and set a stricter upper limit at 5600K. 

\subsection{Additional Clusters}

Minimal measurement errors have made the Hyades and M67 optimal for calibrating and testing mixing calculations, but two temporal points do not provide a complete time line of Li evolution. We therefore apply an analysis similar to $\S$5.1 to the full sample of open clusters described in SR05. They assembled all substantial FGK dwarf Li data sets available prior to 2005, and applied a uniform abundance analysis to examine the timescales of Li evolution. The results were the confirmation of a mass trend in the rate of MS depletion, and the identification of four stages of depletion: depletion on the pre-MS, a stall near the ZAMS, depletion on the MS, and a plateau at late ages. Given the considerations put forth in this paper, we wish to determine if these qualitative and quantitative results are altered by the transformation to anomaly space. To this end, we reanalyze with our methods the data described in $\S$2.1.2, excluding the clusters NGC 2264, since substantial pre-MS Li depletion has yet to occur in this young cluster, and NGC 2547, due to the lack of a quality [Fe/H] measurement. We retain the data sets used in $\S$5.1 for our benchmark clusters.

\subsubsection{Evolution of the Median}

We first compared our results with those of SR05 in absolute space. The data for each cluster were divided into the three \teff\ regions they considered (5600 $\pm$ 100K, 5900 $\pm$ 150K, and 6200 $\pm$ 150K), and the median of each bin was calculated. We then combined the data from similar-aged clusters in the fashion of SR05 (see their Table 3), to maximize the validity of the comparison. Unsurprisingly, our findings are consistent with theirs. A decline in median abundance as a function of age is present in each \teff\ bin, and the rate of depletion increases with decreasing mass. Modest differences between our points and theirs are present, due to the difference between mean and median statistics, and our choice of different data sets for some clusters. Nevertheless, this confirms the similarity of the two analysis processes.

\begin{figure*}
\begin{centering}
\includegraphics[width=7.0in]{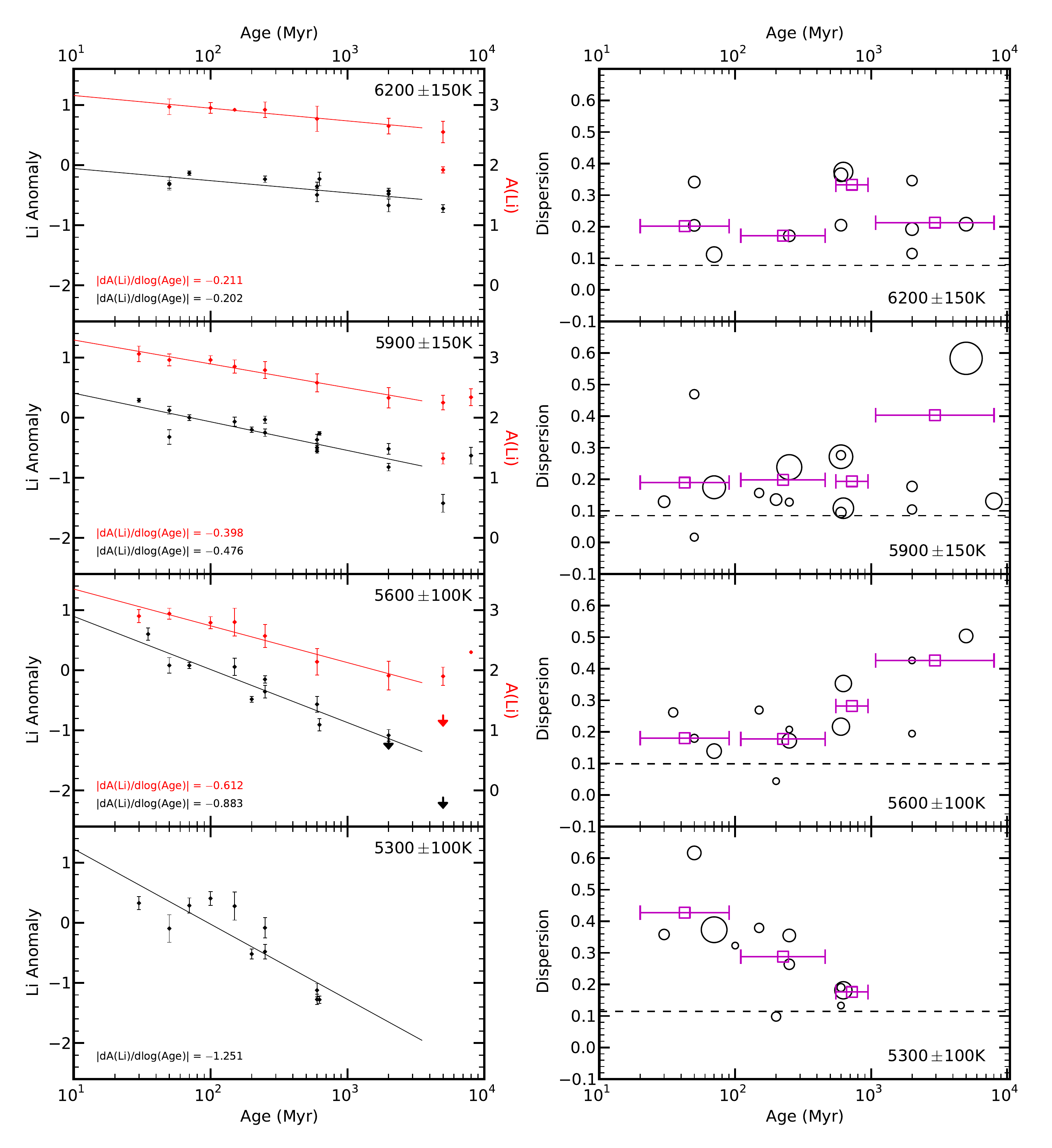}
\caption{The timescales of Li depletion in anomaly space, and the development of Li dispersion at fixed mass. \textit{Left:} Li data from SR05 (red points, right scale) compared to the same data detrended with the methods of $\S$4 (black points). The red and black lines show the best fit power laws calculated up to 3 Gyr, with slopes reported in the lower left of each panel. The detrended clusters show a stronger mass trend than the absolute clusters. \textit{Right:} The dispersion of each cluster considered in the left, with the sample size represented by the radius of the individual points. The dashed lines represents the absolute noise floor, and purple squares represent a weighted average of points within each age bin. Dispersion is found to emerge early in many clusters, and stay relatively constant for late F stars, rise marginally over time in solar analogs, and decrease up to 1 Gyr in cool stars. The scatter of dispersions at fixed age suggest true cosmic variance in cluster Li patterns.}
\end{centering}
\end{figure*}

Next, we detrended each cluster, using the machinery of $\S$2.3 and the systematic corrections of $\S$4, and binned the data as described above. We display each cluster as a single data point instead of applying age bins, allowing for a visual impression of the intrinsic scatter of cluster medians about the mean trend. Finally, we rejected bins with less than 3 members, and plotted the data in the first three panels of the left column of Fig. 12. Black points represent the median anomaly of each cluster in the quoted bins, and the error bars represent the quadrature sum of the standard error of the median ($\frac{\rm MAD}{\sqrt{N}}$), and the uncertainty of the median due to [Fe/H] errors. We also plot the data of SR05 in red, with the relevant scale displayed on the right axis. Finally, we binned and detrended data in the \teff\ range 5300 $\pm$ 100K, and plotted the data in the bottom left panel.

As evinced by this figure, cooler stars possess a higher average depletion rate over their lifetimes in both absolute and anomaly space, in agreement with the findings of $\S$5.1 and SR05. In order to quantify this effect, we calculate the best fit power law for the data in each bin up to 2 Gyr, when the subdivision of M67 in the SR05 data and upper limits begin to complicate the regression process. The red line in each panel shows the best fit for the SR05 data, and the slope is reported in the lower left corner. The 6200K bin depletes slower than the 5900K bin by a factor of 1.9, and slower than the 5600K bin by a factor of 2.9. This trend is preserved in anomaly space, but the mass dependence is found to be marginally steeper: depletion in the the detrended 6200K bin is consistent with the absolute 6200K bin, but is slower by a factor of 2.4 compared to the 5900K bin, and slower by a factor of 4.4 compared to the 5600K bin. The bottom left panel of Fig. 12 shows that this trend continues beyond the \teff\ range considered by SR05; the depletion rate at 5300K is a factor of 6.2 greater than in the warmest bin. The anomalies up to $\sim$150 Myr are positive in this bin, reflecting the over-abundance of cool stars relative to theory (see $\S$6.2), and drop sharply after this age. Therefore, the rate of MS depletion after this age is likely higher than reflected by the best fit power law slope. 

SR05 argued that abundances in each \teff\ bin converge to a plateau after about 2 Gyr. However, this late-time plateau behavior does not satisfactorily describe the median of M67. While the upper envelope of M67 shares the same average abundance as the 2 Gyr and 8 Gyr old clusters in the SR05 data, the large dispersion present in this cluster causes the median abundance to be lower by several tenths of a dex. This can clearly be seen in our data in the 5900K bin, where the median abundance at 5 Gyr is substantially less than the older and younger clusters. This suggests that M67 may represent a different evolutionary pathway for cluster Li patterns than the 2 and 8 Gyr old clusters considered here, and points to cosmic variance affecting cluster evolution. In order to quantify the extent of this variance, we now examine the range of Li dispersions that can develop in open clusters.

\subsubsection{Evolution of the Dispersion}

Some proposed MS mixing mechanisms naturally predict a range of depletion rates at fixed mass, and thus a variable dispersion as a function of time (i.e. rotational mixing), while others predict that all stars of a given mass deplete Li at equal rates on the MS, implying a fixed dispersion (i.e. gravity wave mixing). The evolution of the distribution of Li abundances at fixed \teff\ therefore provides an additional constraint on MS mixing. This quantity is difficult to measure in absolute space, as strong mass trends impart a large difference in average A(Li) at the high and low ends of cool \teff\ bins. This introduces significant scatter about the median abundance of the bin, rendering this measurement impossible without excellent number statistics. However, this issue can be resolved by transforming the data to Li anomaly space. In this plane, the intrinsic scatter due to the presence of pre-MS mass trends is removed, and a robust measurement of dispersion is possible. 

To this end, we calculated the standard deviation of each \teff\ bin from $\S$5.2.1, and plot the results in the right column of Fig. 12. The black circles represent the measured dispersions, and the diameter of the circle reflects the sample size: the smallest circles have 3 members, and the largest has 34. There are three potential sources of noise in the dispersion measurement. The first results from random \teff\ errors, which can cause individual stars to be detrended improperly, altering the inferred anomaly. This is somewhat mitigated by the correlation between \teff\ and A(Li) errors (see $\S$2.2), but the effect may still be important. The second source is uncertainty in the metallicity of the cluster. If we detrend data with a SSM calculated for an improper [Fe/H], the pre-MS mass trend will not be completely removed. This will leave behind a systematic offset within the \teff\ bin, which can introduce additional scatter. Finally, random errors in EW(Li) measurements will impart scatter. To assess the impact of these factors, we randomly generated 1000 stars within each \teff\ bin. The stars were assigned true A(Li)s based on their temperature, by interpolating in a SSM LDP calculated with solar metallicity. We assigned each an observed \teff\ error by randomly sampling a Gaussian with $\sigma = 50$K, and propagated this error into the observed A(Li). When this was done, we assigned each a $\sigma = 10\%$ Gaussian random EW(Li) error, and altered the inferred A(Li) according. We believe 10\% to be a conservative estimate of the relative error of Li EWs within a single sample; although larger systematic errors between different studies may be present, this does not impact a dispersion measurement for a single data set. Finally, we detrended the stars against a SSM calculated with [Fe/H] = $\pm$0.1 dex from the fiducial value. We performed this experiment 1000 times, and calculated the standard deviation within each bin. We found that in no case did the additional dispersion exceed 0.12 dex, implying that this method is reasonably stable to errors in photometry, spectroscopy, and composition. We represent this absolute noise floor by the dashed lines in each right-hand panel of Fig. 12; values above this noise floor can be considered detections of dispersion.

This figure demonstrates that dispersion is a generic feature of cluster LDPs at all ages. It is present in some large samples as early as 100-200 Myr, suggesting an early origin for the dispersion in some clusters. However, dispersion is undetected some very old clusters. This could be because of the small number of stars with known Li abundances in some clusters, or could imply that dispersion is not present in all systems. Clusters of equal age can show significantly different dispersions, and some clusters of dissimilar age show equivalent dispersions; this again could be due to the quality of the data sets, or signify cosmic variance. To examine the mean trends, we grouped the clusters into 4 age bins, calculated the mean dispersion within each, weighing data points by their sample size, and plotted the results as purple squares. The 6200K bin shows that dispersion is, on average, present at all ages, but does not necessarily grow over time. Even the cluster with the greatest dispersion, M67, agrees with the mean trend in this \teff\ range. The 5900K bin shows the most clusters are confined to a small band of dispersions, with only a few outliers. There is again little evidence of dispersion evolution in this range, although some outlying clusters are now apparent. The 5600K bin shows a more convincing rising trend, with the five largest samples rising over time; this is evidence that in this temperature range, the spread can increase along the MS. Finally, in the coolest bin we see a trend of {\it decreasing} dispersion with age. This is related to the large dispersion at fixed temperature seen in young systems, such as the Pleiades, which is suppressed in some intermediate-age systems, such as the Hyades. This implies that the most abundant stars at ZAMS (i.e. the fastest rotating) deplete Li more rapidly on the MS, such that the dispersion decreases over the first Gyr of MS evolution.

These plots reveal great complexity in the development of Li dispersion in open clusters, and raise several questions. First, can cluster Li dispersions really differ at the same age? The most uninteresting explanation for this figure would be that no underlying difference exists, and that the data sets we employ are sampling the same distribution, but with poor enough statistics that noise is dominating. We can clearly rule this out in some instances, such as the comparison between M67 and NGC 188. The younger M67 has both a more depleted upper envelope, and a substantially larger scatter than NGC 188. There are 13 stars in our NGC 188 sample between 5700K and 6000K, all of which are more abundant than A(Li) = 1.8. There are 29 stars in the M67 sample in this range, 13 of which are more abundant than 1.8, and 16 of which are less abundant than this value. It seems highly unlikely that these fractions could be drawn from the same distribution. Furthermore, the total range of abundances is $\sim$ 0.5 dex for NGC 188, and $\gtrsim$ 1.7 dex for M67, again highly inconsistent with being drawn from the same distribution. This strongly implies that the depletion histories of these two clusters, and therefore clusters in general, can differ.

Second, why does the magnitude of dispersion vary from cluster to cluster? The source of this variance may be related to the initial conditions of open clusters. The early onset of dispersion described above is qualitatively consistent with the picture expected from rotational-mixing. In short, the greatest spread in rotation rates occurs in the first 200 Myr of the MS, so the largest degree of differential depletion would occur then. If the initial AM distribution can vary between clusters, then the Li dispersion that ultimately develops will vary as well. Open cluster rotation distributions typically show a narrow, densely populated band of converged rotators, and a sparsely populated tail towards more rapid rotation (e.g. M37 $-$ Hartman \etal 2009; Pleiades $-$ Hartman \etal 2010; Irwin \& Bouvier 2009, and refs. therein). In this paradigm, the tail would produce the most Li-depleted objects, and the converged stars should show smaller, nearly uniform depletion factors. If rotation is truly responsible for the cosmic variance in LDPs, then the fractional size of the rapid-rotator tail must be what differentiates high-dispersion clusters from low-dispersion clusters. However, given the small fraction of stars in the high-velocity tail in some clusters (e.g. Hartman \etal 2010, Fig. 14), large Li data sets may be needed for a robust measure of the dispersion. Another cluster initial condition that could contribute to the scatter is the number density of members. Stellar mergers are more likely in dense environments, and low mass stars that undergo normal depletion on the pre-MS, then merge into a higher mass star during the MS, would appear scattered below the mean trend in a present day LDP. While both of these scenarios are qualitatively sensible, detailed work must be done to establish their predictions for the range of plausible resulting Li patterns.

Finally, does the dispersion of Li correlate with other observables? Another light element that can be destroyed by mixing on the MS is beryllium. One would expect a corresponding spread of this element, which burns at a temperature of 3.5 million K, to be present in clusters with a large Li dispersion. Since Be survives to deeper layers in stars, the expected pattern for a given theoretical scenario will differ between the elements, but should correlate. Therefore, dual data sets would place additional constraints on theoretical models seeking to explain MS dispersion. Additionally, rich rotation data sets in clusters with extensive Li measurements have recently been obtained (M34 $-$ Meibom \etal 2011; Pleiades $-$ Hartman \etal 2010). If the early AM distributions of clusters impact their Li evolution, this may be observable in the present day rotation patterns of clusters with differing LDPs. 

\section{Discussion}

\subsection{Does Metallicity Impact Pre-MS Lithium Depletion?}

In SSMs, the stronger opacity resulting from an increased metal abundance suppresses the efficiency of radiative energy transport. The result is an increased $T$ gradient, which causes the depth of the surface CZ to increase. Consequently, $T_{BCZ}$ increases, and the rate of Li destruction goes up. This causes pre-MS Li depletion in stars to be highly sensitive to metallicity. We have relied on the validity of this strong theoretical prediction in our calculation of the lithium anomaly. However, the existence of this metallicity effect has been questioned by some authors from the observational side. Jeffries \& James (1999) argued that the Li distribution of the open cluster Blanco 1 is identical to that of the similarly aged Pleiades, despite being claimed to be 0.1-0.2 dex richer in metals. Sestito \etal (2003) reached a similar conclusion from a comparison of NGC 6475 and M 34. While puzzling at the time, these results have since been called into question by more recent calculations of the iron abundance of Blanco 1 (+0.04 $\pm$ 0.02; Ford \etal 2005), NGC 6475 (+0.03 $\pm$ 0.02; Villanova \etal 2009), and M 34 (+0.07 $\pm$ 0.04; Schuler \etal 2003). Nevertheless, there remain open clusters of differing [Fe/H] with apparently similar empirical LDPs, particularly the trio of $\sim$ 1.5 Gyr old clusters: NGC 752, NGC 3680, and IC 4651 (Sestito \etal 2004; Anthony-Twarog \etal 2009, AT09 hereafter). Differential depletion imprinted at ZAMS should persist as cluster evolve on the MS, so these authors argue that the similarity of the current Li patterns suggests pre-MS depletion cannot depend on [Fe/H]. Can this result be reconciled with standard stellar theory? In this section, we address this question by comparing these cluster patterns in absolute and anomaly space. We apply this same analysis to the Hyades and NGC 6633, two 600 Myr old clusters with substantial composition differences. If metallicity is an important factor in determining ZAMS abundance, the metal poor cluster patterns should lie above the metal rich patterns in absolute space, and on top of them in anomaly space. 

\begin{figure*}
\begin{centering}
\includegraphics[width=7.0in]{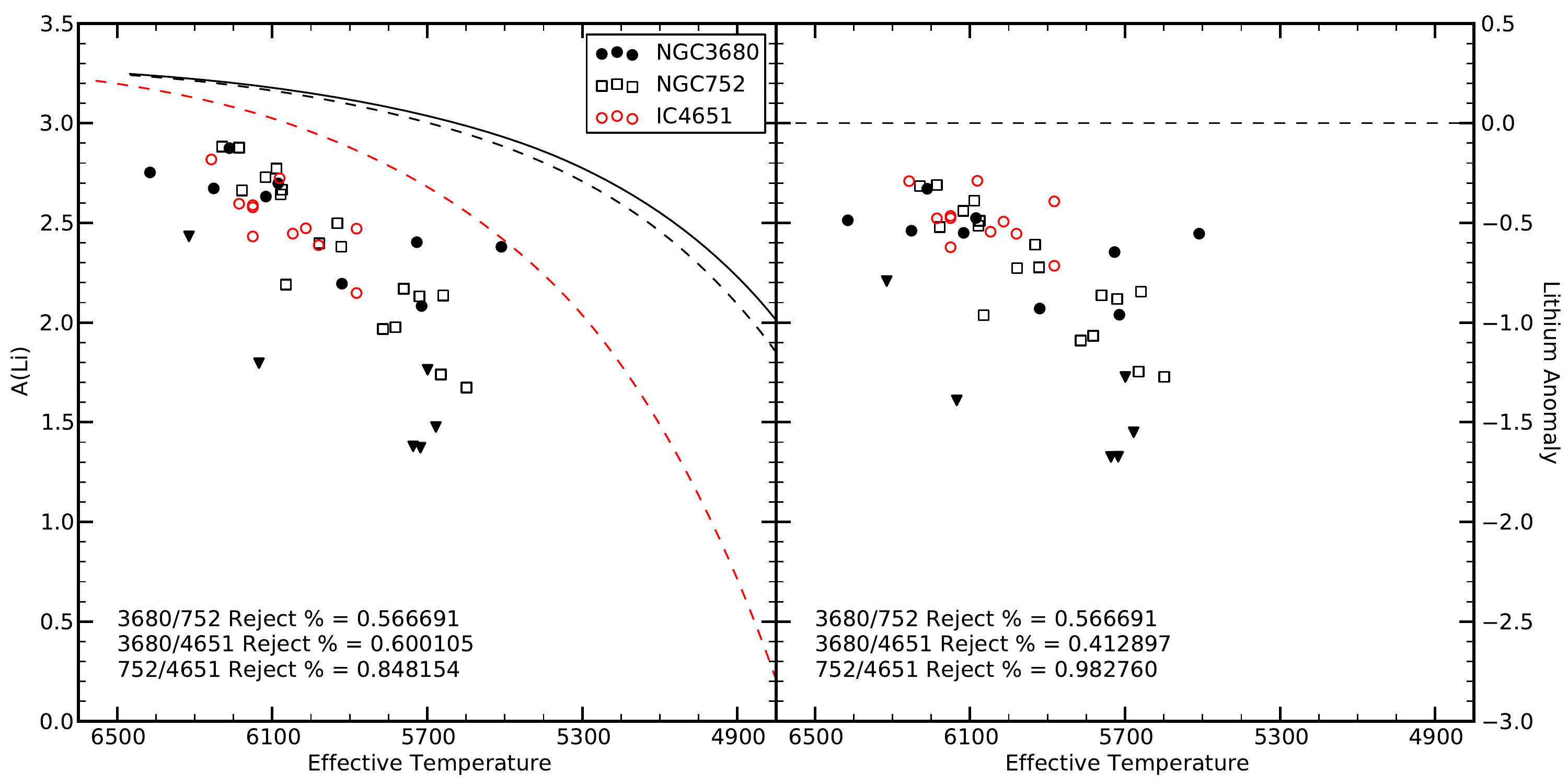}
\caption{A comparison of three 2 Gyr old clusters in A(Li) and Li anomaly space, which demonstrates that they do not rule out metallicity effects on the pre-MS. NGC 3680 (filled black circles), NGC 752 (empty squares), and IC 4651 (empty red circles) data and shown alongside their SSM predictions (solid black, dashed black, and dashed red lines, respectively) in the left panel, and subtracted from their SSM predictions in the right. IC 4651 is $\sim$ 0.2 dex richer in metals, but lines up well with the other clusters in both absolute and anomaly space. The inconclusive rejection probabilities shown in the bottom left of each panel imply that in neither case can the abundance distributions be statistically distinguished.}
\end{centering}
\end{figure*}

\begin{figure*}
\begin{centering}
\includegraphics[width=7.0in]{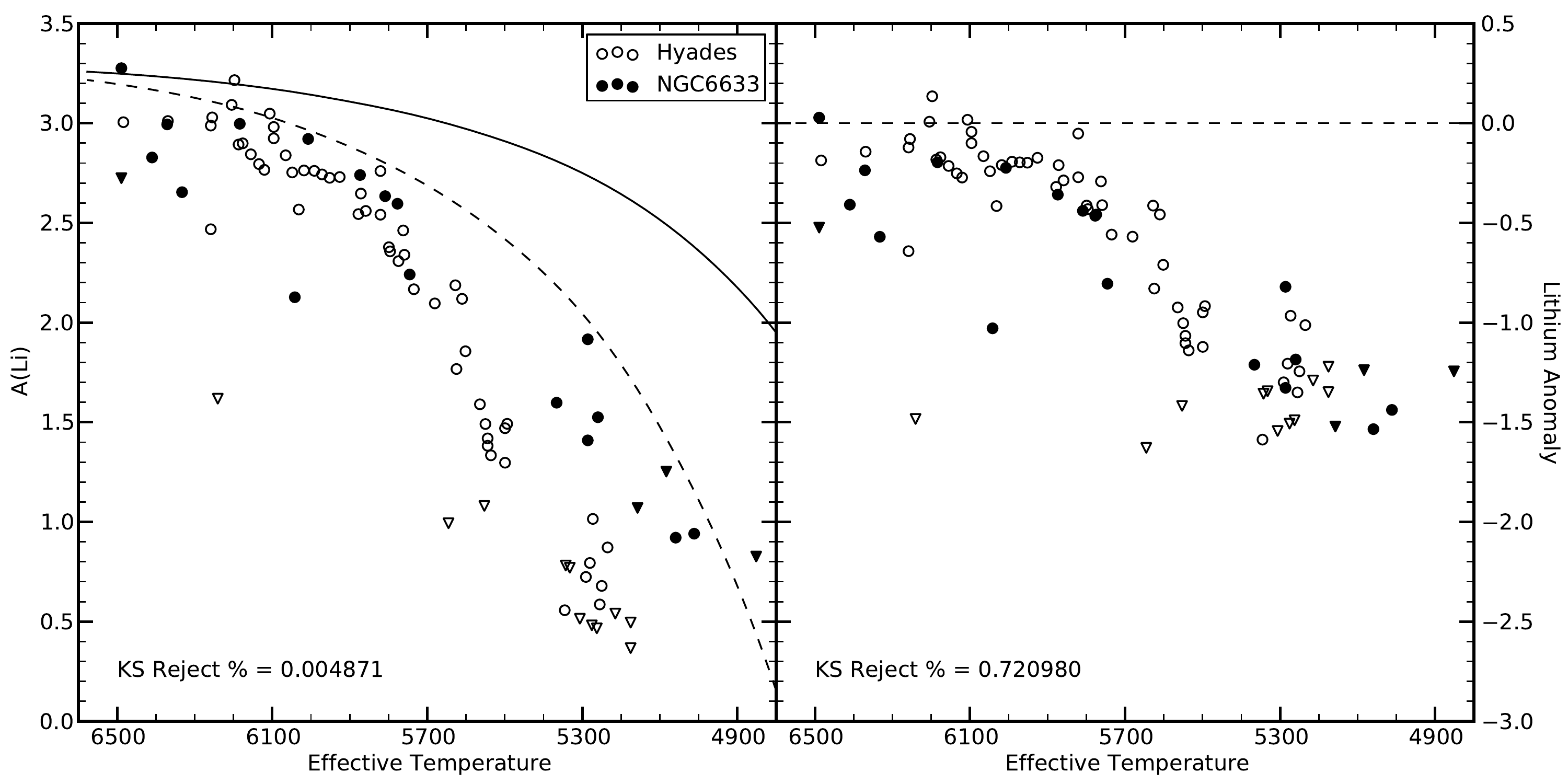}
\caption{A comparison of two 600 Myr old clusters in A(Li) and anomaly space, which demonstrates that metallicity likely impacted their pre-MS abundance distributions. Hyades (empty) and NGC 6633 (filled) data are shown alongside their SSM predictions (dashed black and solid black lines, respectively) in the left, and subtracted from their SSM predictions in the right. The null hypothesis that the Li distributions are equal can be rejected at high confidence in A(Li) space, but cannot be rejected in detrended space. This suggests that SSMs accurately predict the relative depletion of these clusters.}
\end{centering}
\end{figure*}

We first examine the LDPs of the three clusters mentioned above: NGC 752, NGC 3680, and IC 4651. Iron abundances and ages for each are reported by AT09: [Fe/H] = $-$0.05 and age = 1.45 Gyr for NGC 752, [Fe/H] = $-$0.08 and age = 1.75 Gyr for NGC 3680, and [Fe/H] = +0.13 and age = 1.5 Gyr for IC 4651. These abundances are due to high resolution spectroscopy, and appear robust in their relative values (see the discussion in AT09 and references therein). We draw BV photometry and Li EWs from Sestito \etal (2004) for NGC 752, from AT09 for NGC 3680, and from Randich \etal (2000) for IC 4651. We synthesize this data into effective temperatures and A(Li)s using the method described in $\S$2.2, and compute SSM Li predictions using the methodology of $\S$2.3.

These data are presented in the left panel of Fig. 13. NGC 3680 (filled black) and NGC 752 (empty black) have similar Fe abundances, and unsurprisingly show similar median Li trends. IC 4651 (empty red) is $\sim$ 0.2 dex richer in Fe, but lines up well in absolute space with the other clusters. At face value this appears surprising, given the assumption of a strong [Fe/H] dependence of ZAMS Li patterns. However, since IC 4651 data only exists for stars with \teff\ $\gtrsim$ 5800K, a regime that is particularly insensitive to composition ($\Delta$[Fe/H] = 0.2 dex $\rightarrow$ $\Delta$A(Li) = 0.3 dex at 5800K; $\S$3.1), modest scatter may obscure the relative depletion signal. The bottom left shows the results of Kolmogorov-Smirnov (K-S) tests, which demonstrates that in each case, the clusters cannot be statistically distinguished. This supports the visual impression that the cluster LDPs are similar.

The right hand side shows the same data detrended with respect to each clusters' SSM prediction, as described in $\S$4. In the anomaly plane, the relative cluster distributions remain similar. NGC 3680 and NGC 752 have shifted very little with respect to one another, but IC 4651 has trended upward relative to the other clusters. The locus of IC 4651 data appears to lie near the top of the distribution of NGC 3680 and NGC 752, but this visual impression is largely due to a single data point at the top right of the IC 4651 distribution. These clusters still cannot be distinguished by a K-S test, and show no statistical improvement over the comparison in absolute space. This suggests that the cluster LDPs appear similar because they are in a temperature range that is particularly insensitive to metallicity. Furthermore, we cannot rule out differences in the initial cluster abundance of a few tenths of a dex, which could bring these clusters into superior agreement. Cummings (2011) finds a higher initial abundance for more iron rich clusters, so IC 4651 may have begun life at a slightly higher abundance, and suffered slightly higher depletion to wind up at a similar location in anomaly space at 1.5 Gyr. Given the weak dependence of the models on metallicity in this regime, the similarity of the LDPs in anomaly space, and the remaining uncertainties in initial Li abundance and relative [Fe/H], we conclude that these clusters do not convincingly demonstrate that composition is unimportant in determining the ZAMS Li pattern.

Another pair of clusters that are similar in age, and whose relative metallicities have been well-determined, are the Hyades and NGC 6633. The latter is a $\sim$ 600 Myr old (Strobel 1991) open cluster with an iron abundance that is 0.206 $\pm$ 0.40 dex less than the Hyades (Jeffries \etal 2002, J02 hereafter). This is a secure relative value, since both used high resolution spectroscopy, and the analysis of NGC 6633 was carried out in precisely the same way as the analysis that derived the benchmark Hyades data. A substantial Li data set is available for this cluster, due as well to J02. With our prior Hyades [Fe/H] of +0.135, we derive [Fe/H] = $-$0.071 $\pm$ 0.040 for NGC 6633. We then compute SSM Li predictions for this cluster, draw BV photometry and Li EWs from J02, and synthesize the data into effective temperatures and abundances using the procedure described in $\S$2.

The results are presented in Fig. 14. The left panel shows the Hyades (empty circles) and NGC 6633 (filled circles) abundances compared with one another. The Li distributions are similar above 5700K, just as in the above case. However, below 5300K the median NGC 6633 trend is clearly more abundant than the Hyades stars, and a K-S test shows that the probability that these data sets are equivalent is $<$ 0.005. This discrepancy cannot be due to age differences, since if the Hyades and NGC 6633 had identical Li patterns at ZAMS, the Hyades would need to be nearly twice the age of NGC 6633 to produce the difference between median abundances at 5300K. This would imply either NGC 6633 is $\sim$300 Myr old or the Hyades is $\sim$1200 Myr old, both of which are strongly inconsistent with their measured ages. Next, we detrend the clusters and plot them in the right panel of Fig. 14. The agreement between the clusters is significantly improved when viewed in anomaly space, and both the warm and cool NGC 6633 stars appear to lie on top of the Hyades stars. A K-S test reveals that the populations can no longer be statistically distinguished ($\sim$ 0.72 rejection probability), suggesting that the magnitude of differential depletion between these two clusters is accurately predicted by SSMs. This strongly implicates metallicity as the cause of the difference between the Hyades and NGC 6633 LDPs. 

\begin{figure}
\includegraphics[width=3.3in]{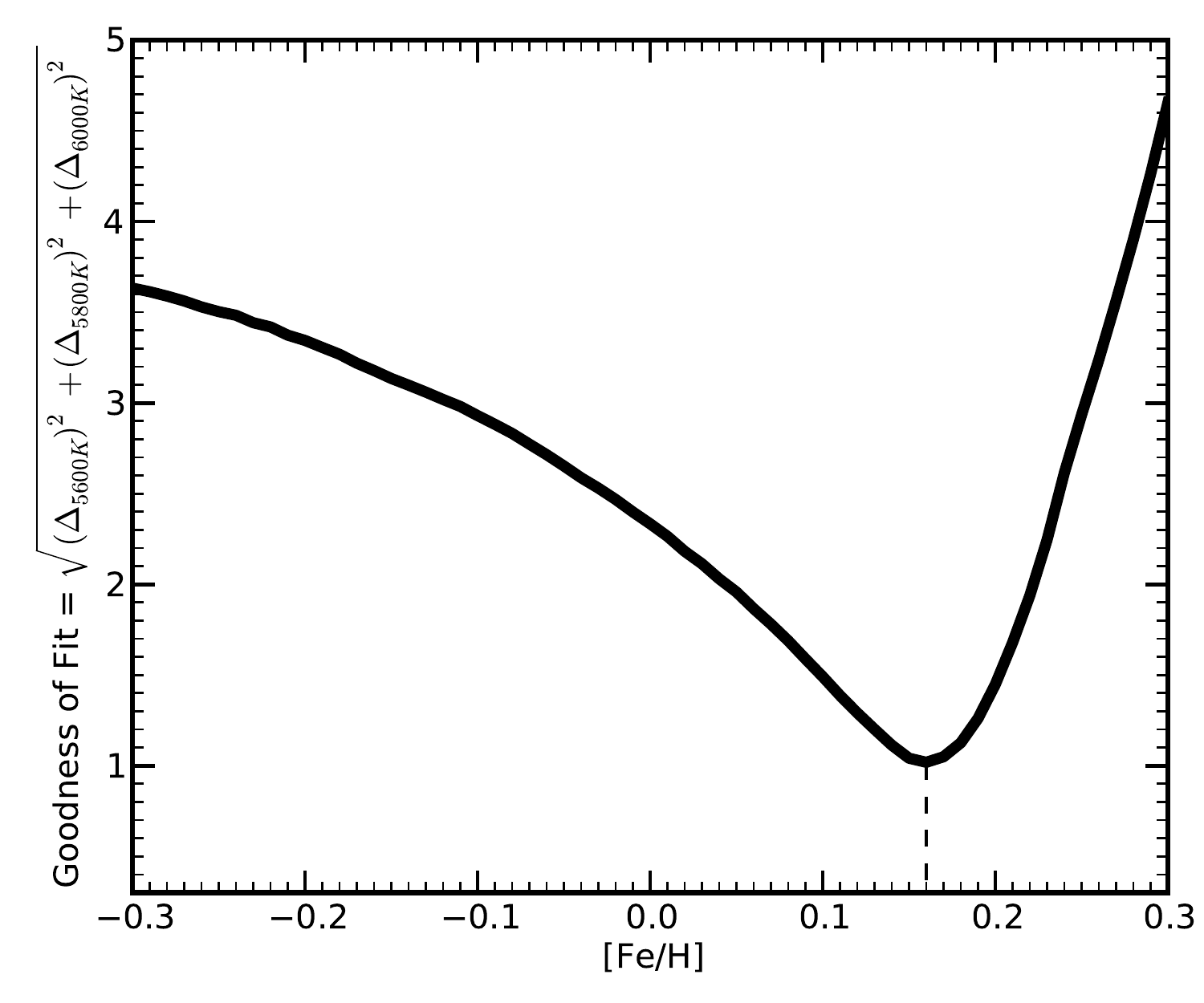}
\caption{The quality of agreement between detrended Hyades Li data and detrended Praesepe Li data, versus the metallicity assumed when detrending Praesepe. The best fit [Fe/H] is at +0.16, implying that this is the most likely metallicity for the Praesepe, according to the method described in $\S$6.1. This is in good agreement with the most recent high-resolution spectroscopic measurement of +0.12 $\pm$ 0.04 (Boesgaard \etal 2013).}
\end{figure}

The similarity of these two clusters in detrended space, and the paucity of secure counter-examples, leads us to conclude that metallicity does impact pre-MS Li destruction. This implies that our calculation of the lithium anomaly is a valid measure of MS depletion. To solidify this result, and to quantify the magnitude of the pre-MS metallicity effect, the LDPs of additional groups of similar-aged clusters must be compared. In particular, their relative compositions must be robustly determined. Recent work by Heiter \etal\ (2013) demonstrates that differences in the analysis method are a larger source of error in [Fe/H] determinations than differences in the quality of observations when using high-resolution spectra. This suggests that a uniform re-analysis of high-resolution Fe observations may be able to resolve this issue. Finally, we have shown that LDP similarity is difficult to establish with solar analogs, so future work should focus on obtaining large samples of Li measurements of \textit{cool} stars. The temperature range is critical, since SSMs predict that a 0.2 dex difference in iron abundance leads to a 0.2 dex difference in Li at 6000K, but a 1.2 dex difference at 5000K. The former may not be detectable, but the latter certainly is.

The extreme sensitivity of Li depletion to metallicity opens the exciting possibility of using Li to infer cluster composition. If the detrended LDPs of equal age clusters are morphologically similar, then the composition of a cluster can be constrained by finding the SSM LDP that produces the best agreement with a cluster of known composition in anomaly space. We tested this concept by estimating the metallicity of the $\sim$ 700 Myr old Praesepe (Salaris \etal 2004) using the well-constrained, and similar aged, Hyades. The [Fe/H] of Praesepe has been a controversial subject: it has been measured as low as [Fe/H] = $+0.04 \pm 0.04$ (Boesgaard \& Friel 1990), and as high as [Fe/H] = $+0.27 \pm 0.10$ (Pace \etal 2008), with the most recent, high-resolution spectroscopic study finding [Fe/H] = $+0.12 \pm 0.04$ (Boesgaard \etal 2013). We first analyzed Praesepe Li EW and BV data from Soderblom \etal (1993b) using the methods described in $\S$2.2, with the metallicity of Boesgaard \etal (2013). Using the machinery of $\S$2.3, we created 0.7 Gyr SSM predictions for [Fe/H]s ranging from -0.30 to 0.30, in steps of 0.01, and detrended the Praesepe data with each. We then compared each detrended pattern with the Hyades anomaly pattern, shown in Fig. 14. To establish a quantitative measure of agreement, we calculated the difference in the median anomalies of the two clusters in three \teff\ bins: 5500-5700K, 5700-5900K, and 5900-6100K. We then weighted each point by the MAD of the data around the median, and added the differences in quadrature. The resulting [Fe/H] vs. goodness-of-fit curve is shown in Fig. 15, with lower goodness-of-fit values demonstrating superior agreement. As can be seen, our best fit [Fe/H] = +0.16, consistent at 1$\sigma$ with the most recent high-resolution spectroscopy estimate for Praesepe. 

We note a few caveats to this method. First, there is a partial degeneracy between cluster [Fe/H] and the initial Li abundance of its members. We assessed the impact of this effect by changing the initial Li abundance of our Praesepe models, and recalculating the best fit [Fe/H]. We found that the best fit [Fe/H] changed to +0.10 when the initial Praesepe abundance was decreased by 0.1 dex, and +0.20 when the initial Praesepe abundance was increased by 0.1 dex. We expect that this is a minor effect, since clusters must be coeval for this method to work. Furthermore, an uncertainty of $\pm$ 0.05 dex in [Fe/H], from an uncertainty of $\pm$ 0.1 dex in initial Li abundance, is small compared to quoted errors on most spectroscopic metallicities. Second, we did not re-derive the stellar parameters of Praesepe members for each [Fe/H]. The \teff\ scale is somewhat sensitive to [Fe/H], so a more careful calculation should take this effect into account when determining goodness of fit. Nevertheless, this proof-of-concept demonstrates that this method merits further attention.

In summary, we have shown that the metal abundance of stars is a crucial factor in shaping pre-MS Li abundance patterns. A trio of 2 Gyr old clusters with different compositions but similar LDPs were shown to be consistent with the prediction of additional depletion in metal-rich stars. Since the stars in these samples are all in an \teff\ regime where the metallicity effect is weak, their apparent similarity in A(Li) space does not constitute evidence against the standard picture. On the other hand, two 600 Myr old clusters of differing composition were shown to have significantly different average Li abundances in absolute space. Once detrended into anomaly space, the clusters became statistically indistinguishable, demonstrating that SSMs accurately predicted the difference between their patterns. Finally, we have shown that the strong dependence of LDPs on composition can be used to constrain the metallicity of open clusters, by comparing them to an equal-age clusters with well known metallicity.

\subsection{Suppressed Convection and the Radius Anomaly as the Origin of the Lithium Dispersion in Young Cool Dwarfs?}

An intriguing feature of the Pleiades Li pattern that standard stellar theory cannot explain is the large abundance dispersion in stars cooler than 5500K. While SSMs predict that Li content is uniquely determined by age, mass, and composition, a spread of $\sim$1.5 dex in A(Li) was discovered in the cool Pleiads by Duncan \& Jones (1983), and later confirmed by S93. The latter authors further showed that in this temperature regime, fast rotating stars are on average \textit{less} depleted than their slow rotating counterparts (S93, Fig. 2). This is counter-intuitive, because rapid rotation is expected to, if anything, drive \textit{additional} mixing, and therefore deplete more Li (Pinsonneault \etal 1989). 

In the years since this discovery, several authors have attempted to explain this dispersion through surface effects that impact $\lambda$6708 \ion{Li}{1} EWs. One possibility is chromospheric activity affecting the physics of line formation (S93; Carlsson \etal 1994; Stuik \etal 1997; Jeffries 1999). These authors argued that such an effect would impact other lines as well, such as the $\lambda$7699 \ion{K}{1} resonance line. S93 initially found a commensurate spread in the $\lambda$7699 \ion{K}{1} and $\lambda$6707 \ion{Li}{1} features in cool Pleiads, supporting the notion that the Li spread is spurious. However, high resolution spectra obtained by King \etal (2010; JK10 hereafter) found no such scatter in the K I line, while confirming a spread of $\gtrsim$ a factor of 2 in the \ion{Li}{1} line. Another potential effect is rotational broadening of the Li absorption line in rapid rotators (Margheim \etal 2002), which could naturally explain why fast spinning stars appear more abundant. However, Margheim (2007) later demonstrated that this cannot account for the full dispersion, since abundances derived though EW analysis and abundances derived through rotationally-broadened spectral synthesis agreed for the majority of the stars. Finally, if cool Pleiads have a large spot filling factor, the neutral \ion{Li}{1} line could be enhanced, as the ionization fraction would be lower on large swaths of the surface. However, JK10 demonstrated that this cannot account for the full effect, as the maximum filling factors necessary to develop the observed EW spread would result in a V-band magnitude dispersion of $\sim$ 1 mag, a feature not observed in Pleiades HR diagrams. It appears that a combination of these effects may contribute to the total dispersion, but that a true underlying Li spread must exist (JK10).

Several mechanisms have been proposed as the cause of this effect. These include episodic accretion during the proto-stellar phase altering the interior temperature (Baraffe \& Chabrier 2010), star-disk coupling inducing strong internal mixing shears (Eggenberger \etal 2012), and increased stellar radii impacting depletion efficiency (K10). In this section, we investigate a version of this final possibility. The rate of pre-MS Li depletion in a star is exquisitely sensitive to $T_{cen}$ during the proto-stellar epoch ($\S$1). If some mechanism reduced $T_{cen}$ below its standard predictions during this time period, the Li depletion rate would be severely suppressed, and the ZAMS abundance would be higher than anticipated. As described in the introduction, numerous studies have identified pre-MS and MS stars whose radii are inflated $\approx$ 5-15\% relative to standard theory. The presence of this radius anomaly would reduce the pressure in the central regions, thus decreasing the temperature required to maintain equilibrium, and slowing the rate of Li burning.

In $\S$6.2.1, we briefly discuss potential underlying causes of the radius anomaly, and describe how such anomalies could induce the observed Li pattern. In $\S$6.2.2, we present inflated stellar models of the Pleiades, and conclude that radius anomalies of the observed magnitude can suppress Li depletion by the required amount. In $\S$6.2.3, we extend this analysis to six additional young clusters, and show that radius anomalies can explain the general pattern of LDP evolution on the pre-MS. In $\S$6.2.4, we discuss implications of radius anomalies for pre-MS evolutionary tracks, stellar initial mass function (IMF) measurements, and the ages of young open clusters. Finally, in $\S$6.2.5 we summarize the results of this section and suggest directions for future study.

\subsubsection{The Radius Anomaly}

Radius anomalies have been observed in a large number of systems (see $\S$1). While the underlying cause is not yet understood, several explanations have been put forward. Accretion from a circumstellar disk may impact the proto-stellar radius by adding thermal energy to the envelope (Palla \& Stahler 1992), but the existence of radius anomalies at a few 100 Myr (e.g. YY Gem; Torres \& Ribas 2002), long after the T Tauri phase, makes this unlikely to be the sole culprit. Theoretical expectations for a given system can also be greatly impacted by errors in metallicity, as stellar radii are sensitive to opacity. This is particularly apparent in interferometric measurements of M dwarfs (Berger \etal 2006), and may indicate missing opacity sources in very cool stars. Nevertheless, radius anomalies persist in many systems with very well measured composition (e.g. CM Dra; Terrien \etal 2012), so this cannot fully explain the phenomenon.

\begin{figure*}
\includegraphics[width=7.0in]{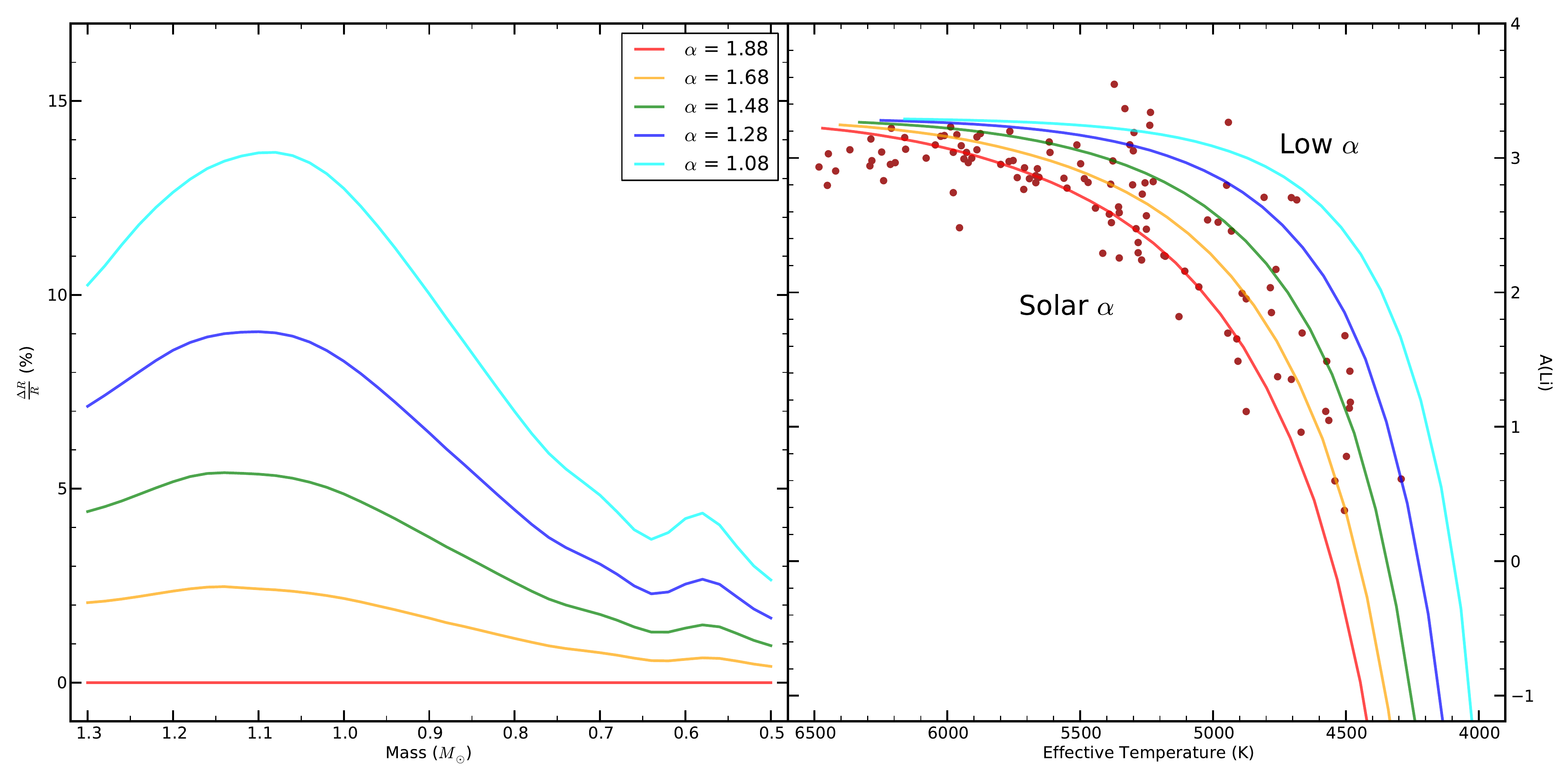}
\caption{The radius anomalies and Li abundances resulting from inflation on the pre-MS. The left panel shows the fractional change in the radius of stellar models with reduced mixing-lengths, relative to the solar-calibrated mixing-length model (red line), as a function of mass. Radius anomalies of $\sim$ 5-15\% on the pre-MS can inhibit SSM depletion enough to account for most cool stars in the Pleiades, as seen in the right panel. This paradigm predicts a large spread developing at fixed effective temperature around 10 Myr.}
\end{figure*}

The current leading explanation is the presence of magnetic fields in low-mass stars (Mullan \& MacDonald 2001; Chabrier \etal 2007; Morales \etal 2008; MacDonald \& Mullan 2012; Feiden \& Chaboyer 2013), which can impact the radius in two ways. First, strong magnetic activity can increase the coverage of spots on the stellar surface, which reduces \teff\ and puffs out the envelope (Andronov \& Pinsonneault 2004). Second, it can inhibit the efficiency of thermal convection, which creates a stronger radiative energy gradient to compensate for the reduced convective energy flux. This enhances the temperature gradient, which leads to a lower surface temperature and causes the radius to expand. This theory is supported by the discovery of a correlation between the radius anomaly and the strength of surface magnetic field proxies. Stars with stronger coronal activity, often inferred as the ratio of X-ray (or H$\alpha$) to bolometric luminosity, show a larger fractional disagreement between their measured radii and SSM predictions (L\'opez-Morales 2007; Clausen \etal 2009; Stassun \etal 2012). Coronal activity also correlates with rotation on the MS (e.g. Wilson 1966; Kraft 1967; Fleming \etal 1989; Bouvier 1990), implying that stars with the most inflated radii may also the most rapidly rotating. Young clusters, such as the Orion Nebula Cluster, possess a large range of X-ray luminosities (Preibisch \etal 2005) and rotation rates (Stassun \etal 1999; Herbst \etal 2001; Herbst \etal 2002) at fixed mass, suggesting that a range of radius anomalies, as large as 0-15\%, may be present in young clusters. 

This presents a plausible explanation for the Li spread in the Pleiades: the most rapidly rotating stars are puffed up relative to standard models on the pre-MS, causing their central temperatures to decrease. This greatly inhibits Li depletion during the pre-MS burning phase, and so they lie significantly above the SSM LDP at the age of the Pleiades. The slowest rotating stars have no radius anomaly, burn Li at a rate consistent with standard predictions, and therefore lie close to the SSM LDP. The result is a range of abundances between the standard prediction for a given \teff, and the Li abundance produced by a star with the maximal radius anomaly. Moreover, the most inflated stars are also the most rapidly rotating, creating an observable correlation between Li abundance and surface rotation rate at 120 Myr. This is an elegant solution to this problem, as an empirically observed radius effect may simultaneously explain the abundance spread, why the median lies above the standard prediction, and why fast rotating stars are less depleted.

\subsubsection{Lithium and Radius in the Pleiades}

Before this possibility can be considered further, we must first show that radius anomalies of the observed magnitude can suppress Li burning by the required amount? To do this, we calculate standard stellar models with inflated radii. This is achieved by decreasing the mixing-length ($\alpha$) in our calculations, which inhibits the efficiency of convection and puffs up the stellar envelope, similar to the effect of strong magnetic fields (Chabrier \etal 2007). We run the models to 120 Myrs, and display them in Fig. 16. The left panel quantifies the radius anomaly induced by each choice of $\alpha$, as a function of mass. The red line represents the solar-calibrated SSM ($\frac{\Delta R}{R} = 0$, by definition), and the yellow, green, blue, and cyan lines represent calculations where the mixing-length is progressively lower in steps of 0.2. The largest anomalies for these models are $\sim$ 10-15\% above 0.9\msun, and $\sim$ 5\% for lower masses, consistent with the range of observed anomalies (Stassun \etal 2012). 

\begin{figure*}
\includegraphics[width=7.0in]{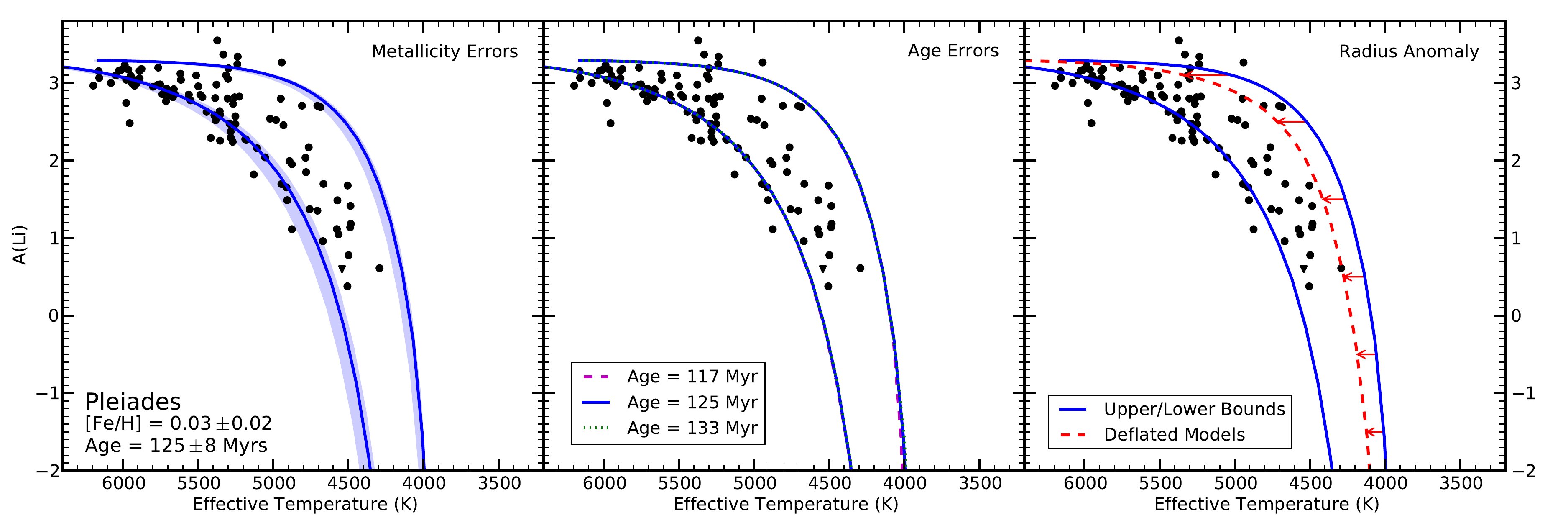}
\caption{The effects of three sources of error on the agreement between the Li predictions of inflated stellar models and the empirical Pleiades LDP. \textit{Left:} Blue lines represent inflated (upper) and standard (lower) models calculated with the best-guess cluster parameters for the Pleiades. The blue shaded bands represent the locations these models could lie given 1$\sigma$ [Fe/H] errors. \textit{Center:} Blue lines again represent best-guess Pleiades models. The red dashed and green dotted lines represent the locations of the models given 1$\sigma$ age errors. \textit{Right:} Blue lines represent the best guess models for the Pleiades, and the red dashed line represents the location of the upper envelope, once the models have deflated to their standard radius predictions.}
\end{figure*}

The right panel of Fig. 16 shows the resulting LDPs, plotted alongside the Pleiades data from Fig. 1. These models were calculated with [Fe/H] = +0.03 (Table 1), and have been corrected for theoretical systematics using the method described in $\S$4. In the absence of additional mixing mechanisms, the qualitative features of the Pleiades LDP are well reproduced by inflated models. The solar mixing-length LDP neatly traces the lower envelope of the data below $\sim$ 6000K, and the most inflated LDP closely brackets the upper envelope of the distribution. Similar to the empirical data, the dispersion in the models is tight above 1 solar mass, and widens considerably towards cooler stars: $\Delta$A(Li) $= 0.21$~dex at 6000K, 0.52~dex at 5500K, 1.18~dex at 5000K, and 2.83~dex at 4500K. Given errors in \teff\ and A(Li), the upper envelope of the distribution appears somewhere between the models with $\alpha$ = 1.28 and 1.08. This corresponds to radius anomalies $\sim$ 2-4\% at 0.6~\msun, 4-7\% at 0.8~\msun, and 8-12\% at 1.0 and 1.2~\msun, well within the observed range of anomalies in rapidly rotating systems.

While this qualitative agreement is excellent, there are a few additional factor to account for. First, the correlation between rotation rate and radius anomaly suggests that stars converge to their standard model radii as they spin down on the MS. When the initially inflated models deflate to their solar-calibrated radii, the surface temperature will increase, and the stars near the top of the Li distribution will move towards the left in our plots. This may reduce the apparent Li dispersion in Fig. 16. Second, errors in the age and metallicity of the Pleiades will alter the location of the theoretical LDPs. To address these issues, we recast our models in Fig. 17. The lower blue line in each panel reflects the standard model LDP at the fiducial metallicity and age, and the upper blue line reflects inflated models calculated with the mixing-length $\alpha$ = $\alpha_{\odot} - 0.8$. The shaded blue regions in the left panel represent the range of locations these bounds could lie in, given the quoted 1$\sigma$ [Fe/H] errors. The predicted bounds still neatly bracket the available data, confirming the qualitative suggestion of Fig. 16. The central panel shows two additional upper and lower LDPs, calculated for 1$\sigma$ age errors. There is no substantial difference between the theoretical prediction at these three ages, suggesting that the relevant Pleiades stars have entered the MS, and no longer change on Myr timescales. Finally, the right panel of Fig. 17 shows an LDP that was inflated on the pre-MS, but whose radii have converged to standard predictions (red line). SSM Li depletion is completed before MS spin-down commences in all stars, so the Li abundance is not affected when stars deflate. The LDP therefore shifts only in \teff, and not in A(Li), as represented by the red arrows. The deflated models still trace the upper envelope of the Pleiades well, confirming that the cool star Li dispersion persists after substantial MS spin-down has occurred.

With these effects included, our models self-consistently explain many features of the Pleiades LDP: 1) the small dispersion in hot stars, 2) the median abundance of hot stars, 3) the dispersion in cool stars, 4) the locations of the upper and lower envelopes of the cool star distribution, and 5) the rotation-Li correlation seen below 5500K. Given that these models were calibrated only to reproduce the empirically-observed radius anomaly, without any regard for the Li predictions, the excellent agreement between data and theory strongly suggests that pre-MS inflation is the cause for the Pleiades Li dispersion.

\begin{figure*}
\includegraphics[width=7.0in]{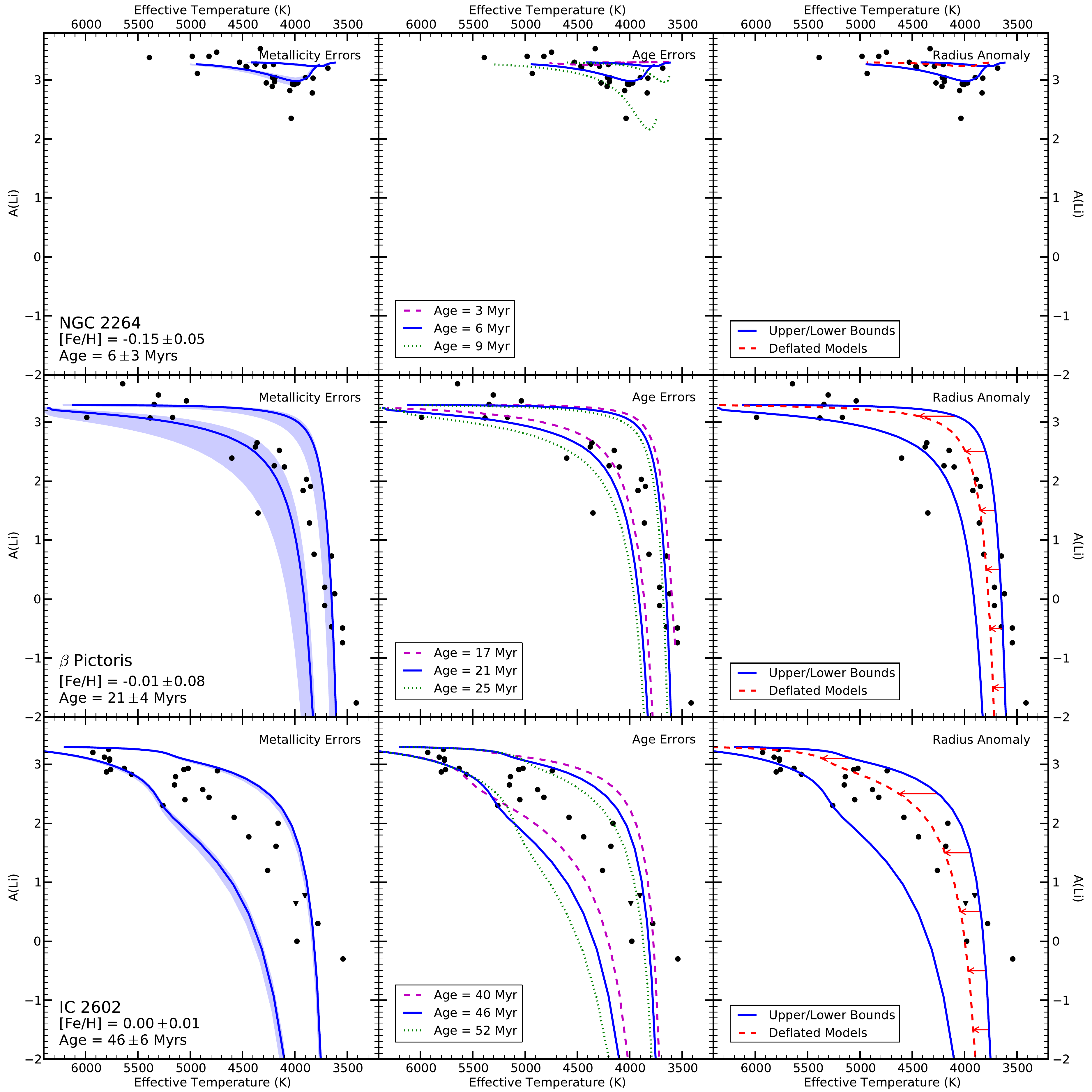}
\caption{The same as in Fig. 17 for NGC 2264 (top), $\beta$ Pictoris (middle), and IC 2602 (bottom).}
\end{figure*}

\begin{figure*}
\includegraphics[width=7.0in]{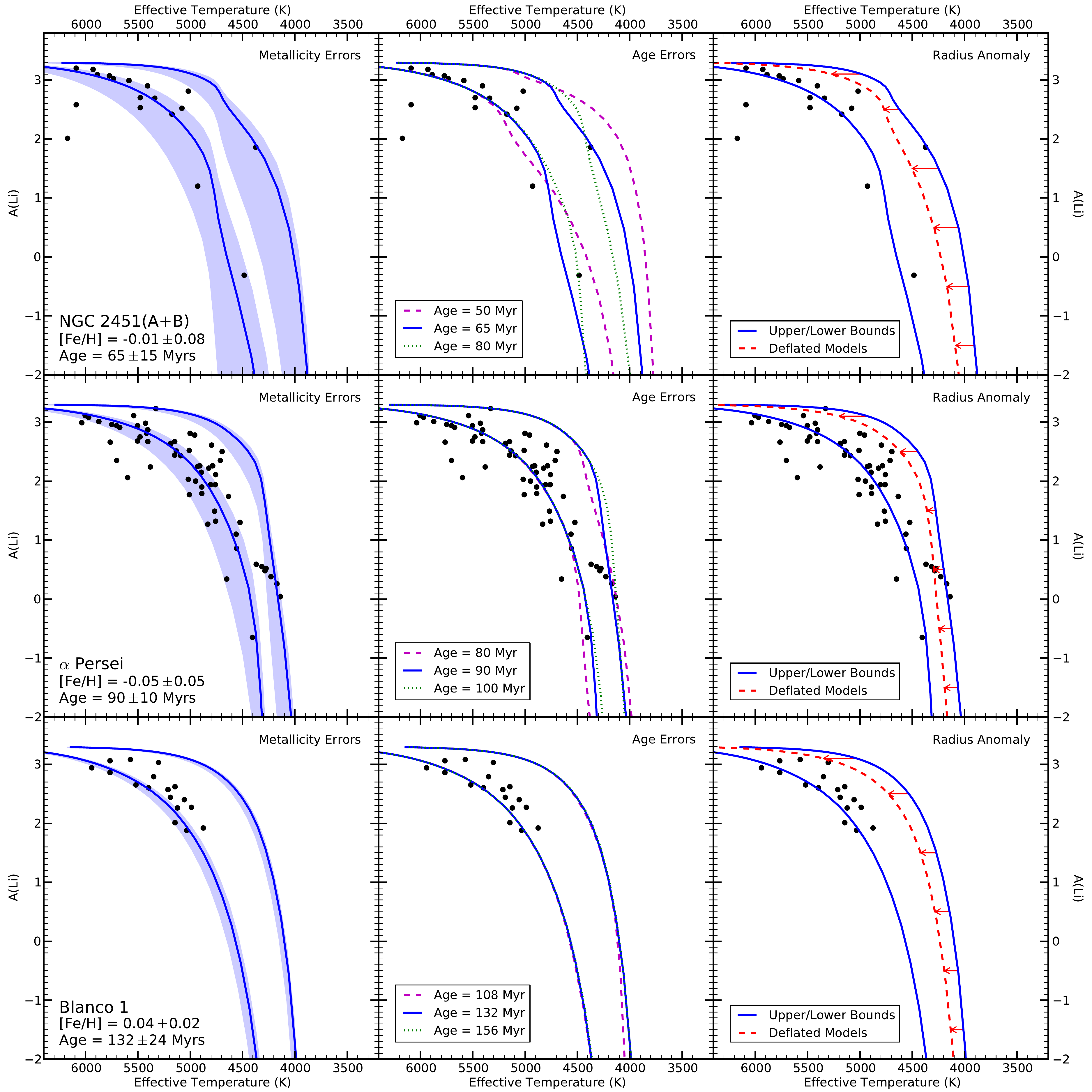}
\caption{The same as in Fig. 17 for NGC 2451 A+B (top), $\alpha$ Persei (middle), and Blanco 1 (bottom).}
\end{figure*}

\subsubsection{Additional Clusters}

As a test of the generality of this picture, we extend this full analysis to six additional young clusters: NGC 2264, $\beta$ Pictoris, IC 2602, NGC 2451 A+B, $\alpha$ Persei, and Blanco 1 (see $\S$2.1.2). These data are shown in Figs. 18 and 19, along with models corresponding to the quoted age and metallicity of each cluster. The meaning of the figures in each row is the same as in Fig. 17. Each has been corrected for theoretical systematics by scaling relative to the calibrated Pleiades LDP described in $\S$4, except NGC 2264, which has not yet suffered enough Li depletion for our systematic corrections to be meaningful.

The agreement between the data and models is good for each cluster. The dispersion present in the Pleiades is clearly a general phenomenon which develops between 6 and 20 Myr, and persists onto the ZAMS. This is consistent with an early origin of the Li spread, which our models predict develops between 8 and 15 Myr, and the subsequent termination of additional depletion until MS mechanisms kick in. There is a dip present at 4000K in NGC 2264, which agrees well with the non-inflated models. At this same temperature, there are no data points that are consistent with the inflated upper envelope. This could reflect a time delay in the development of the dispersion: at this young age, proto-stars are still contracting and spinning up, and therefore may not yet possess their ultimate radius anomaly. In the left columns, each cluster older than NGC 2264 falls within the bounds of our models for a $<1\sigma$ [Fe/H] value. The Pleiades, $\beta$ Pictoris, NGC 2451, and Blanco 1 are all fit well by models generated with the fiducial [Fe/H], but $\alpha$ Per is best fit by a metallicity close to solar. This agrees well with the $\alpha$ Per iron abundance measured by Boesgaard \& Friel (1990), who found an [Fe/H] nearly identical to that of the near-solar Pleiades. The only clusters whose lower distribution is substantially different from the models is IC 2602, which appears to lie a few 100K to the right of the predicted lower bound. While this could simply be due to the small number statistics of this data set, it could also be accounted for by assuming a slightly younger age for the cluster, as the bottom middle panel in Fig. 18 demonstrates. The agreement between these clusters and our predictions further support the arguments of $\S$6.1, where we concluded that metallicity is a key factor in shaping cluster LDPs.

The central columns shows that the early-time LDP is quite sensitive to age. Li patterns rapidly evolve both in A(Li) and in \teff\ during their early stages, before stabilizing around 100 Myr. Each cluster appears consistent with its fiducial age, except IC 2602, whose LDP appears closer to 40~Myr old than 46~Myr old. This suggests that if the composition of a cluster is well known, and the distribution of pre-MS radii is taken into account, LDPs may be a viable method for inferring young cluster ages. This possibility has been doubted in the past, as the cause of the Li dispersion was not understood (i.e. Jeffries \etal 2009). However, these figures suggest that strong age constraints can be obtained through comparison with theory, even with our qualitative models. Furthermore, young cluster Li patterns are a useful way to measure the ages of clusters in the range 5-15 Myr, when our models predict that the dispersion develops. Bell \etal (2013) have recently proposed a revised age scale for several clusters younger than 20 Myr, which suggested new ages approximately double their previous literature values. Li observations could strongly discriminate between their new ages and previous ages in several cases. For example, Bell \etal found an age of 12 Myr for NGC 2362, up from 6.3 Myr (D'Antona \& Mazzitelli 1997). The former age implies significant depletion in some members, similar to the pattern of $\beta$ Pictoris, while the latter suggests only mild depletion and small dispersion, similar to the pattern of NGC 2264. This provides a strong test of the age scale of young clusters.

Finally, the right columns shows that even after the radius anomaly has vanished, the dispersion persists in all clusters older than 10~Myr. One might expect the red lines in the right column to bracket the upper envelope only in older clusters, and not in young clusters, whose members are still rapidly rotating. Indeed, the red line presents a superior qualitative representation of the upper envelope of NGC 2451(A+B), $\alpha$ Per and Blanco 1, suggesting that, by 65 Myr, stars have spun down enough to suppress their radius anomalies. A single star in NGC 2451 lies substantially closer to the still-inflated upper envelope, as do several stars in $\alpha$ Per; this is not surprising, as lower mass stars spend more time on the pre-MS and spin down slower, and thus could still be rapidly rotating at this age. These clusters are likely in transition from an inflated to a deflated upper envelope. By contrast, the red line is clearly a worse fit to the upper envelopes of $\beta$ Pic and IC 2602 than the blue line, implying that the most inflated stars are still rotating quite rapidly at these young ages. The general picture that arises from this plot is that stars develop a Li dispersion around 10 Myr due to a large range in radii at fixed mass, and converge on SSM predictions of radius and \teff\ sometime between 50 and 100 Myr. 

It is worth re-emphasizing that these models were not calibrated to reproduce the Li patterns seen in Figs. 16-19. The mixing-lengths used in our inflated models were chosen to qualitatively reproduce the range of radius anomalies seen in rapidly rotating systems. This implies that the Li patterns presented above are purely predictions of the model, not calibration points. The success of these models in reproducing the depletion patterns of many young and ZAMS clusters is very encouraging, and strongly implicates radius anomalies as a key ingredient in shaping young, cool star abundances. We note that the upper envelopes in Figs. 18 and 19 represent near-maximal inflation, and thus are almost certainly an optimistic prediction of the true upper envelope of the distribution. However, as described at the beginning of this section, several observational effects may artificially increase the measured Li abundance in these stars (e.g. line formation physics, rapid rotation, spots). If one were able to deconvolve the actual abundances from errors brought on by observational effects and Poisson noise, the maximum radius anomaly needed to reflect the upper envelope would likely decrease. Therefore, even in the event that 10-15\% is an optimistic upper limit, this effect may produce a sufficient abundance spread.

It is also worth noting that several open clusters contain a population of \teff\ $\sim$ 5500-6000K stars that lie significantly below the mean trend. These can be seen in the Pleiades in Fig. 17, NGC 2451 A+B and $\alpha$ Per in Fig. 19, the Hyades in Fig. 1, and in many other clusters (i.e. IC 4665 $-$ Jeffries \etal 2009; NGC 3532 $-$ S03). These stars cannot be explained through depletion mechanisms considered in this work, since the one non-standard effect we have included can only suppress Li depletion. Non-member contamination cannot be completely ruled out, even when objects are consistent with membership, which may explain a portion of these objects. However, if some are indeed cluster members, additional physical processes must have influenced their depletion history, the exploration of which is beyond the scope of this paper. We therefore call attention to these objects as possibly reflecting a distinct Li depletion pathway, but do not speculate about their nature. Further observational work, particularly a robust confirmation of cluster membership, should be undertaken to investigate these objects.

\subsubsection{Consequences of the Radius Anomaly in Young Clusters}

\begin{figure}
\includegraphics[width=3.3in]{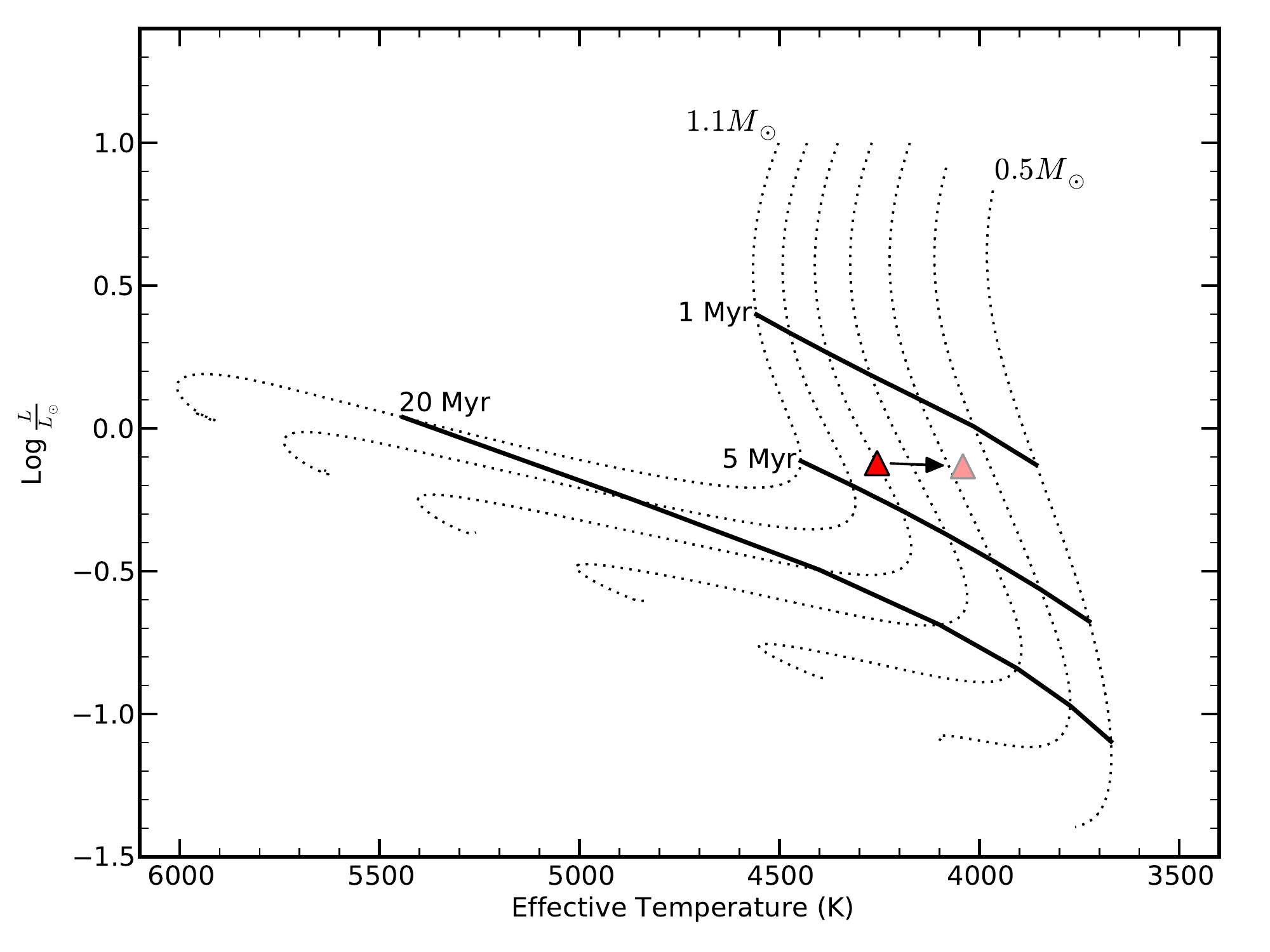}
\caption{The impact of the radius anomaly for the location of pre-MS stars on the HR diagram. Black dotted lines represent the location of SSM evolutionary tracks for M = 1.1-0.5\msun, and the black solid lines show SSM isochrones at 3 ages. The solid red triangle shows the HR diagram location of a 0.9\msun, 3 Myr old SSM, while the faded triangle shows the same model with a radius inflation of $\sim$ 10\%. Underestimates of 30\% and 40\% in the mass and age would result if the stellar parameters of the inflated star were inferred with these standard pre-MS models.}
\end{figure}

The presence of radius anomalies in young systems has consequences for some astrophysical measurements. Fig. 20 shows pre-MS tracks and isochrones for standard stellar models. The solid red triangle represents the HR diagram location of a 0.9~\msun, 3 Myr old model with standard radius, and the faded red triangle represents the HR diagram location of a star with the same mass and age, but with a radius anomaly of $\sim$ 10\%. As can be seen, a puffed-up envelope reduces the \teff\ of a stellar model at fixed mass, while the luminosity remains nearly invariant. This effect shift pre-MS isochrones towards cooler temperatures. The Kelvin-Helmholtz time of inflated stars is also longer, since their initial gravitational potential energy is larger. This shifts the inflated isochrones to younger ages. The result is that while the faded triangle is 0.9~\msun\ and 3 Myr old, isochrone matching would infer a mass of $\sim$ 0.65~\msun\ and an age of $\sim$ 1.7~Myr, resulting in errors of $\sim$ 30\% and 40\% respectively for a 10\% radius anomaly. This is an important effect, since the fundamental parameters of young stars are routinely measured by placement on a grid of pre-MS isochrones. If this method is used to derive the masses and ages of individual stars, inflated objects will be systematically \textit{older} and \textit{more massive} than inferred. 

This could have important consequences for several calculations. First, many studies use masses derived for pre-MS stars to measure the initial mass function (IMF) of the stellar distribution. These studies rely on the one-to-one mapping between absolute magnitudes and mass predicted by standard theory to derive stellar properties. However, the range of surface activity in young systems implies the presence of a range of radius anomalies. This breaks down the one-to-one mapping between observables and stellar parameters, and can severely bias determinations of the cluster mass function. The presence of inflated stars will tend to scatter individual mass measurements towards low values relative to the true mass, causing the inferred IMFs to be bottom-heavy. The impact of this effect can be minimized by studying older clusters, whose rotation distribution has converged to the rotation-mass-age relationship predicted by gyrochronology (e.g. Lachaume \etal 1999; Barnes 2007; Epstein \& Pinsonneault 2014). Other studies use the locus of data of pre-MS stars on the HR diagram to estimate the age of clusters. If a range of radius anomalies is present, the data will not fall on a single isochrone, but instead be spread out between a standard isochrone and a maximally inflated isochrone. If a single isochrone is fit to the mean data trend for low mass stars on the pre-MS, the age could be underestimated by up to 40\%. This error could be minimized by determining the SSM isochrone that best fits the hot side of the data (i.e. the slowest rotating stars), as radius anomalies scatter data towards cooler effective temperatures. Fortunately, this behavior does not impact measurements that use high-mass stars to determine ages. 

Inflated radii may also bias cluster age measurements obtained through the LDB technique. To test this, we run standard and inflated stellar models of fully convective stars to the ZAMS. We find that if a star is inflated on the pre-MS, the onset of Li depletion will be delayed compared to a SSM star of equal mass, and thus systematically bias ages to \textit{younger} values. In particular, if a range of radius anomalies are present in a young cluster, the transition from Li-rich to Li-poor will be fuzzed out over a characteristic mass range, since the transition mass at fixed age depends on the degree of inflation. Our findings are consistent with the studies of Burke \etal (2004), and MacDonald \& Mullan (2010). However, they are not consistent with Yee \& Jensen (2010), who argue that pre-MS inflation may lead to LDB estimates that are \textit{older} than the true ages. They reach this conclusion by noting that inflated stars have a lower \teff, which causes FCS masses to be systematically underestimated by SSM isochrones. This will in turn lower the inferred transition mass, leading to an older inferred age. However, their analysis assumed that the onset of Li destruction is not altered by inflation, which our models strongly rule out. Delaying Li burning drives age estimates to lower values, canceling out to some degree the impact of the effect they describe. This demonstrates a weakness in using \teff\ as a mass proxy: unless you know the radius anomaly of a given star, you cannot use its temperature to derive the mass. There is a similar weakness that arises when luminosity is used as a mass proxy: changes to the mass-radius relationship, and a possible increase of the spot filling-factor through stronger magnetic activity, can impact bolometric corrections. A full investigation of the impact of inflation on LDB ages must therefore take three important effects into account: the delay in the onset of Li destruction in inflated stars, the decreased \teff\ accompanying a larger radius, and changing bolometric corrections due to radius anomalies and magnetic activity. We suggest that future LDB studies aim to identify inflated stars by their CMD locations, and use both absolute and differential photometry to provide the best constraints on their masses. Finally, we note that we have used LDB ages for the clusters in this section. If these ages are incorrect, it could impact the agreement between open cluster data and our models. However, we find that a reduction of $\alpha$ of 0.8 produces a maximum age error of $\sim$10\% which, given the minor sensitivity of LDPs to age (center columns, Figs. 16-18), does not change the qualitative conclusions of this section.

\subsubsection{Summary and Future Prospects}

We have proposed a explanation for the cool star Li abundance spread present in young clusters. This dispersion would arise naturally in the presence of a radius dispersion at fixed mass during the pre-MS. Interferometric observations of single stars and precision tests of stellar models using detached eclipsing binaries evince the existence of inflated radii in some chromospherically active stars. Activity also correlates with rotation, so the most rapidly rotating stars likely possess the largest radius anomalies. If these stars are inflated on the pre-MS, the temperature at the base of their CZs will be less than their slowly rotating counterparts, strongly inhibiting the rate of pre-MS Li depletion, and causing them to be over-abundant in Li on the ZAMS.

We imposed a radius anomaly on our models of $\sim$ 5-15\%, consistent with the upper envelope of observed anomalies, by reducing the efficiency of convection. Li depletion is dramatically inhibited in these models, leading to a $>$~1~dex spread at some temperatures. These models succeeded in qualitatively predicting the upper envelope of the Li distribution of the Pleiades. Furthermore, this theory naturally predicts the rotation-abundance correlation seen in cool Pleiads. We then extended our models to 6 additional young clusters, and found good agreement between the empirical evolution of Li patterns from 0-120 Myr and our models. Since the inflated models were not calibrated to reproduce the Li patterns of these clusters, and instead were calibrated to match the observed upper envelope of radius anomalies in rapidly rotating systems, their success in reproducing the LDPs of several young clusters strongly supports the validity of this theory. Finally, we ended with a discussion of some effects that inflated radii could have on pre-MS isochrones, stellar IMF measurements, and ages determined through the LDB technique. 

Improved measurements of Li and rotation in clusters shortly before and after the epoch of SSM burning will provide a crucial test of this theory. Equal mass stars with different rotation rates will show similar Li abundances before 5 Myr, and radically different abundances by $\sim$ 15 Myr, with the slower rotating stars showing larger degrees of depletion. The 13 Myr old cluster h Per is a prime target for studying the late-time dispersion evolution, as a rich rotation data set has recently become available (Moraux \etal 2013). Fast and slow rotating stars of equal mass can be targeted, providing a measure of both the Li spread at fixed \teff, and of the correlation between Li abundance and rotation at 13 Myr. Further theoretical work remains to solidify the predictions of this theory. In particular, future work should explicitly tie the radius anomaly to stellar rotation rates through empirical correlations (i.e. Stassun \etal 2012), and account for the simultaneous effects of inflation and rotational mixing. In Paper II, we will present calculation of non-inflated, rapidly rotating stellar models that greatly over-predict Li depletion on the early pre-MS. If true pre-MS stars are in fact inflated as described above, this extra mixing may be suppressed such that they enter the ZAMS with the correct abundances. Predictions of the rotating ZAMS Li pattern hinge on the interplay between the suppression of depletion by the inflated radius and the enhancement of depletion by meridional flows and shear instabilities. The relative strengths of these counteracting effects are difficult to anticipate without detailed modeling, but are crucial for establishing the validity of this theory.

\section{Conclusions}

We have attempted to reconcile SSMs with observations at the ZAMS, and use their predictions to uncover the behavior of Li on the MS. Through our investigation, we have reached several conclusions about SSM Li predictions, the progression of Li destruction on the MS, and the development of dispersion in cluster Li patterns.

$\bullet$ Pre-MS Li depletion in solar analogs is accurately predicted by standard stellar models.

SSM Li depletion on the pre-MS is highly sensitive to the assumed input physics. Adopting alternate choices of physical inputs, such as the cluster metal content, the equation of state tables, the opacity tables, and the $^7$Li$(p,\alpha)\alpha$ interaction cross-section, induces substantial changes in the predicted Li pattern at the ZAMS. We measured the impact of each physical input on pre-MS Li destruction, and derived an error band. Within this band, the distribution of solar analogs in the near-ZAMS Pleiades agrees with standard predictions. This allows us to perform an $\textit{ad hoc}$ calibration of the physics in our models, by forcing them to accurately predict the Li pattern of the Pleiades. With these calibrated models, we can predict the ZAMS distribution of other clusters by applying the metallicity scalings predicted by standard theory. This allows us to detrend the metallicity-dependent pre-MS depletion out of empirical cluster data, leaving behind the depletion occurring on the MS, which is insensitive to metallicity and input physics.

$\bullet$ A radius dispersion on the pre-MS is the cause of the cool star Li dispersion in young clusters.

Some chromospherically active stars are known to possess inflated radii relative to standard predictions. We present calculations that demonstrate this radius inflation will decrease the temperature at the base of the CZ, leading to a severe suppression of the rate of Li depletion during the pre-MS. The wide range of chromospheric activity at fixed mass in the cool Pleiads then naturally explains the spread in Li abundances. Correlations between chromospheric activity and rotation have also been reported, demonstrating that the most abundant stars are also the most rapidly rotating. We calculate inflated stellar models, and show that they accurately reproduce the width of the empirical Li spread in the Pleiades, as well as 6 additional young clusters. This strongly implies that a radius dispersion on the pre-MS is the underlying cause of the spread in lithium. We further explored some consequences of this effect for pre-MS isochrones and age measurements of young clusters. Finally, we suggest directions for future theoretical work, and propose dual Li and rotation data sets of $\sim$8-15 Myr old clusters as an observational test.

$\bullet$ Dispersion is a general feature of Li patterns on the MS.

We detrend the strong pre-MS mass trends out of the full suite of clusters considered by SR05, and infer the dispersion within \teff\ bins centered at 6200K, 5900K, 5600K, and 5300K. Within each bin, a dispersion is found to emerge in some clusters during the first 200 Myr. However, the evolution of this dispersion over time depends on the bin. The dispersion present at 100 Myr shows little evolution for the 6200K bin, but a marginal rising trend is seen in the 5900K bin, a more convincing rising trend is seen in the 5600K bin, and a decreasing trend is seen in the 5300K bin. The degree of dispersion can differ between clusters of equal age. This complicated picture likely points to true cosmic variance in the initial conditions of open clusters. Finally, we show that the depletion mass trends identified by SR05 are stronger in anomaly space than in absolute space.

$\bullet$ The rate of MS Li depletion is a strong function of mass.

We isolated the MS depletion signal from two well-studied MS clusters, the Hyades and M67, by subtracting the predicted ZAMS depletion factors from the empirical data of the clusters. Then, we measured the average rate of MS depletion over their lifetimes by comparing the median abundances of the clusters at several temperatures. This is the first measure of MS Li depletion that is not biased by [Fe/H]-dependent, differential pre-MS depletion. We find that low mass stars deplete Li significantly faster than high mass stars on the MS, placing strong constraints on propose mechanisms of MS mixing. We also find that the rate of MS depletion decays over time, in agreement with previous work. These are both robust conclusions, since they persist even in the presence of significant differences in the initial Li abundance of our benchmark clusters. 

$\bullet$ Cluster metallicity is crucial in shaping ZAMS Li abundance patterns.

Some authors have argued that empirical studies of open clusters rule out a strong dependence of pre-MS Li depletion on metallicity. We show that an outstanding case for this argument, involving three 2 Gyr old clusters with dissimilar composition but similar Li patterns, does not rule out the theoretically predicted pre-MS metallicity dependence. We then analyze two $\sim$ 600 Myr old clusters that differ in [Fe/H] by 0.2 dex, and show substantial differences in their cool star Li patterns. Their absolute Li distributions are shown to be different at high confidence. When the metallicity dependence of SSMs is corrected for, they become statistically indistinguishable. Our models accurately predicted the difference in abundances of these clusters, strongly implicating metallicity as an important factor. Since cluster LDPs are sensitive to metallicity, sizable Li data sets can be used to estimate the metallicity of their host clusters by comparison with a similar aged cluster of well known composition. We apply a simple implement ion of this method to the Praesepe cluster, and find a best-fit [Fe/H] = +0.16, in 1$\sigma$ agreement with the recent high-resolution spectroscopic measurement of +0.12 $\pm$ 0.04 (Boesgaard \etal 2013). 

The authors thank Keivan Stassun for his helpful comments on the manuscript.

\input{ms.bbl}
\end{document}

%% file: Tab01.txt
\begin{deluxetable*}{ccccccccc}
\tabletypesize{\scriptsize}
\tablecaption{Benchmark Cluster Data \label{tab:obs}}
\tablewidth{0\textwidth}
\tablehead{
\colhead{Cluster}
&\colhead{Number}
&\colhead{EW(Li)}
&\colhead{B-V}
&\colhead{Age}
&\colhead{[Fe/H]}
&\colhead{E[B-V]}
&\colhead{Parameter Sources} \\
\colhead{}
&\colhead{of stars}
&\colhead{Source}
&\colhead{Source}
&\colhead{}
&\colhead{}
&\colhead{}
&\colhead{(Age, [Fe/H], E[B-V])}
} 
\startdata
Pleiades & 115 & (1) & (1) & 125 $\pm$ 5 Myr & +0.03 $\pm$ 0.02 & Variable$^*$ & (2), (3), (1)  \\
Hyades & 65 & (4) & (5) & 625 $\pm$ 25 Myr & +0.135 $\pm$ 0.005 & 0.01 & (6), (7), (8)\\
M67 & 56 & (9) & (10) & 3.9 $\pm$ 0.6 Gyr & +0.01 $\pm$ 0.03 & 0.041 & (11), (11), (12)
\enddata
\tablecomments {Sources: (1) Soderblom \etal 1993a (S93); (2) Stauffer \etal (1998); (3) Soderblom \etal (2009); (4) Thorburn \etal (1993); (5) Johnson \& Knuckles (1955); (6) Perryman \etal (1998); (7) Cummings \etal (2012); (8) Lyng\aa\ Catalogue (1987 $-$ fifth edition; Lyng\aa\ 1985); (9) Pasquini \etal (2008); (10) Montgomery \etal (1993); (11) Castro \etal (2011); (12) Taylor \etal (2007). 
*Due to significant differential extinction in the Pleiades, S93 de-reddened each star individually. We adopt their values.
}

\end{deluxetable*}

%% file: Tab02.txt
\begin{deluxetable*}{ccccccccc}
\tabletypesize{\scriptsize}
\tablecaption{Additional Cluster Data \label{tab:obs}}
\tablewidth{0\textwidth}
\tablehead{
\colhead{Cluster}
&\colhead{A(Li)/\teff}
&\colhead{Age}
&\colhead{[Fe/H]}
&\colhead{Parameter Sources} \\
\colhead{}
&\colhead{Source}
&\colhead{(Myr)}
&\colhead{}
&\colhead{(Age, [Fe/H])}
} 
\startdata
NGC 2264         & (1)  &   6 $\pm$ 3  & -0.15            & (2), (3)   \\
$\beta$ Pictoris & (4)  &  21 $\pm$ 4  &  0.01 $\pm$ 0.08 & (5), (6)   \\
IC 2602          & (7)  &  46 $\pm$ 6  &  0.00 $\pm$ 0.01 & (8), (9)   \\
NGC 2451 A+B     & (10) &  65 $\pm$ 15 & -0.01 $\pm$ 0.08 & (11), (11) \\
$\alpha$ Persei  & (12) &  90 $\pm$ 10 & -0.05 $\pm$ 0.05 & (13), (14) \\
Blanco 1         & (15) & 132 $\pm$ 24 &  0.04 $\pm$ 0.02 & (16), (17) \\
\enddata
\tablecomments {Sources: (1) Soderblom \etal (1999); (2) King \etal (2000); (3) Dahm (2008); (4) Torres \etal (2006); (5) Binks \& Jeffries (2013); (6) Viana Almeida \etal (2009); (7) Randich \etal (2001); (8) Dobbie \etal (2010); (9) D'Orazi \etal (2009); (10) H\"unsch \etal (2004); (11) H\"unsch \etal (2003); (12) Balachandran \etal (2011); (13) Stauffer \etal (1999); (14) Boesgaard \& Friel (1990); (15) Jeffries \& James (1999); (16) Cargile \etal (2010); (17) Ford \etal (2005)
}

\end{deluxetable*}

%% file: Tab03.txt
\begin{deluxetable*}{ccccccccc}
\tabletypesize{\scriptsize}
\tablecaption{Pre-MS Systematic Errors \label{tab:obs}}
\tablewidth{0\textwidth}
\tablehead{
\colhead{Error Source}
&\colhead{Fiducial}
&\colhead{Alternate}
&\colhead{Alternate}
&\colhead{}
&\colhead{}
&\colhead{$\Delta$A(Li) at...}
&\colhead{}
&\colhead{}\\
\cline{5-9} 
\colhead{}
&\colhead{Choice}
&\colhead{Choice}
&\colhead{Ref.}
&\colhead{6500K}
&\colhead{6000K}
&\colhead{5500K}
&\colhead{5000K}
&\colhead{4500K}
}
\startdata
Equation of State                        & OPAL 2006  & SCV        & (1) & +0.03     & +0.11         & +0.31         & +0.64         & +0.83         \\
Model Atmosphere                         & Kurucz     & Allard     & (2) & $-$0.01   & $-$0.02       & +0.01         & +0.16         & +0.71         \\
High-Temp Opacity                        & OP17       & OPAL17     & (3) & +0.01     & +0.02         & +0.03         & $-$0.04       & $-$0.40       \\
Low-Temp Opacity                         & Alex 2006  & Alex 1995  & (4) & $-$0.03   & $-$0.08       & $-$0.17       & $-$0.32       & $-$0.67       \\
$^7$Li$(p,\alpha)\alpha$ Cross Section   & Lamia 2012 & $\pm$ 9\%  &     &$\pm$0.01  & $\pm$0.03     & +0.07/$-$0.08 & +0.17/$-$0.19 & +0.45/$-$0.52 \\
Initial Solar Composition                & GS98       & Asp09      & (5) & +0.04     & +0.14         & +0.37         & +0.90         & +2.01         \\
$\Delta$Y/$\Delta$Z                      & 1.13       & $\pm$ 0.2  &     &$\pm$0.00 & +0.01/$-$0.01 & +0.03/$-$0.03 & +0.09/$-$0.08 & +0.29/$-$0.26  \\
&&&&&&& \\
Total $+$ Error                          &            &            &     & +0.05     & +0.18         & +0.49         & +1.13         & +2.35         \\
Total $-$ Error                          &            &            &     & $-$0.03   & $-$0.09       & $-$0.19       & $-$0.38       & $-$0.97       
\enddata
\tablecomments {Citations are as follow: (1) Saumon \etal (1995); (2) Allard \etal (1997); (3) Iglesias \& Rogers (1996); (4) Alexander \& Ferguson (1994); (5) Asplund \etal (2009)
}

\end{deluxetable*}

%% file: Tab04.txt
\begin{deluxetable*}{cccccccc}
\tabletypesize{\scriptsize}
\tablecaption{Benchmark Cluster Lithium Anomalies \label{tab:obs}}
\tablewidth{0\textwidth}
\tablehead{
\colhead{Curves of}
&\colhead{Cluster}
&\multicolumn{5}{c}{Anomaly At...} \\
\cline{3-7}
\colhead{Growth}
&\colhead{}
&\colhead{6200K}
&\colhead{6000K}
&\colhead{5800K}
&\colhead{5600K}
&\colhead{5250K}
}
\startdata
\textit{S93} & Pleiades              & $-$0.157 $\pm$ 0.039 & $-$0.012 $\pm$ 0.052 &    0.033 $\pm$ 0.042 &    0.021 $\pm$ 0.047  &        0.291 $\pm$ 0.177 \\
             & Hyades                & $-$0.177 $\pm$ 0.119 & $-$0.195 $\pm$ 0.030 & $-$0.320 $\pm$ 0.041 & $-$0.964 $\pm$ 0.103  & $<$ $-$1.345             \\
             & M67                   & $-$0.627 $\pm$ 0.089 & $-$0.913 $\pm$ 0.075 & $-$1.697 $\pm$ 0.134 & $<$ $-$2.415          & $-$                      \\ \\
             & Pleiades $\to$ Hyades & 0.020 $\pm$ 0.125    & 0.183 $\pm$ 0.060    & 0.353 $\pm$ 0.058    &     0.984 $\pm$ 0.113 & $>$ 1.636                \\
             & Pleiades $\to$ M67    & 0.471 $\pm$ 0.097    & 0.901 $\pm$ 0.092    & 1.730 $\pm$ 0.141    & $>$ 2.436             & $-$                      \\
             & Hyades $\to$ M67      & 0.451 $\pm$ 0.149    & 0.718 $\pm$ 0.081    & 1.377 $\pm$ 0.140    & $>$ 1.451             & $-$                      \\ \\
\hline \\
\textit{S03} & Pleaides              & $-$0.141 $\pm$ 0.038 &    0.001 $\pm$ 0.052 &    0.039 $\pm$ 0.041 &    0.018 $\pm$ 0.046  &        0.259 $\pm$ 0.174 \\
             & Hyades                & $-$0.162 $\pm$ 0.127 & $-$0.177 $\pm$ 0.031 & $-$0.306 $\pm$ 0.042 & $-$1.085 $\pm$ 0.124  & $<$ $-$1.549             \\
             & M67                   & $-$0.623 $\pm$ 0.095 & $-$0.950 $\pm$ 0.083 & $-$1.891 $\pm$ 0.127 & $<$ $-$1.980          & $-$                      \\ \\
             & Pleaides $\to$ Hyades &    0.021 $\pm$ 0.132 &    0.177 $\pm$ 0.060 &    0.345 $\pm$ 0.059 &     1.103 $\pm$ 0.133 & $>$ 1.808                \\
             & Pleaides $\to$ M67    &    0.482 $\pm$ 0.103 &    0.950 $\pm$ 0.098 &    1.930 $\pm$ 0.133 & $>$ 1.998             & $-$                      \\
             & Hyades $\to$ M67      &    0.461 $\pm$ 0.159 &    0.773 $\pm$ 0.089 &    1.585 $\pm$ 0.133 & $>$ 0.895             & $-$                      \\
\enddata
\tablecomments {Absolute and relative MS Li depletion factors. The upper values were calculated with the curves of growth of Soderblom \etal (1993a), and the lower values with the curves of growth of Steinhauer (2003).
}

\end{deluxetable*}

%% file: ms.bbl
\begin{thebibliography}{1}
\expandafter\ifx\csname natexlab\endcsname\relax\def\natexlab#1{#1}\fi

\bibitem[Alexander 
\& Ferguson(1994)]{1994ApJ...437..879A} Alexander, D.~R., \& Ferguson, J.~W.\ 1994, \apj, 437, 879 

\bibitem[Allard et 
al.(1997)]{1997ARA&A..35..137A} Allard, F., Hauschildt, P.~H., Alexander, D.~R., \& Starrfield, S.\ 1997, \araa, 35, 137

\bibitem[An et al.(2007)]{2007ApJ...655..233A} An, D., Terndrup, D.~M., 
Pinsonneault, M.~H., et al.\ 2007, \apj, 655, 233 

\bibitem[Anders 
\& Grevesse(1989)]{1989GeCoA..53..197A} Anders, E., \& Grevesse, N.\ 1989, \gca, 53, 197 

\bibitem[Andronov 
\& Pinsonneault(2004)]{2004ApJ...614..326A} Andronov, N., \& Pinsonneault, M.~H.\ 2004, \apj, 614, 326 

\bibitem[Anthony-Twarog et al.(2009)]{2009AJ....138.1171A} Anthony-Twarog, 
B.~J., Deliyannis, C.~P., Twarog, B.~A., Croxall, K.~V., 
\& Cummings, J.~D.\ 2009, \aj, 138, 1171 

\bibitem[Asplund et 
al.(2009)]{2009ARA&A..47..481A} Asplund, M., Grevesse, N., Sauval, A.~J., \& Scott, P.\ 2009, \araa, 47, 481 

\bibitem[Balachandran(1995)]{1995ApJ...446..203B} Balachandran, S.\ 1995, 
\apj, 446, 203 

\bibitem[Balachandran et al.(2011)]{2011MNRAS.410.2526B} Balachandran, 
S.~C., Mallik, S.~V., \& Lambert, D.~L.\ 2011, \mnras, 410, 2526 

\bibitem[Baraffe 
\& Chabrier(2010)]{2010A&A...521A..44B} Baraffe, I., \& Chabrier, G.\ 2010, \aap, 521, A44 

\bibitem[Barnes(2007)]{2007ApJ...669.1167B} Barnes, S.~A.\ 2007, \apj, 669, 
1167 

\bibitem[Basri et al.(1996)]{1996ApJ...458..600B} Basri, G., Marcy, G.~W., 
\& Graham, J.~R.\ 1996, \apj, 458, 600 

\bibitem[Bell et al.(2013)]{2013MNRAS.434..806B} Bell, C.~P.~M., Naylor, 
T., Mayne, N.~J., Jeffries, R.~D., 
\& Littlefair, S.~P.\ 2013, \mnras, 434, 806 

\bibitem[Berger et al.(2006)]{2006ApJ...644..475B} Berger, D.~H., Gies, 
D.~R., McAlister, H.~A., et al.\ 2006, \apj, 644, 475 

\bibitem[Bildsten et al.(1997)]{1997ApJ...482..442B} Bildsten, L., Brown, 
E.~F., Matzner, C.~D., \& Ushomirsky, G.\ 1997, \apj, 482, 442 

\bibitem[Binks 
\& Jeffries(2013)]{2013MNRAS.tmpL.192B} Binks, A.~S., \& Jeffries, R.~D.\ 2013, \mnras, L192

\bibitem[Boesgaard 
\& Steigman(1985)]{1985ARA&A..23..319B} Boesgaard, A.~M., \& Steigman, G.\ 1985, \araa, 23, 319 

\bibitem[Boesgaard 
\& Tripicco(1986)]{1986ApJ...302L..49B} Boesgaard, A.~M., \& Tripicco, M.~J.\ 1986, \apjl, 302, L49 

\bibitem[Boesgaard 
\& Friel(1990)]{1990ApJ...351..467B} Boesgaard, A.~M., \& Friel, E.~D.\ 1990, \apj, 351, 467

\bibitem[Boesgaard et al.(2013)]{2013ApJ...775...58B} Boesgaard, A.~M., 
Roper, B.~W., \& Lum, M.~G.\ 2013, \apj, 775, 58 

\bibitem[Bouvier(1990)]{1990AJ.....99..946B} Bouvier, J.\ 1990, \aj, 99, 
946 

\bibitem[Bouvier et 
al.(1997)]{1997A&A...323..139B} Bouvier, J., Rigaut, F., \& Nadeau, D.\ 1997, \aap, 323, 139 

\bibitem[Boyajian et al.(2008)]{2008ApJ...683..424B} Boyajian, T.~S., 
McAlister, H.~A., Baines, E.~K., et al.\ 2008, \apj, 683, 424 

\bibitem[Boyajian et al.(2012)]{2012ApJ...757..112B} Boyajian, T.~S., von 
Braun, K., van Belle, G., et al.\ 2012, \apj, 757, 112 

\bibitem[Burke et al.(2004)]{2004ApJ...604..272B} Burke, C.~J., 
Pinsonneault, M.~H., \& Sills, A.\ 2004, \apj, 604, 272 

\bibitem[Cargile et al.(2010)]{2010ApJ...725L.111C} Cargile, P.~A., James, 
D.~J., \& Jeffries, R.~D.\ 2010, \apjl, 725, L111 

\bibitem[Carlsson et 
al.(1994)]{1994A&A...288..860C} Carlsson, M., Rutten, R.~J., Bruls, J.~H.~M.~J., \& Shchukina, N.~G.\ 1994, \aap, 288, 860 

\bibitem[Casagrande et al.(2007)]{2007MNRAS.382.1516C} Casagrande, L., 
Flynn, C., Portinari, L., Girardi, L., 
\& Jimenez, R.\ 2007, \mnras, 382, 1516 

\bibitem[Casagrande et 
al.(2010)]{2010A&A...512A..54C} Casagrande, L., Ram{\'{\i}}rez, I., Mel{\'e}ndez, J., Bessell, M., \& Asplund, M.\ 2010, \aap, 512, A54 

\bibitem[Castro et 
al.(2011)]{2011A&A...526A..17C} Castro, M., Do Nascimento, J.~D., Jr., Biazzo, K., Mel{\'e}ndez, J., \& de Medeiros, J.~R.\ 2011, \aap, 526, A17 

\bibitem[Chaboyer et al.(1995)]{1995ApJ...441..876C} Chaboyer, B., 
Demarque, P., \& Pinsonneault, M.~H.\ 1995, \apj, 441, 876 

\bibitem[Chabrier et 
al.(2007)]{2007A&A...472L..17C} Chabrier, G., Gallardo, J., \& Baraffe, I.\ 2007, \aap, 472, L17 

\bibitem[Clausen et 
al.(2009)]{2009A&A...502..253C} Clausen, J.~V., Bruntt, H., Claret, A., et al.\ 2009, \aap, 502, 253 

\bibitem[Cummings(2011)]{2011PhDT.......192C} Cummings, J.\ 2011, 
Ph.D.~Thesis, 

\bibitem[Cummings et al.(2012)]{2012AJ....144..137C} Cummings, J.~D., 
Deliyannis, C.~P., Anthony-Twarog, B., Twarog, B., 
\& Maderak, R.~M.\ 2012, \aj, 144, 137 

\bibitem[Cyburt et al.(2004)]{2004PhRvD..69l3519C} Cyburt, R.~H., Fields, 
B.~D., \& Olive, K.~A.\ 2004, \prd, 69, 123519 

\bibitem[Cyburt et al.(2008)]{2008JCAP...11..012C} Cyburt, R.~H., Fields, 
B.~D., \& Olive, K.~A.\ 2008, JCAP, 11, 12 

\bibitem[D'Antona 
\& Mazzitelli(1994)]{1994ApJS...90..467D} D'Antona, F., \& Mazzitelli, I.\ 1994, \apjs, 90, 467 

\bibitem[D'Antona 
\& Mazzitelli(1997)]{1997MmSAI..68..807D} D'Antona, F., \& Mazzitelli, I.\ 1997, \memsai, 68, 807 

\bibitem[D'Orazi 
\& Randich(2009)]{2009A&A...501..553D} D'Orazi, V., \& Randich, S.\ 2009, \aap, 501, 553

\bibitem[Dahm(2008)]{2008hsf1.book..966D} Dahm, S.~E.\ 2008, Handbook of 
Star Forming Regions, Volume I, 966

\bibitem[Demory et 
al.(2009)]{2009A&A...505..205D} Demory, B.-O., S{\'e}gransan, D., Forveille, T., et al.\ 2009, \aap, 505, 205 

\bibitem[Dobbie et al.(2010)]{2010MNRAS.409.1002D} Dobbie, P.~D., Lodieu, 
N., \& Sharp, R.~G.\ 2010, \mnras, 409, 1002 

\bibitem[Duncan 
\& Jones(1983)]{1983ApJ...271..663D} Duncan, D.~K., \& Jones, B.~F.\ 1983, \apj, 271, 663 

\bibitem[Eggenberger et 
al.(2012)]{2012A&A...539A..70E} Eggenberger, P., Haemmerl{\'e}, L., Meynet, G., \& Maeder, A.\ 2012, \aap, 539, A70

\bibitem[Epstein 
\& Pinsonneault(2014)]{2014ApJ...780..159E} Epstein, C.~R., \& Pinsonneault, M.~H.\ 2014, \apj, 780, 159 

\bibitem[Feiden 
\& Chaboyer(2012)]{2012ApJ...757...42F} Feiden, G.~A., \& Chaboyer, B.\ 2012, \apj, 757, 42 

\bibitem[Feiden 
\& Chaboyer(2013)]{2013ApJ...779..183F} Feiden, G.~A., \& Chaboyer, B.\ 2013, \apj, 779, 183 

\bibitem[Ferguson et al.(2005)]{2005ApJ...623..585F} Ferguson, J.~W., 
Alexander, D.~R., Allard, F., et al.\ 2005, \apj, 623, 585 

\bibitem[Fleming et al.(1989)]{1989ApJ...340.1011F} Fleming, T.~A., Gioia, 
I.~M., \& Maccacaro, T.\ 1989, \apj, 340, 1011 

\bibitem[Ford et al.(2005)]{2005MNRAS.364..272F} Ford, A., Jeffries, R.~D., 
\& Smalley, B.\ 2005, \mnras, 364, 272 

\bibitem[Fran{\c c}ois et 
al.(2013)]{2013A&A...552A.136F} Fran{\c c}ois, P., Pasquini, L., Biazzo, K., Bonifacio, P., \& Palsa, R.\ 2013, \aap, 552, A136 

\bibitem[Grevesse 
\& Sauval(1998)]{1998SSRv...85..161G} Grevesse, N., \& Sauval, A.~J.\ 1998, \ssr, 85, 161 

\bibitem[Hartman et al.(2009)]{2009ApJ...691..342H} Hartman, J.~D., Gaudi, 
B.~S., Pinsonneault, M.~H., et al.\ 2009, \apj, 691, 342 

\bibitem[Hartman et al.(2010)]{2010MNRAS.408..475H} Hartman, J.~D., Bakos, 
G.~{\'A}., Kov{\'a}cs, G., \& Noyes, R.~W.\ 2010, \mnras, 408, 475 

\bibitem[Heiter et al.(2013)]{2013arXiv1311.2306H} Heiter, U., Soubiran, 
C., Netopil, M., \& Paunzen, E.\ 2013, arXiv:1311.2306 

\bibitem[Herbig(1965)]{1965ApJ...141..588H} Herbig, G.~H.\ 1965, \apj, 141, 
588 

\bibitem[Herbst et al.(2001)]{2001ApJ...554L.197H} Herbst, W., 
Bailer-Jones, C.~A.~L., \& Mundt, R.\ 2001, \apjl, 554, L197 

\bibitem[Herbst et 
al.(2002)]{2002A&A...396..513H} Herbst, W., Bailer-Jones, C.~A.~L., Mundt, R., Meisenheimer, K., \& Wackermann, R.\ 2002, \aap, 396, 513 

\bibitem[H{\"u}nsch et 
al.(2003)]{2003A&A...402..571H} H{\"u}nsch, M., Weidner, C., \& Schmitt, J.~H.~M.~M.\ 2003, \aap, 402, 571 

\bibitem[H{\"u}nsch et 
al.(2004)]{2004A&A...418..539H} H{\"u}nsch, M., Randich, S., Hempel, M., Weidner, C., \& Schmitt, J.~H.~M.~M.\ 2004, \aap, 418, 539 

\bibitem[Iben(1965)]{1965ApJ...141..993I} Iben, I., Jr.\ 1965, \apj, 141, 
993 

\bibitem[Iglesias 
\& Rogers(1996)]{1996ApJ...464..943I} Iglesias, C.~A., \& Rogers, F.~J.\ 1996, \apj, 464, 943

\bibitem[Irwin 
\& Bouvier(2009)]{2009IAUS..258..363I} Irwin, J., \& Bouvier, J.\ 2009, IAU Symposium, 258, 363 

\bibitem[Irwin et al.(2011)]{2011ApJ...727...56I} Irwin, J., Berta, Z.~K., 
Burke, C.~J., et al.\ 2011, \apj, 727, 56 

\bibitem[Jeffries(1999)]{1999MNRAS.304..821J} Jeffries, R.~D.\ 1999, 
\mnras, 304, 821 

\bibitem[Jeffries 
\& James(1999)]{1999ApJ...511..218J} Jeffries, R.~D., \& James, D.~J.\ 1999, \apj, 511, 218 

\bibitem[Jeffries(2000)]{2000ASPC..198..245J} Jeffries, R.~D.\ 2000, 
Stellar Clusters and Associations: Convection, Rotation, and Dynamos, 198, 
245 

\bibitem[Jeffries et al.(2002)]{2002MNRAS.336.1109J} Jeffries, R.~D., 
Totten, E.~J., Harmer, S., \& Deliyannis, C.~P.\ 2002, \mnras, 336, 1109 

\bibitem[Jeffries et al.(2009)]{2009MNRAS.400..317J} Jeffries, R.~D., 
Jackson, R.~J., James, D.~J., \& Cargile, P.~A.\ 2009, \mnras, 400, 317 

\bibitem[Johnson 
\& Knuckles(1955)]{1955ApJ...122..209J} Johnson, H.~L., \& Knuckles, C.~F.\ 1955, \apj, 122, 209 

\bibitem[Jones et al.(1999)]{1999AJ....117..330J} Jones, B.~F., Fischer, 
D., \& Soderblom, D.~R.\ 1999, \aj, 117, 330 

\bibitem[Koenigl(1991)]{1991ApJ...370L..39K} Koenigl, A.\ 1991, \apjl, 370, 
L39 

\bibitem[King et al.(2000)]{2000AJ....119..859K} King, J.~R., 
Krishnamurthi, A., \& Pinsonneault, M.~H.\ 2000, \aj, 119, 859 

\bibitem[King et al.(2010)]{2010ApJ...710.1610K} King, J.~R., Schuler, 
S.~C., Hobbs, L.~M., \& Pinsonneault, M.~H.\ 2010, \apj, 710, 1610 

\bibitem[Kraft(1967)]{1967ApJ...150..551K} Kraft, R.~P.\ 1967, \apj, 150, 
551 

\bibitem[Kraus et al.(2011)]{2011ApJ...728...48K} Kraus, A.~L., Tucker, 
R.~A., Thompson, M.~I., Craine, E.~R., 
\& Hillenbrand, L.~A.\ 2011, \apj, 728, 48 

\bibitem[Kurucz(1979)]{1979ApJS...40....1K} Kurucz, R.~L.\ 1979, \apjs, 40, 
1 

\bibitem[L{\'o}pez-Morales 
\& Ribas(2005)]{2005ApJ...631.1120L} L{\'o}pez-Morales, M., \& Ribas, I.\ 2005, \apj, 631, 1120 

\bibitem[L{\'o}pez-Morales(2007)]{2007ApJ...660..732L} L{\'o}pez-Morales, 
M.\ 2007, \apj, 660, 732 

\bibitem[Lachaume et 
al.(1999)]{1999A&A...348..897L} Lachaume, R., Dominik, C., Lanz, T., \& Habing, H.~J.\ 1999, \aap, 348, 897 

\bibitem[Lamia et 
al.(2012)]{2012A&A...541A.158L} Lamia, L., Spitaleri, C., La Cognata, M., Palmerini, S., \& Pizzone, R.~G.\ 2012, \aap, 541, A158

\bibitem[Lyng\aa et al. (1985)]{lynga}Lyng\aa, G. 1985, IAUS, 106, 143

\bibitem[Macdonald 
\& Mullan(2010)]{2010ApJ...723.1599M} Macdonald, J., \& Mullan, D.~J.\ 2010, \apj, 723, 1599

\bibitem[MacDonald 
\& Mullan(2012)]{2012MNRAS.421.3084M} MacDonald, J., \& Mullan, D.~J.\ 2012, \mnras, 421, 3084 

\bibitem[Margheim et al.(2002)]{2002AAS...20112403M} Margheim, S.~J., 
Deliyannis, C.~P., King, J.~R., 
\& Steinhauer, A.\ 2002, Bulletin of the American Astronomical Society, 34, \#124.03 

\bibitem[Margheim(2007)]{2007PhDT.........2M} Margheim, S.~J.\ 2007, 
Ph.D.~Thesis,  

\bibitem[Meibom et al.(2011)]{2011ApJ...733..115M} Meibom, S., Mathieu, 
R.~D., Stassun, K.~G., Liebesny, P., \& Saar, S.~H.\ 2011, \apj, 733, 115 

\bibitem[Mendoza et al.(2007)]{2007MNRAS.378.1031M} Mendoza, C., Seaton, 
M.~J., Buerger, P., et al.\ 2007, \mnras, 378, 1031 

\bibitem[Montalb{\'a}n 
\& Schatzman(2000)]{2000A&A...354..943M} Montalb{\'a}n, J., \& Schatzman, E.\ 2000, \aap, 354, 943 

\bibitem[Montgomery et al.(1993)]{1993AJ....106..181M} Montgomery, K.~A., 
Marschall, L.~A., \& Janes, K.~A.\ 1993, \aj, 106, 181 

\bibitem[Morales et 
al.(2008)]{2008A&A...478..507M} Morales, J.~C., Ribas, I., \& Jordi, C.\ 2008, \aap, 478, 507

\bibitem[Moraux et 
al.(2013)]{2013A&A...560A..13M} Moraux, E., Artemenko, S., Bouvier, J., et al.\ 2013, \aap, 560, A13 

\bibitem[Mullan 
\& MacDonald(2001)]{2001ApJ...559..353M} Mullan, D.~J., \& MacDonald, J.\ 2001, \apj, 559, 353 

\bibitem[Pace et 
al.(2008)]{2008A&A...489..403P} Pace, G., Pasquini, L., \& Fran{\c c}ois, P.\ 2008, \aap, 489, 403 

\bibitem[Pace et 
al.(2012)]{2012A&A...541A.150P} Pace, G., Castro, M., Mel{\'e}ndez, J., Th{\'e}ado, S., \& do Nascimento, J.-D., Jr.\ 2012, \aap, 541, A150 

\bibitem[Palla 
\& Stahler(1992)]{1992ApJ...392..667P} Palla, F., \& Stahler, S.~W.\ 1992, \apj, 392, 667

\bibitem[Pasquini et 
al.(2008)]{2008A&A...489..677P} Pasquini, L., Biazzo, K., Bonifacio, P., Randich, S., \& Bedin, L.~R.\ 2008, \aap, 489, 677 

\bibitem[Perryman et 
al.(1998)]{1998A&A...331...81P} Perryman, M.~A.~C., Brown, A.~G.~A., Lebreton, Y., et al.\ 1998, \aap, 331, 81

\bibitem[Pinsonneault et al.(1989)]{1989ApJ...338..424P} Pinsonneault, 
M.~H., Kawaler, S.~D., Sofia, S., \& Demarque, P.\ 1989, \apj, 338, 424 

\bibitem[Pinsonneault et al.(1990)]{1990ApJS...74..501P} Pinsonneault, 
M.~H., Kawaler, S.~D., \& Demarque, P.\ 1990, \apjs, 74, 501 

\bibitem[Pinsonneault(1997)]{1997ARA&A..35..557P} Pinsonneault, M.\ 1997, \araa, 35, 557 

\bibitem[Popper(1997)]{1997AJ....114.1195P} Popper, D.~M.\ 1997, \aj, 114, 
1195 

\bibitem[Preibisch et al.(2005)]{2005ApJS..160..401P} Preibisch, T., Kim, 
Y.-C., Favata, F., et al.\ 2005, \apjs, 160, 401 

\bibitem[Press(1981)]{1981ApJ...245..286P} Press, W.~H.\ 1981, \apj, 245, 
286 

\bibitem[Prisinzano 
\& Randich(2007)]{2007A&A...475..539P} Prisinzano, L., \& Randich, S.\ 2007, \aap, 475, 539

\bibitem[Randich et 
al.(2000)]{2000A&A...356L..25R} Randich, S., Pasquini, L., \& Pallavicini, R.\ 2000, \aap, 356, L25 

\bibitem[Randich et 
al.(2001)]{2001A&A...372..862R} Randich, S., Pallavicini, R., Meola, G., Stauffer, J.~R., \& Balachandran, S.~C.\ 2001, \aap, 372, 862

\bibitem[Randich et 
al.(2009)]{2009A&A...496..441R} Randich, S., Pace, G., Pastori, L., \& Bragaglia, A.\ 2009, \aap, 496, 441 

\bibitem[Rebull et al.(2006)]{2006ApJ...646..297R} Rebull, L.~M., Stauffer, 
J.~R., Megeath, S.~T., Hora, J.~L., \& Hartmann, L.\ 2006, \apj, 646, 297

\bibitem[Ribas(2003)]{2003A&A...398..239R} Ribas, I.\ 2003, \aap, 398, 239

\bibitem[Richer 
\& Michaud(1993)]{1993ApJ...416..312R} Richer, J., \& Michaud, G.\ 1993, \apj, 416, 312

\bibitem[Rogers 
\& Nayfonov(2002)]{2002ApJ...576.1064R} Rogers, F.~J., \& Nayfonov, A.\ 2002, \apj, 576, 1064

\bibitem[Rogers et al.(1996)]{1996ApJ...456..902R} Rogers, F.~J., Swenson, 
F.~J., \& Iglesias, C.~A.\ 1996, \apj, 456, 902

\bibitem[Ryan(2000)]{2000MNRAS.316L..35R} Ryan, S.~G.\ 2000, \mnras, 316, 
L35 

\bibitem[Sacco et 
al.(2007)]{2007A&A...462L..23S} Sacco, G.~G., Randich, S., Franciosini, E., Pallavicini, R., \& Palla, F.\ 2007, \aap, 462, L23 

\bibitem[Salaris et 
al.(2004)]{2004A&A...414..163S} Salaris, M., Weiss, A., \& Percival, S.~M.\ 2004, \aap, 414, 163

\bibitem[Saumon et al.(1995)]{1995ApJS...99..713S} Saumon, D., Chabrier, 
G., \& van Horn, H.~M.\ 1995, \apjs, 99, 713

\bibitem[Schuler et al.(2003)]{2003AJ....125.2085S} Schuler, S.~C., King, 
J.~R., Fischer, D.~A., Soderblom, D.~R., 
\& Jones, B.~F.\ 2003, \aj, 125, 2085

\bibitem[Sestito et 
al.(2003)]{2003A&A...407..289S} Sestito, P., Randich, S., Mermilliod, J.-C., \& Pallavicini, R.\ 2003, \aap, 407, 289

\bibitem[Sestito et 
al.(2004)]{2004A&A...426..809S} Sestito, P., Randich, S., \& Pallavicini, R.\ 2004, \aap, 426, 809

\bibitem[Sestito 
\& Randich(2005)]{2005A&A...442..615S} Sestito, P., \& Randich, S.\ 2005, \aap, 442, 615

\bibitem[Shen et al.(2005)]{2005ApJ...635..608S} Shen, Z.-X., Jones, B., 
Lin, D.~N.~C., Liu, X.-W., \& Li, S.-L.\ 2005, \apj, 635, 608

\bibitem[Skumanich(1972)]{1972ApJ...171..565S} Skumanich, A.\ 1972, \apj, 
171, 565

\bibitem[Soderblom et al.(1993)]{1993AJ....106.1059S} Soderblom, D.~R., 
Jones, B.~F., Balachandran, S., et al.\ 1993, \aj, 106, 1059

\bibitem[Soderblom et al.(1993)]{1993AJ....106.1080S} Soderblom, D.~R., 
Fedele, S.~B., Jones, B.~F., Stauffer, J.~R., 
\& Prosser, C.~F.\ 1993, \aj, 106, 1080

\bibitem[Soderblom et al.(1999)]{1999AJ....118.1301S} Soderblom, D.~R., 
King, J.~R., Siess, L., Jones, B.~F., \& Fischer, D.\ 1999, \aj, 118, 1301

\bibitem[Soderblom et al.(2009)]{2009AJ....138.1292S} Soderblom, D.~R., 
Laskar, T., Valenti, J.~A., Stauffer, J.~R., 
\& Rebull, L.~M.\ 2009, \aj, 138, 1292 

\bibitem[Spite 
\& Spite(1982)]{1982A&A...115..357S} Spite, F., \& Spite, M.\ 1982, \aap, 115, 357

\bibitem[Stassun et al.(1999)]{1999AJ....117.2941S} Stassun, K.~G., 
Mathieu, R.~D., Mazeh, T., \& Vrba, F.~J.\ 1999, \aj, 117, 2941

\bibitem[Stassun et al.(2006)]{2006ApJ...649..914S} Stassun, K.~G., van den 
Berg, M., Feigelson, E., \& Flaccomio, E.\ 2006, \apj, 649, 914

\bibitem[Stassun et al.(2007)]{2007ApJ...664.1154S} Stassun, K.~G., 
Mathieu, R.~D., \& Valenti, J.~A.\ 2007, \apj, 664, 1154

\bibitem[Stassun et al.(2012)]{2012ApJ...756...47S} Stassun, K.~G., 
Kratter, K.~M., Scholz, A., \& Dupuy, T.~J.\ 2012, \apj, 756, 47

\bibitem[Stauffer et al.(1984)]{1984ApJ...280..202S} Stauffer, J.~R., 
Hartmann, L., Soderblom, D.~R., \& Burnham, N.\ 1984, \apj, 280, 202 

\bibitem[Stauffer et al.(1998)]{1998ApJ...499L.199S} Stauffer, J.~R., 
Schultz, G., \& Kirkpatrick, J.~D.\ 1998, \apjl, 499, L199

\bibitem[Stauffer et al.(1999)]{1999ApJ...527..219S} Stauffer, J.~R., 
Barrado y Navascu{\'e}s, D., Bouvier, J., et al.\ 1999, \apj, 527, 219

\bibitem[Steinhauer(2003)]{2003PhDT.......252S} Steinhauer, A.\ 2003, 
Ph.D.~Thesis,

\bibitem[Strobel(1991)]{1991AN....312..177S} Strobel, A.\ 1991, 
Astronomische Nachrichten, 312, 177

\bibitem[Stuik et 
al.(1997)]{1997A&A...322..911S} Stuik, R., Bruls, J.~H.~M.~J., \& Rutten, R.~J.\ 1997, \aap, 322, 911

\bibitem[Swenson 
\& Faulkner(1992)]{1992ApJ...395..654S} Swenson, F.~J., \& Faulkner, J.\ 1992, \apj, 395, 654

\bibitem[Takeda et al.(2013)]{2013PASJ...65...53T} Takeda, Y., Honda, S., 
Ohnishi, T., et al.\ 2013, \pasj, 65, 53

\bibitem[Taylor(2007)]{2007AJ....133..370T} Taylor, B.~J.\ 2007, \aj, 133, 
370 

\bibitem[Terndrup et al.(2002)]{2002ApJ...576..950T} Terndrup, D.~M., 
Pinsonneault, M., Jeffries, R.~D., et al.\ 2002, \apj, 576, 950 

\bibitem[Terrien et al.(2012)]{2012ApJ...760L...9T} Terrien, R.~C., 
Fleming, S.~W., Mahadevan, S., et al.\ 2012, \apjl, 760, L9

\bibitem[Thorburn et al.(1993)]{1993ApJ...415..150T} Thorburn, J.~A., 
Hobbs, L.~M., Deliyannis, C.~P., 
\& Pinsonneault, M.~H.\ 1993, \apj, 415, 150

\bibitem[Tognelli et 
al.(2012)]{2012A&A...548A..41T} Tognelli, E., Degl'Innocenti, S., \& Prada Moroni, P.~G.\ 2012, \aap, 548, A41

\bibitem[Torres 
\& Ribas(2002)]{2002ApJ...567.1140T} Torres, G., \& Ribas, I.\ 2002, \apj, 567, 1140

\bibitem[Torres et 
al.(2006)]{2006A&A...460..695T} Torres, C.~A.~O., Quast, G.~R., da Silva, L., et al.\ 2006, \aap, 460, 695

\bibitem[Torres et 
al.(2010)]{2010A&ARv..18...67T} Torres, G., Andersen, J., \& Gim{\'e}nez, A.\ 2010, \aapr, 18, 67

\bibitem[Viana Almeida et 
al.(2009)]{2009A&A...501..965V} Viana Almeida, P., Santos, N.~C., Melo, C., et al.\ 2009, \aap, 501, 965

\bibitem[Villanova et 
al.(2009)]{2009A&A...504..845V} Villanova, S., Carraro, G., \& Saviane, I.\ 2009, \aap, 504, 845 

\bibitem[Wilson(1966)]{1966ApJ...144..695W} Wilson, O.~C.\ 1966, \apj, 144, 
695

\bibitem[Yee 
\& Jensen(2010)]{2010ApJ...711..303Y} Yee, J.~C., \& Jensen, E.~L.~N.\ 2010, \apj, 711, 303

\bibitem[Zahn(1992)]{1992A&A...265..115Z} Zahn, J.-P.\ 1992, \aap, 265, 115 

\bibitem[Zappala(1972)]{1972ApJ...172...57Z} Zappala, R.~R.\ 1972, \apj, 
172, 57

\end{thebibliography}
